\pdfoutput=1

\documentclass[twocolumn,epjc3]{svjour3} 
\journalname{Eur. Phys. J. C}

\RequirePackage{comment}

\def\makeheadbox{{
\hbox to0pt{\vbox{\baselineskip=10dd\hrule\hbox
to\hsize{\vrule\kern3pt\vbox{\kern3pt
\hbox{}
\hbox{}
\kern3pt}\hfil\kern3pt\vrule}\hrule}
\hss}}}

\usepackage{amsmath}
\usepackage{amssymb}
\usepackage{graphicx}
\usepackage{lmodern}
\usepackage{makecell}
\usepackage{multirow}
\usepackage{tikz}

\usepackage{listings}
\usepackage{float}
\newfloat{mycode}{tbph}{lol}
\floatname{mycode}{Listing}

\usepackage[colorlinks=true, citecolor=blue, linkcolor=blue, urlcolor=blue, pdfborder={0 0 0}, breaklinks=true]{hyperref} 

\usepackage[switch,columnwise]{lineno}

\newcommand*\patchAmsMathEnvironmentForLineno[1]{
  \expandafter\let\csname old#1\expandafter\endcsname\csname #1\endcsname
  \expandafter\let\csname oldend#1\expandafter\endcsname\csname end#1\endcsname
  \renewenvironment{#1}
     {\linenomath\csname old#1\endcsname}
     {\csname oldend#1\endcsname\endlinenomath}}
\newcommand*\patchBothAmsMathEnvironmentsForLineno[1]{
  \patchAmsMathEnvironmentForLineno{#1}
  \patchAmsMathEnvironmentForLineno{#1*}}
\AtBeginDocument{
\patchBothAmsMathEnvironmentsForLineno{equation}
\patchBothAmsMathEnvironmentsForLineno{align}
\patchBothAmsMathEnvironmentsForLineno{flalign}
\patchBothAmsMathEnvironmentsForLineno{alignat}
\patchBothAmsMathEnvironmentsForLineno{gather}
\patchBothAmsMathEnvironmentsForLineno{multline}
}

\begin{document}

\newcommand{\mgamc}{MG5aMC}
\newcommand{\as}{$\alpha_{s}$}
\newcommand{\eemumu}{\mbox{$e^+\!e^-\!\!\rightarrow\!\mu^+\!\mu^-$}}
\newcommand{\ggtt}{\mbox{$gg\!\rightarrow\! t\bar{t}$}}
\newcommand{\ggttg}{\mbox{$gg\!\rightarrow\! t\bar{t}g$}}
\newcommand{\ggttgg}{\mbox{$gg\!\rightarrow\! t\bar{t}gg$}}
\newcommand{\ggttggg}{\mbox{$gg\!\rightarrow\! t\bar{t}ggg$}}
\newcommand{\ggttgggg}{\mbox{$gg\!\rightarrow\! t\bar{t}gggg$}}
\newcommand{\ggttggggg}{\mbox{$gg\!\rightarrow\! t\bar{t}ggggg$}}
\newcommand{\twotosix}{\mbox{$2\!\rightarrow\!6$}}
\newcommand{\twotoseven}{\mbox{$2\!\rightarrow\!7$}}
\newcommand{\sk}{{\tt sigmaKin}}

\title{New GPU developments
in the Madgraph CUDACPP plugin:
kernel splitting,
helicity streams,
cuBLAS color sums}

\author{Andrea Valassi\thanksref{e1,addr1}}
\thankstext{e1}{andrea.valassi@cern.ch}
\institute{CERN, Experimental Physics Department,
CH-1211 Geneva 23, Switzerland\label{addr1}}
\journalname{Eur. Phys. J. C}

\date{
{\bfseries Version 3} -- Wednesday 10 December 2025
}

\maketitle

\begin{abstract}
The first production release
of the CUDACPP plugin for the
Madgraph5\_aMC@NLO generator,
which speeds up matrix element 
(ME) calculations
for leading-order (LO) 
processes
using a data parallel
approach on vector CPUs 
and GPUs, was delivered
in October 2024.
This was 
described
in previous publications
by the team behind that effort.
In this paper, I describe
my work on
some additional developments
and optimizations of CUDACPP,
mainly but not exclusively for GPUs.
The new approach, which
represents a major
restructuring of the CUDACPP
computational engine, 
primarily consists in splitting
the ME calculation, 
previously performed
using a single large GPU kernel,
into many 
smaller kernels.
A first batch of changes,
involving the move to 
separate ``helicity streams''
and the optional offloading
of QCD color sums to BLAS,
was recently
merged into
a new CUDACPP release,
in collaboration with 
my colleagues.
Since then, I have 
completed a second 
batch of changes,
involving 
the possibility to split 
the calculation
into 
groups of Feynman diagrams
in separate source code files.
This new feature
makes it possible to compute 
QCD matrix elements
for physics processes
with a larger number
of final state gluons:
in particular,
I present the first
performance results
from CUDACPP
for the \twotosix\ process 
\ggttgggg\ on 
CPUs and GPUs
and the \twotoseven\ process 
\ggttggggg\ on CPUs,
which involve over 15k
and 
230k Feynman
diagrams, respectively.
I also take this opportunity 
to describe 
in detail some 
previously undocumented
features of the 
CUDACPP software, 
both in the GPU
and vector CPU implementations.
\end{abstract}

\newcommand{\mctwo}[2]{\makecell{\underline{\em\normalsize #1}\\{\scriptsize #2}}}
\newcommand{\mcthr}[3]{\makecell{\underline{\em\normalsize #1}\\{\em\normalsize #2}\\{\scriptsize #3}}}

\newcommand{\figtkz}[1]{
\begin{figure}[#1]
\centering
\hspace*{-4mm}
\begin{tikzpicture}[
bdot/.style={circle, fill=blue, 
minimum size=0.5mm, inner sep=0pt},
rdot/.style={circle, fill=red, 
minimum size=0.5mm, inner sep=0pt},
gsquarednode/.style={rectangle, 
draw=green!100, fill=green!20, very thick, 
minimum width=38mm,
minimum height=10mm},
rsquarednode/.style={rectangle, 
draw=red!80, fill=red!20, very thick, 
minimum width=38mm,
minimum height=10mm},
osquarednode/.style={rectangle, 
draw=orange!80, fill=orange!20, very thick, 
minimum width=38mm,
minimum height=10mm},
]
\node[gsquarednode](ihel0) at (0,-0.8)
{\mcthr{ihel0}
{\bf cudacpp v1.00.02/3.6.3}
{\makecell{one single kernel for all helicities\\(monolithic \sk\ kernel)}}
};
\node[gsquarednode](ihel1) at (0,-2.5)
{\mctwo{ihel1}{\makecell{one kernel per helicity;\\one CUDA Stream per helicity}}};
\node[gsquarednode](ihel2) at (0,-4)
{\mctwo{ihel2}{\makecell{two kernels per helicity:\\color sum \& diagrams}}};
\node[] at (-2.7,-4.45){\makecell{QCD\\color sum}};
\node[] at (2.6,-5){\makecell{Feynman\\ diagrams}};
\node[gsquarednode](ihel3) at (-2.4,-5.75)
{\mctwo{ihel3}{\makecell{GPU color sum:\\one kernel per helicity\\OR\\one BLAS per helicity}}};
\node[rsquarednode](ihel4) at (2,-6)
{\mctwo{ihel4}{\makecell{one kernel per diagram\\in each helicity stream}}};
\node[bdot](v1v2west) at (-4.7,-6.8) {}; 
\node[bdot](v1v2east) at (4.1,-6.8) {};
\draw[dotted,thick,blue] (v1v2west) -- (v1v2east);
\node[gsquarednode](ihel3p1) at (-2.4,-8.05)
{\mcthr{ihel3p1}
{\bf cudacpp v1.01.01/3.6.4}
{\makecell{GPU color sum:\\one kernel per helicity\\OR\\one single BLAS for all helicities}}
};
\node[osquarednode](csm) at (-2.4,-10.0) 
{\mctwo{csm}
{\makecell{CPU color sum:\\FPTYPE=m\\SIMD optimizations}}
};
\node[rsquarednode](ihel4p1) at (2,-7.7)
{\mctwo{ihel4p1}{\makecell{one kernel per diagram\\in each helicity stream;\\CUDA Graph orchestration}}};
\node[rsquarednode](ihel5) at (2,-9.5)
{\mctwo{ihel5}{\makecell{one kernel per diagram group\\in each helicity stream;\\separate source code files}}};
\node[rdot](ihel5and3p1) at (0,-10.55) {};
\node[rsquarednode](ihel6p1) at (0,-12.55)
{\mctwo{ihel6p1}{\makecell{(DCDIAG=0) one kernel per diagram group\\in each helicity stream,\\exchange wavefunctions in GPU global memory,\\exchange color amplitudes in GPU global memory\\OR\\(DCDIAG=1) one single kernel for all diagram groups\\in each helicity stream,\\exchange wavefunctions in GPU registers,\\exchange color amplitudes in GPU registers}}};
\node[rsquarednode](ihel6p2) at (0,-15)
{\mctwo{ihel6p2}{\makecell{remove templates from\\helicity amplitude methods}}};
\draw[->] 
(ihel0.south) -- (ihel1.north);
\draw[->] 
(ihel1.south) -- (ihel2.north);
\draw[->] 
(ihel2.south) -- (ihel3.north);
\draw[->] 
(ihel2.south) -- (ihel4.north);
\draw[->,dashed] 
(ihel3.east) -- (ihel4.west);
\draw[->] 
(ihel3.south) -- (ihel3p1.north);
\draw[->] 
(ihel4.south) -- (ihel4p1.north);
\draw[->] 
(ihel4p1.south) -- (ihel5.north);
\draw[->] 
(ihel3p1.south) -- (csm.north);
\draw[->] 
(ihel3p1.south east) -- (ihel5and3p1);
\draw[->,dashed] 
(ihel5.south) -- (ihel5and3p1);
\draw[->] 
(ihel5and3p1) -- (ihel6p1.north);
\draw[->] 
(ihel6p1.south) -- (ihel6p2.north);
\end{tikzpicture}
\vspace*{-0.2cm}
\caption{
Schematic representation
of the architectural evolution
of matrix element calculations
in the \mgamc\ CUDACPP plugin,
in the context of the
kernel splitting developments
presented in this paper.
The steps above the blue dotted line,
up to ihel3 and ihel4 included,
have been presented in 
the first
preprint~\cite{bib:preprintV1}
of this paper, while those below it 
are only 
described here.
The steps in the green boxes
(up to ihel3p1 included)
have been merged upstream
and included in production
releases of CUDACPP;
the steps in the red boxes
(from ihel4 to ihel6p2 included)
and the step in the orange box (csm),
conversely, 
are included 
in two pull requests
that I recommend
for merging upstream.
More details about each step
are provided in the text
and in other figures.
}
\vspace*{-0.3cm}
\label{fig:tikz}
\end{figure}}

\newcommand{\figanatomy}[1]{
\begin{figure*}[#1]
\vspace*{-0.30cm}
\centering
\hspace*{0.00\textwidth}
\includegraphics[width=0.99\textwidth,clip]{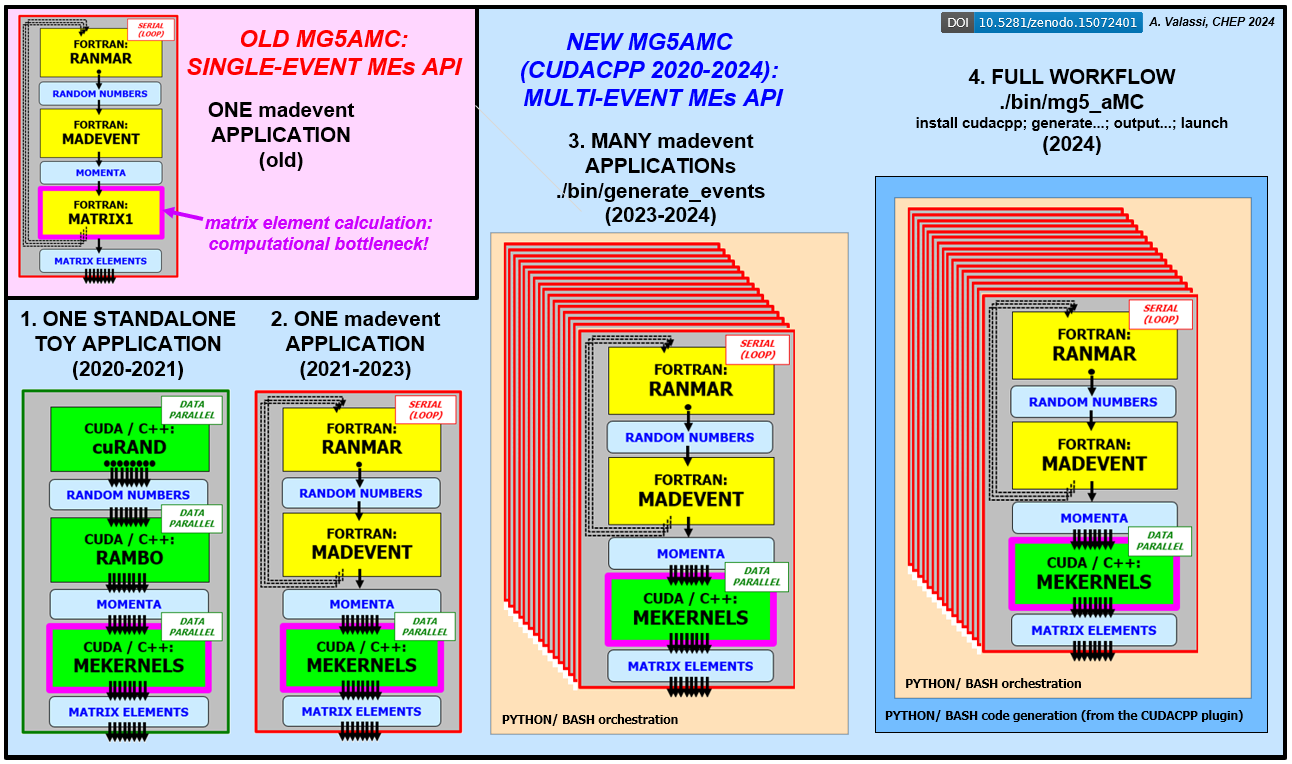}\\
\vspace*{-0.05cm}
\caption{
Schematic representation
of the architectural evolution of 
the work on 
the \mgamc\ CUDACPP plugin
between 2020 and 2024.
This plot was prepared for CHEP2024
and is taken as-is from its 
proceedings~\cite{bib:chep2024},
where additional details
can be found.
}
\vspace*{-0.25cm}
\label{fig:anatomy}
\end{figure*}
}

\newcommand{\figrddmfsimd}[1]{
\begin{figure*}[#1]
\centering
\hspace*{-0.005\textwidth}
\includegraphics[width=0.89\textwidth,clip]{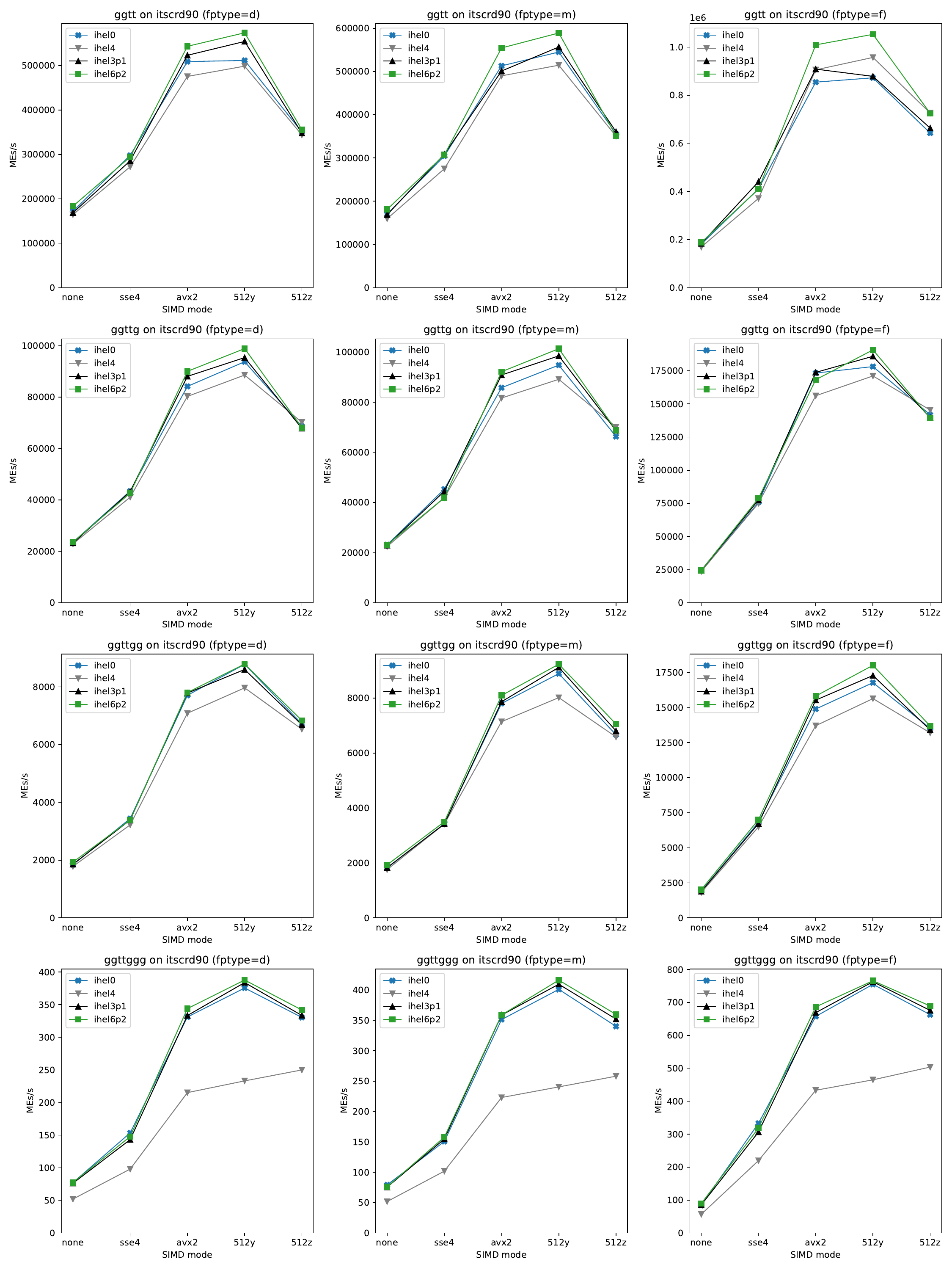}\\
\caption{
Throughputs (ME/s) 
as a function 
of CUDACPP SIMD build mode
for a single core of an Intel
Xeon Silver 4216 CPU
at CERN
(using {\tt gcc} 11.5).
Higher is better.
The 12 plots correspond to 
4 physics processes in 
3 floating point precisions.
Each plot compares
different scenarios
considered in this paper:
(ihel0)  
release v1.00.02,
before kernel splitting;
(ihel3p1) 
release v1.01.01;
(ihel4) Feynman diagrams
as individual kernels;
(ihel6p2) 
diagram groups,
using code generated with
a single diagram group.
}
\label{fig:rd90dmfsimd}
\end{figure*}
}

\newcommand{\figrddmf}[1]{
\begin{figure*}[#1]
\centering
\hspace*{-0.01\textwidth}
\includegraphics[width=0.875\textwidth,clip]{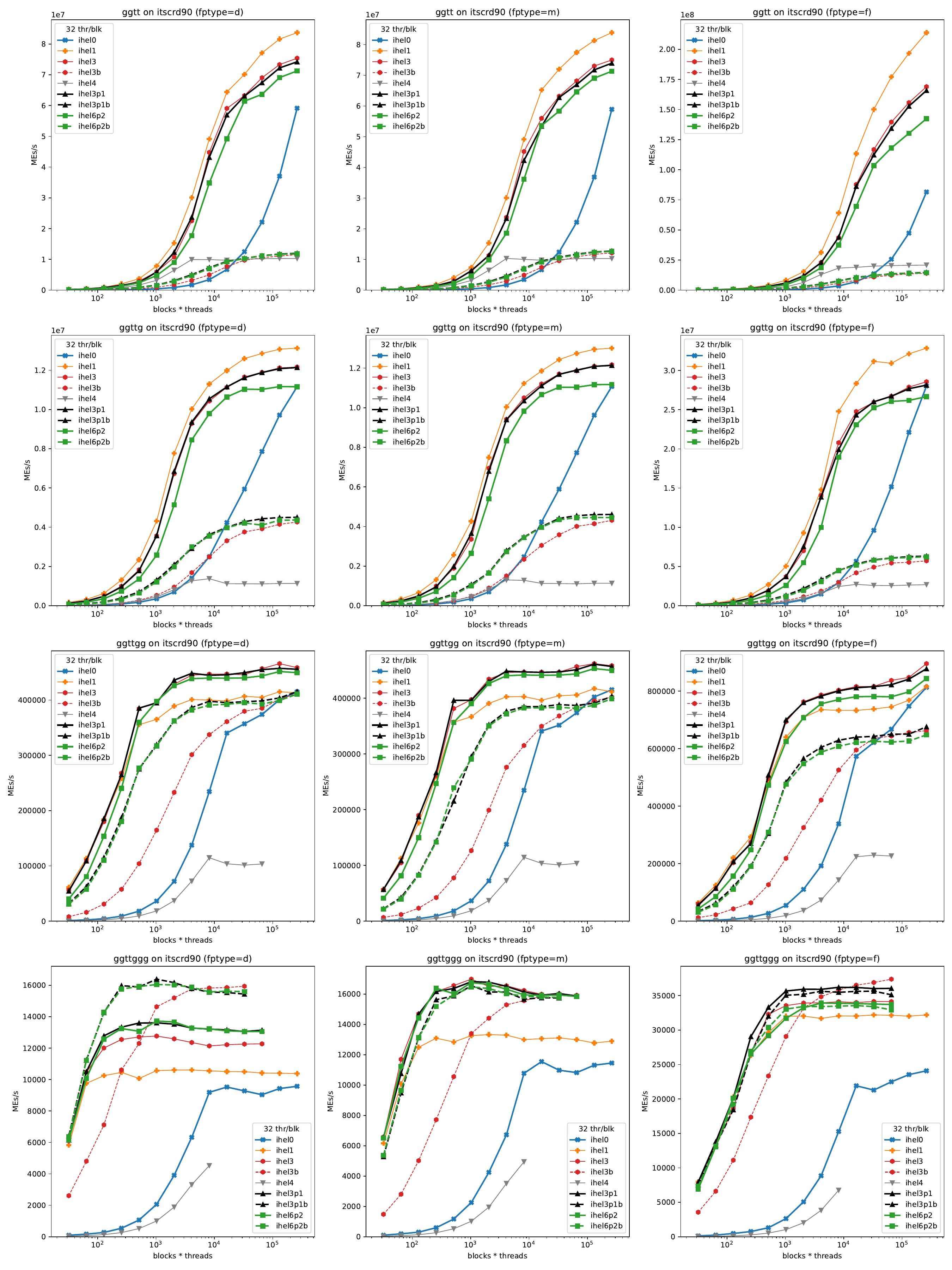}\\
\caption{
Throughputs (ME/s) 
as a function of grid size
for an NVidia V100 GPU
at CERN,
on a node equipped
with Intel
Xeon Silver 4216 CPUs.
Code built 
using CUDA 12.0 and {\tt gcc} 11.5.
Higher is better.
The 12 plots correspond to 
4 physics processes in 
3 floating point precisions.
The number of threads per block
is fixed to 32 
(NVidia GPU warp size); 
the grid size is varied
by changing the number of blocks.
Each plot compares
different scenarios
considered in this paper:
(ihel0)  
release v1.00.02,
before kernel splitting;
(ihel1) helicity streams;
(ihel3) color sum kernel;
(ihel3b) cuBLAS color sum;
(ihel3p1) 
release v1.01.01,
color sum kernel;
(ihel3b) 
release v1.01.01,
cuBLAS color sum;
(ihel4) Feynman diagrams
as individual kernels;
(ihel6p2) 
diagram groups,
color sum kernel;
(ihel6p2b) 
diagram groups,
cuBLAS color sum.
For ihel6p2
and ihel6p2b,
all four processes
were generated with 
a single diagram group,
executed 
as a kernel at runtime
({\tt DCDIAG=0}),
without graphs.
}
\label{fig:rd90dmf}
\end{figure*}
}

\newcommand{\figlumidmf}[1]{
\begin{figure*}[#1]
\centering
\hspace*{-0.02\textwidth}
\includegraphics[width=0.92\textwidth,clip]{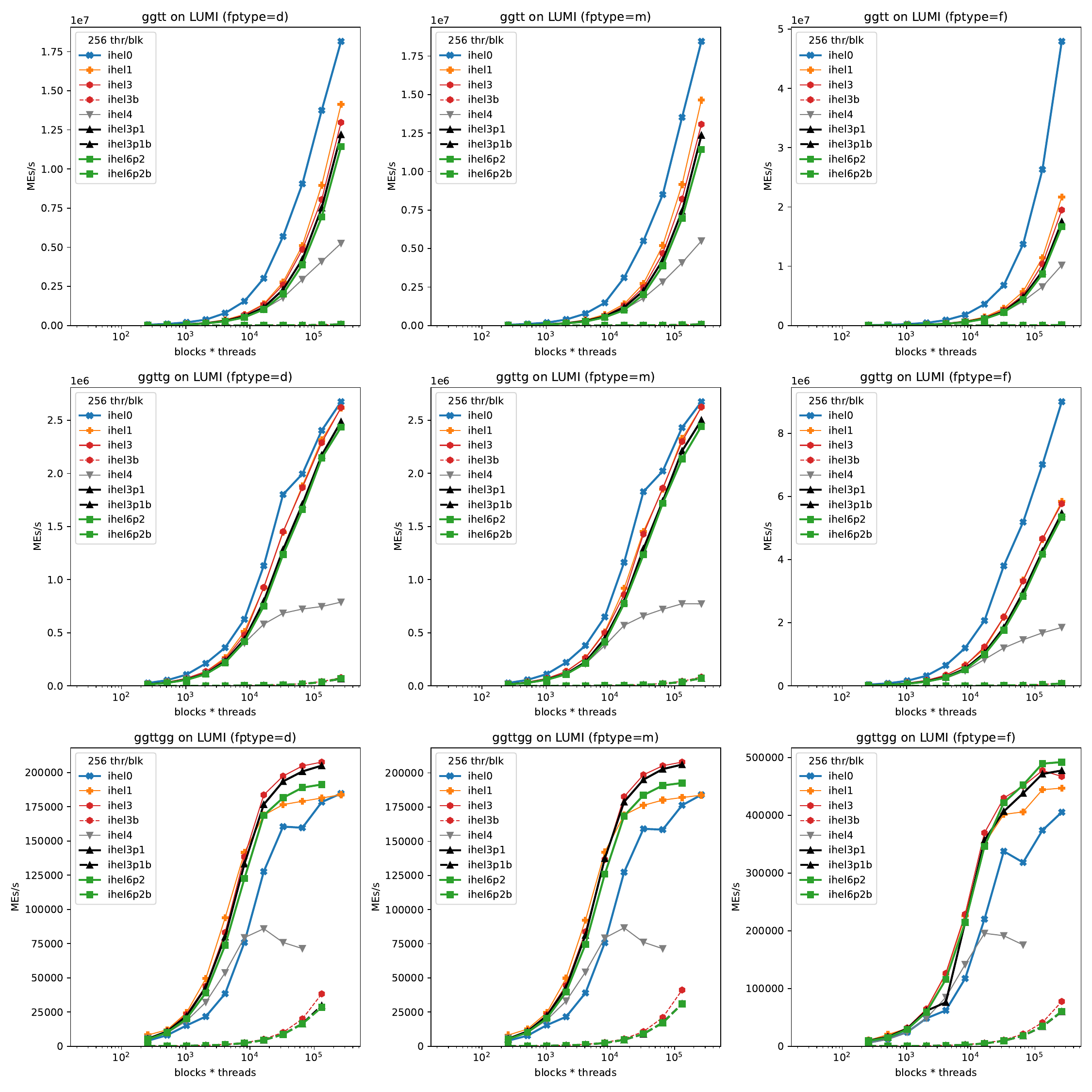}\\
\vspace*{-0.1cm}
\caption{
Throughputs (ME/s) 
as a function of grid size
for an AMD MI200 GPU 
at LUMI,
on a node equipped
with AMD EPYC 7A53 CPUs.
Code built 
using ROCm 6.0 and {\tt gcc} 13.2.
Higher is better.
The 9 plots correspond to 
3 physics processes in 
3 floating point precisions.
The number of threads per block
is fixed to 256; 
the grid size is varied
by changing the number of blocks.
Each plot compares
different scenarios
considered in this paper:
(ihel0)  
release v1.00.02,
before kernel splitting;
(ihel1) helicity streams;
(ihel3) color sum kernel;
(ihel3b) hipBLAS color sum;
(ihel3p1) 
release v1.01.01,
color sum kernel;
(ihel3b) 
release v1.01.01,
hipBLAS color sum;
(ihel4) Feynman diagrams
as individual kernels;
(ihel6p2) 
diagram groups,
color sum kernel;
(ihel6p2b) 
diagram groups,
hipBLAS color sum.
For ihel6p2
and ihel6p2b,
all four processes
were generated with 
a single diagram group,
which was executed 
as a kernel at runtime
({\tt DCDIAG=0}),
without graphs.
}
\vspace*{-0.15cm}
\label{fig:lumidmf}
\end{figure*}
}

\newcommand{\figgraphs}[1]{
\begin{figure*}[#1]
\centering
\hspace*{-0.05\textwidth}
\includegraphics[width=0.80\textwidth,clip]{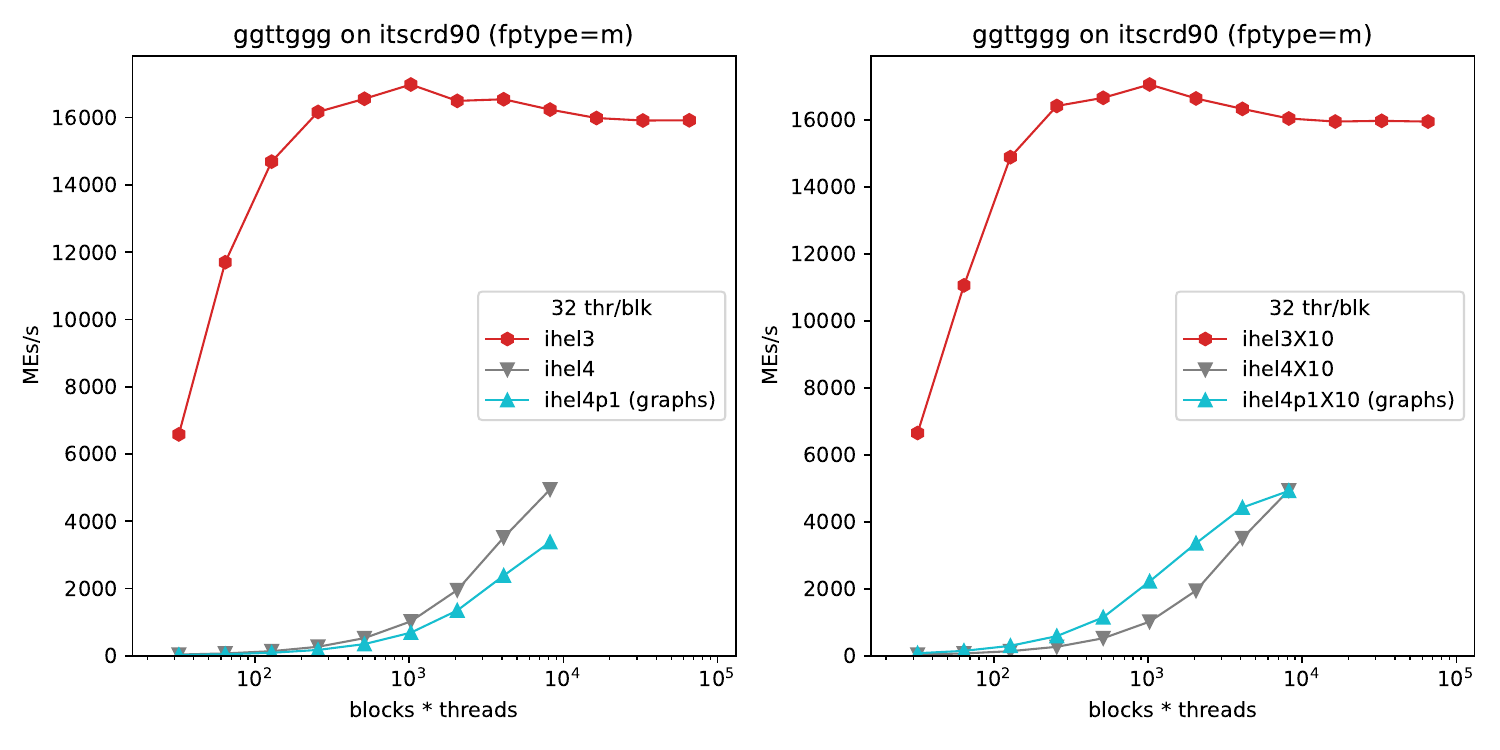}\\
\caption{
Throughputs (ME/s) 
as a function of grid size
for an NVidia V100 GPU
(using CUDA 12.0 and {\tt gcc} 11.5).
Higher is better.
The 2 plots correspond to 
throughputs derived
by executing the 
{\tt check.exe} application
through 1 (left)
or 10 (right) cycles
of a GPU grid.
The number of threads per block
is fixed to 32 
(warp size);
the grid size is varied
by changing the number of blocks.
Each plot compares
different scenarios
considered in this paper:
(ihel3) all Feynman diagrams
in a single kernel;
(ihel4) Feynman diagrams
as individual kernels,
without CUDA graphs;
(ihel4p1) Feynman diagrams
as individual kernels,
with CUDA graphs.
}
\label{fig:graphs}
\end{figure*}
}

\newcommand{\figdpgscan}[1]{
\begin{figure*}[#1]
\centering
\hspace*{-0.03\textwidth}
\includegraphics[width=0.94\textwidth,clip]{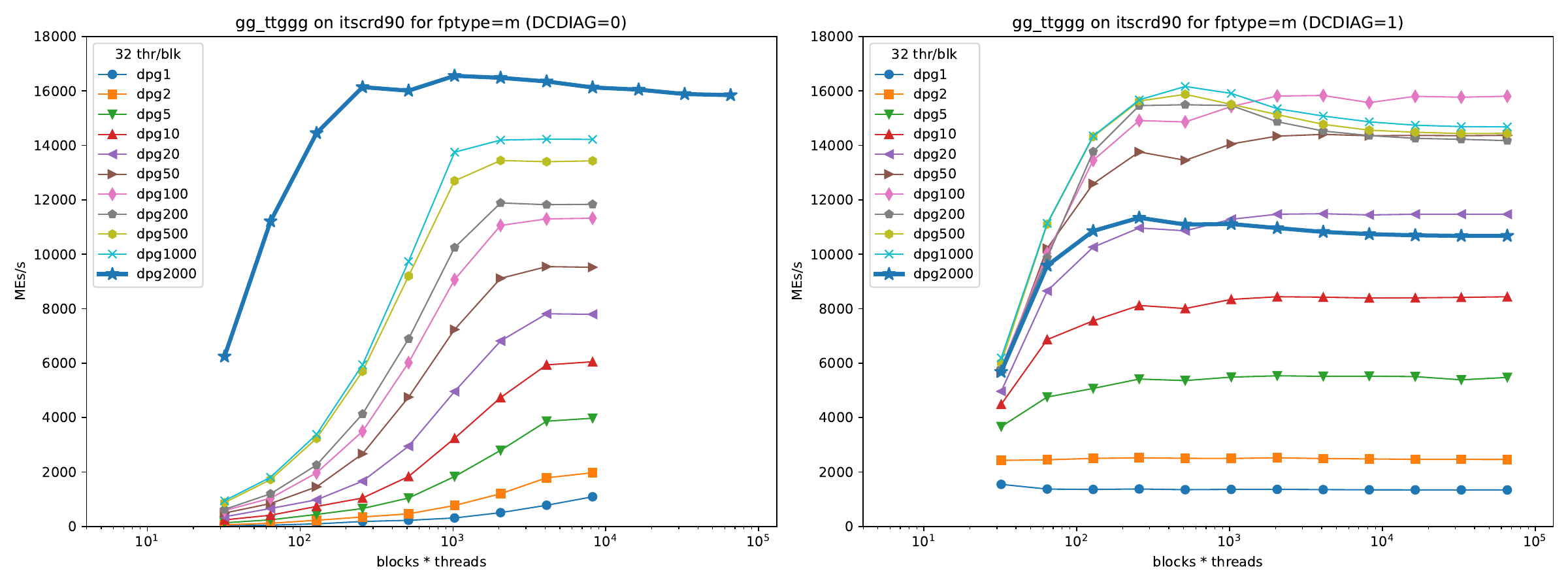}\\
\caption{
Throughputs (ME/s) 
as a function of grid size
for \ggttggg\ on 
an NVidia V100 GPU
(using CUDA 12.0 and {\tt gcc} 11.5).
Higher is better.
The number of threads per block
is fixed to 32 
(NVidia GPU warp size);
the grid size is varied
by changing the number of blocks.
Each plot compares
code generated 
with different numbers
of diagrams per group
({\tt dpg}),
ranging from 1 to 2000.
The thick solid line
({\tt dpg2000}) corresponds
to the case where
all 1240 diagrams
are in a single diagram group,
i.e. to the default 
scenario for \ggttggg\ elsewhere in 
this paper.
The 2 plots correspond 
to code built
in the {\tt DCDIAG=0}
(left: each diagram
group is a kernel)
and {\tt DCDIAG=1}
(right: there is a single 
kernel, calling all diagram
groups as device functions)
build modes.
}
\label{fig:dpgscan}
\end{figure*}
}

\newcommand{\figihelzott}[1]{
\begin{figure*}[#1]
\centering
\hspace*{-0.02\textwidth}
\includegraphics[width=0.96\textwidth,clip]{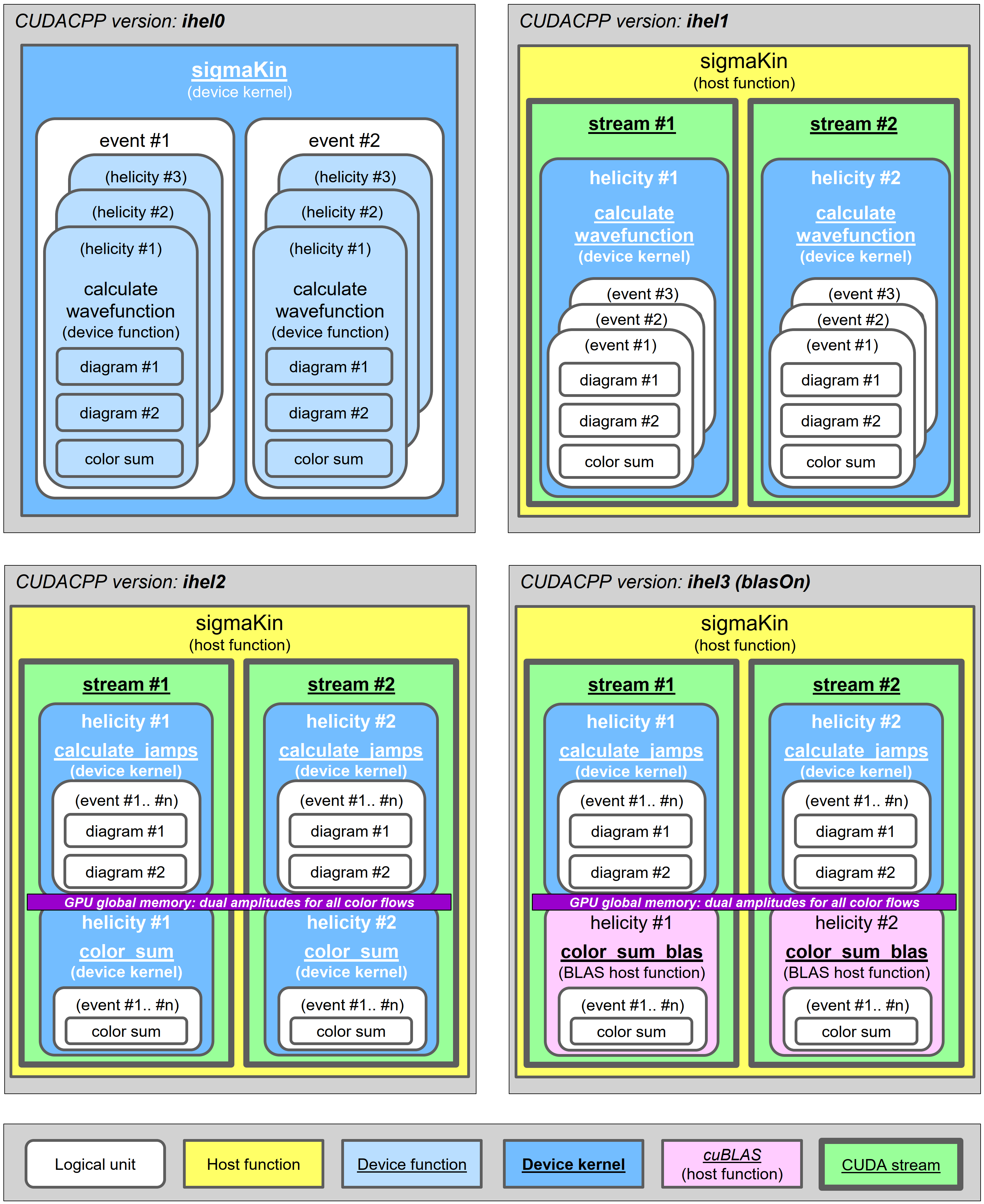}
\caption{
Schematic representation
of the CUDACPP engine
for computing MEs (\sk),
and of its evolution
through the first four
scenarios described 
in this paper:
(ihel0) current version
before kernel splitting;
(ihel1) helicity streams;
(ihel2) color sum kernel;
(ihel3b) color sum on BLAS
via host dispatcher.
For the ihel3 software,
only the (non-default) case
with BLAS enabled 
at runtime is illustrated:
by default, the ihel3 software
has BLAS disabled at runtime,
which is essentially
the same as what is shown
for the ihel2 scenario
(the only difference is
that in the ihel3 scenario
the kernel is named
{\tt color\_sum\_kernel}
and is invoked by a 
{\tt color\_sum\_gpu} host function,
which could also dispatch
the calculation 
to the {\tt color\_sum\_blas}
BLAS host function).}
\label{fig:ihel0123}
\end{figure*}
}

\newcommand{\figihelfour}[1]{
\begin{figure}[#1]
\centering
\hspace*{-0.02\textwidth}
\includegraphics[width=0.48\textwidth,clip]{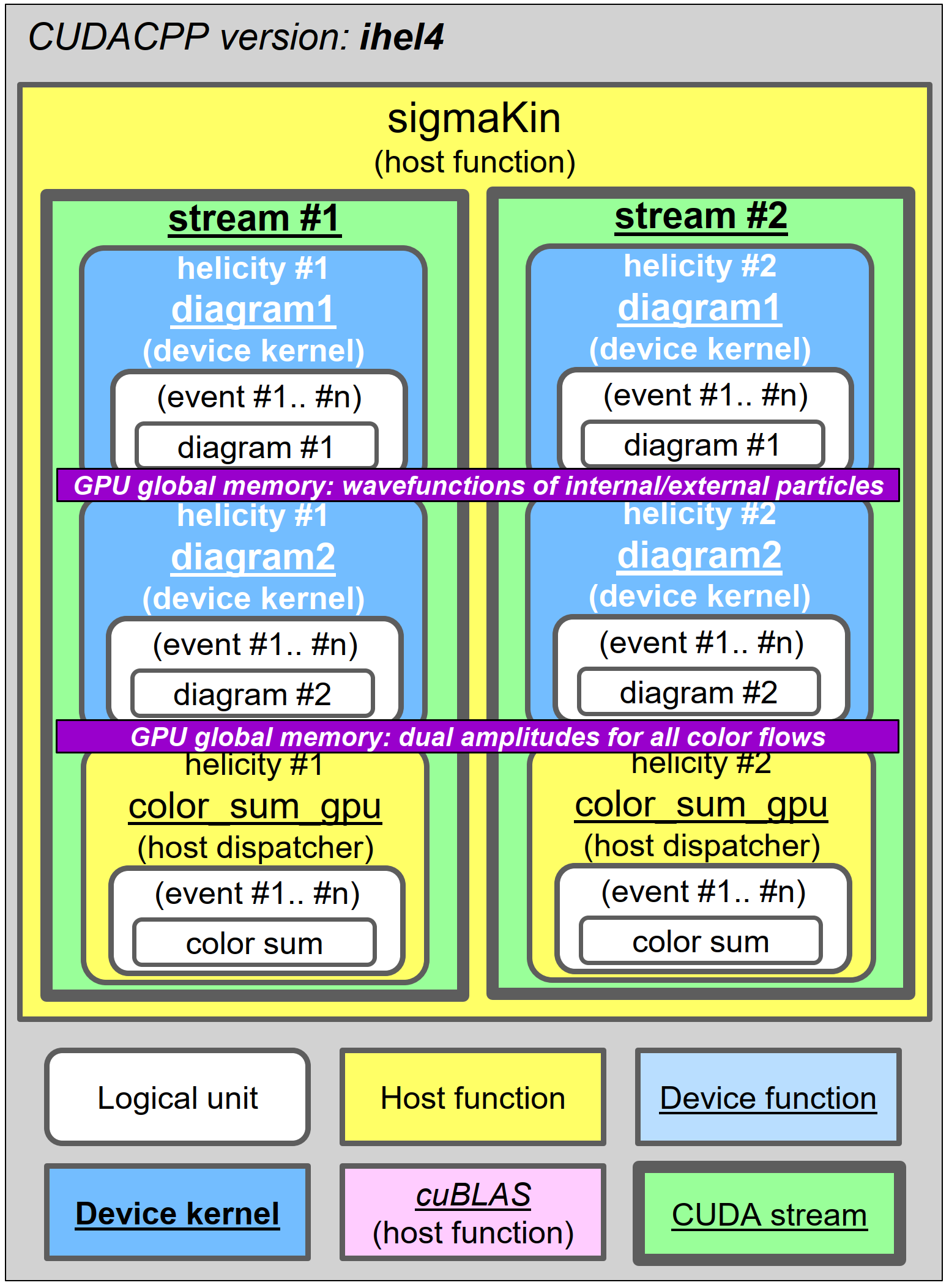}
\caption{
Schematic representation
of the CUDACPP ME engine (\sk),
in the fourth kernel splitting
scenario described in this paper:
(ihel4) Feynman diagrams
as individual kernels.
}
\label{fig:ihel4}
\end{figure}
}

\newcommand{\figihelsix}[1]{
\begin{figure*}[#1]
\vspace*{-0.1cm}
\centering
\hspace*{-0.015\textwidth}
\includegraphics[width=0.94\textwidth,clip]{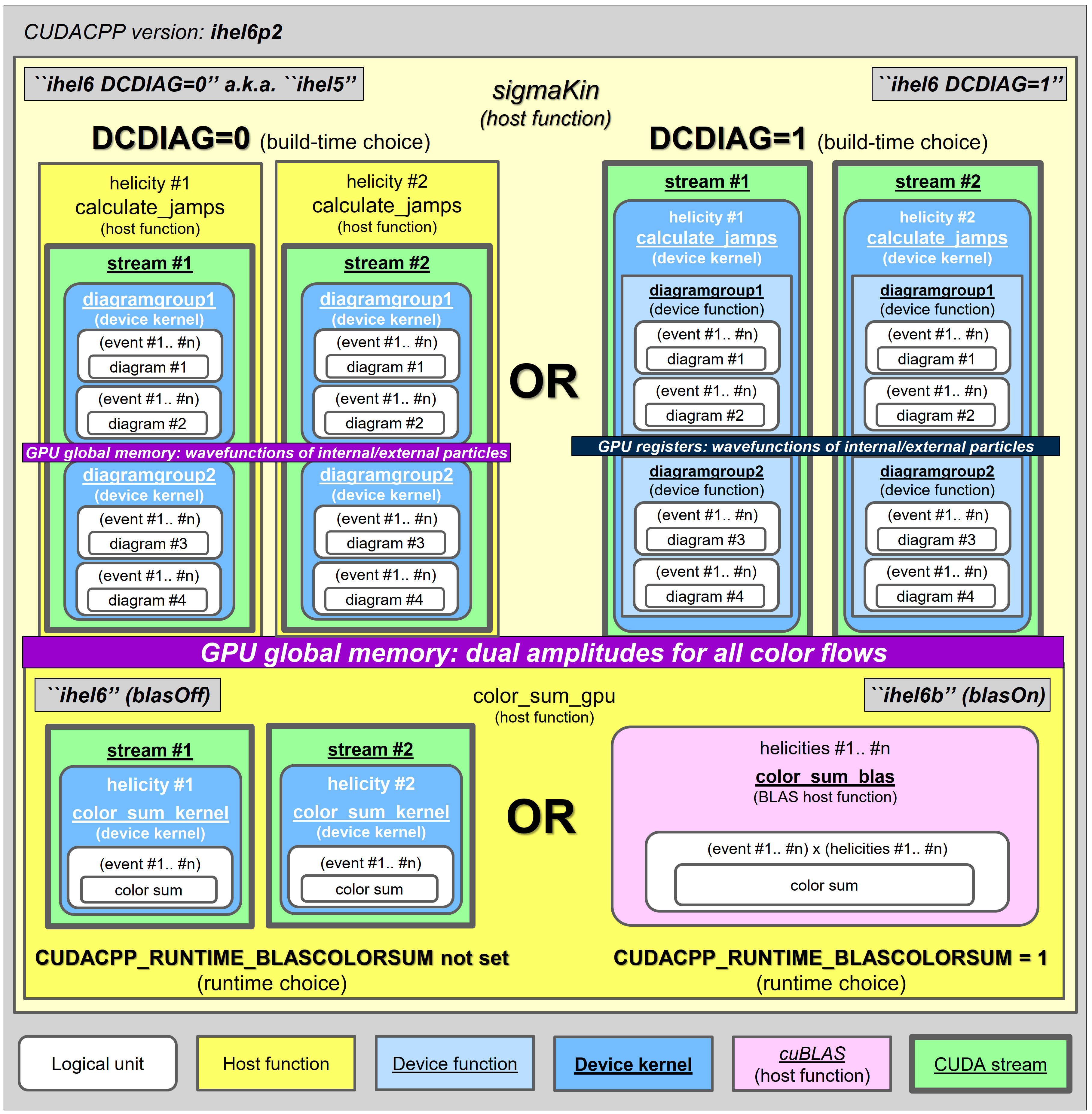}
\caption{
Schematic representation
of the CUDACPP ME engine (\sk),
in the last two kernel splitting
scenarios described in this paper:
(ihel5, and ihel6 
with {\tt DCDIAG=0})
Feynman diagram
groups as individual kernels;
(ihel6 with {\tt DCDIAG=1}) 
Feynman diagram
groups as device functions
in a single kernel.
The representation of color sums
is valid for 
all versions
from ihel3p1 onwards.
}
\vspace*{-0.3cm}
\label{fig:ihel6}
\end{figure*}
}

\newcommand{\tabcomplexity}[1]{
\begin{table*}[#1]
\hspace*{-0.8cm}
\centering
\begin{tabular}{r|l|c|c|c|c|c|c|c|}
\cline{2-8} 
(1) &
\emph{Physics process} &
\ggtt &
\ggttg &
\ggttgg &
\ggttggg &
\ggttgggg &
\ggttggggg \\
\cline{2-8}
(2) &
\emph{Feynman diagrams} & 
3 & 16 & 123 & 1240 & 15495 & 231280 \\
(3) &
\emph{SDE channels} & 
3 & 15 & 105 & 945 & N/A 
& N/A \\
(4) &
\emph{Leading colors} & 
2 & 6 & 24 & 120 & 720 & 5040 \\
(5) &
\emph{Color matrix} & 
2 x 2 & 6 x 6 & 24 x 24 & 120 x 120 & 720 x 720 & 5040 x 5040 \\
(6) &
\emph{Helicities} & 
16 & 32 & 64 & 128 & 256 & 512 \\
(7) &
\emph{Wave functions} &
5 & 12 & 26 & 121 & 750 & 5834 \\
\cline{2-8}
(8) &
\emph{Code generation codebase} & 
ihel6p2 & ihel6p2 & ihel6p2 & ihel6p2 & ihel6p2 & ihel6p1 \\
(9) &
\emph{Code generation mode} & 
mad & mad & mad & mad & sa & sa \\
(10) &
\emph{Code generation d.p.g.} & 
any & any & any & any & $\leq$100 & $\leq$1000 \\
(11) &
\emph{Code generation d.p.f.} & 
any & any & any & any & $\geq$100 & $\geq$1000 \\
(12) &
\emph{Code generation time} & 
2s & 2s & 4s & 32s & 13m & 30h \\
(13) &
\emph{Feynman diagram groups} & 
1 & 1 & 1 & 1 & 155 & 232 \\
(14) &
\emph{Feynman diagram files} & 
1 & 1 & 1 & 1 & 155 & 232 \\
(15) &
\emph{LOC (color\_sum.cc)} &
438 & 442 & 460 & 556 & 1156 & 5.4k \\
(16) &
\emph{LOC (diagramsXYZ.cc)} &
135 & 307 & 1986 & 27k & 652k & 18.7M \\
(17) &
\emph{Size (color\_sum.cc)} &
24 kB & 24 kB & 28 kB & 92 kB & 2.6 MB & 137 MB \\
(18) &
\emph{Size (diagramsXYZ.cc)} &
8 kB & 16 kB & 84 kB & 1.0 MB & 28.2 MB & 721 MB \\
\cline{2-8}
(19) &
\emph{Build time (five C++ modes)} &
6s & 6s & 6s & 23s & 3m & 55m \\
(20) &
\emph{Build time (CUDA {\tt DCDIAG=0})} &
11s & 11s & 14s & 6m & 9m & failed \\
(21) &
\emph{Build time (CUDA {\tt DCDIAG=1})} &
13s & 13s & 15s & 3m & 6m & failed \\
\cline{2-8}
\end{tabular}
\caption{Comparison 
of the computational complexity 
of the six physics processes
considered in the tests
described in this paper.
The rows represent 
the following:
(1) the physics process;
(2) the number 
of Feynman diagrams;
(3) the number 
of distinct SDE channels,
i.e. of the diagrams
used in the MadEvent
single diagram 
enhancement algorithm;
(4) the number of leading
QCD color flows;
(5)~the size 
of the color matrix;
(6) the number 
of combinations
of helicities 
for initial and 
final state particles
(for 
these six processes,
this coincides 
with the number
of good helicities
with non-zero
contributions to MEs);
(7)~the number 
of wavefunctions
to be computed
for all external
(initial and final state)
and internal
(propagator) particles;
(8) the codebase used
for code generation;
(9) the code generation mode,
``mad'' for full generation
including Fortran {\tt madevent}
and ``sa'' for standalone only;
(9-10) the maximum number
of diagrams per group ({\tt dpg})
and the minimum number
of diagrams per file ({\tt dpf})
during code generation;
(12) the code generation time
on a typical CPU;
(13-14) the resulting
numbers of Feynman 
diagram groups
and source code files;
(15-17) the total 
number of lines of code
and 
total size
in the CUDAPP generated
{\tt color\_sum.cc}
and the sum of all
{\tt diagramsXYZ.cc} files;
(19-21) the time 
for a parallel ({\tt -j})
build of 
the software 
with a cold cache 
({\tt CCACHE\_RECACHE=1}),
either for all five C++ modes
on an Intel Xeon Gold 6326, 
or for CUDA in each of two 
{\tt DCDIAG} modes
on an AMD EPYC 7313. 
All of these numbers
are needed to describe 
the complexity of
the ME calculation
in a single standalone
or {\tt madevent}
application,
which is the main focus
of this paper;
the number of SDE channels,
in addition,
is one of the factors
that determines how many
instances of a
{\tt madevent} application
must be launched
in a full event
generation workflow.
The number of helicities
determines how many 
GPU streams are used,
as of the ihel1 software.
The number of leading colors
determines the number
of dual amplitudes
that are computed 
using Feynman diagrams
and used as input 
to the color sum;
as of ihel2,
these are stored into 
and retrieved from
GPU global memory.
The size of the color matrix
determines the complexity
of the color sum
calculation
(addressed with 
a separate GPU kernel
as of ihel2,
and optionally
via BLAS 
as of ihel3).
The number 
of wavefunctions
is particularly relevant 
to the ihel4-ihel6 
software versions,
which handle their
storage and retrieval
in GPU registers
and/or GPU global memory
in subtly different ways.
}
\label{tab:complexity}
\vspace*{-5mm}
\end{table*}
}

\newcommand{\tabggttthreegcpuCS}[1]{
\begin{table*}[#1]
\centering
\begin{tabular}{|l|l|c|c|c|c|c|c|}
\cline{2-8}
\multicolumn{1}{c|}{}
& \multicolumn{7}{c|}{\ggttggg\ --- current code (master/ihel3p1)}\\
\cline{2-8}
\multicolumn{1}{c|}{}
& \emph{C++ SIMD build mode} &
{\tt \makecell{none\\novec-cs}} &
{\tt none} & {\tt sse4} & {\tt avx2} & {\tt 512y} & {\tt 512z} \\
\hline
\multirow{3}{*}{FPTYPE=d}
&\emph{Time to compute 128 MEs} & 
--- & 1.293s & 0.680s &
0.305s & 0.268s & 0.152s \\
&\emph{$\rightarrow$ Time in {\tt calculate\_jamps}} &
--- & 1.233s & 0.626s &
0.280s & 0.243s & 0.138s\\
&\emph{$\rightarrow$ Time in {\tt color\_sum\_cpu}} &
--- & 0.061s & 0.054s &
0.025s & 0.025s & 0.013s \\
\hline
\multirow{3}{*}{FPTYPE=m}
&\emph{Time to compute 128 MEs} & 
--- & 1.308s & 0.646s &
0.289s & 0.252s & 0.143s \\
&\emph{$\rightarrow$ Time in {\tt calculate\_jamps}} &
--- & 1.232s & 0.616s &
0.275s & 0.238s & 0.136s\\
&\emph{$\rightarrow$ Time in {\tt color\_sum\_cpu}} &
--- & 0.076s & 0.031s &
0.014s & 0.014s & 0.008s \\
\hline
\multirow{3}{*}{FPTYPE=f}
&\emph{Time to compute 128 MEs} & 
--- & 1.242s & 0.304s &
0.152s & 0.133s & 0.075s \\
&\emph{$\rightarrow$ Time in {\tt calculate\_jamps}} &
--- & 1.206s & 0.276s &
0.139s & 0.121s & 0.069s\\
&\emph{$\rightarrow$ Time in {\tt color\_sum\_cpu}} &
--- & 0.037s & 0.028s &
0.012s & 0.013s & 0.006s \\
\hline
\multicolumn{8}{c}{}\\
\cline{2-8}
\multicolumn{1}{c|}{}
& \multicolumn{7}{c|}{\ggttggg\ --- new code (csm)}\\
\cline{2-8}
\multicolumn{1}{c|}{}
& \emph{C++ SIMD build mode} &
{\tt \makecell{none\\novec-cs}} &
{\tt none} & {\tt sse4} & {\tt avx2} & {\tt 512y} & {\tt 512z} \\
\hline
\multirow{3}{*}{FPTYPE=d}
&\emph{Time to compute 128 MEs} & 
1.336s & 1.286s & 0.680s &
0.305s & 0.267s & 0.152s \\
&\emph{$\rightarrow$ Time in {\tt calculate\_jamps}} &
1.229s & 1.230s & 0.626s &
0.280s & 0.242s & 0.139s\\
&\emph{$\rightarrow$ Time in {\tt color\_sum\_cpu}} &
0.108s & 0.056s & 0.054s &
0.025s & 0.025s & 0.013s \\
\hline
\multirow{3}{*}{FPTYPE=m}
&\emph{Time to compute 128 MEs} & 
1.340s & 1.264s & 0.647s &
0.288s & 0.251s & 0.144s \\
&\emph{$\rightarrow$ Time in {\tt calculate\_jamps}} &
1.232s & 1.230s & 0.617s &
0.275s & 0.238s & 0.137s\\
&\emph{$\rightarrow$ Time in {\tt color\_sum\_cpu}} &
0.107s & 0.034s & 0.029s &
0.014s & 0.014s & 0.007s \\
\hline
\multirow{3}{*}{FPTYPE=f}
&\emph{Time to compute 128 MEs} & 
1.313s & 1.240s & 0.305s &
0.152s & 0.133s & 0.076s \\
&\emph{$\rightarrow$ Time in {\tt calculate\_jamps}} &
1.206s & 1.207s & 0.276s &
0.139s & 0.121s & 0.069s\\
&\emph{$\rightarrow$ Time in {\tt color\_sum\_cpu}} &
0.107s & 0.033s & 0.028s &
0.013s & 0.013s & 0.006s \\
\hline
\end{tabular}
\vspace*{1mm}
\caption{
Profiling of Feynman diagrams
and color sums 
for \ggttggg\ on CPU for 
different SIMD modes
and 
floating point precisions,
in the current version 
(top: ihel3p1)
and an improved version 
(bottom: csm)
of CUDACPP.
The code is
instrumented with 
{\tt rdtsc} timers
for color sum profiling.
Results for a single core
of an Intel Xeon Gold 6326 CPU
at CERN
(using {\tt gcc} 11.5).
}
\label{tab:ggtt3gcpuCS}
\vspace*{-3mm}
\end{table*}
}

\newcommand{\tabggttfourgcpuCS}[1]{
\begin{table*}[#1]
\centering
\begin{tabular}{|l|c|c|c|c|c|}
\hline
\multicolumn{6}{|c|}{\ggttgggg\ with 
155 diagram groups 
(100 {\tt dpg})}\\
\hline
\emph{C++ SIMD build mode} &
{\tt none} & {\tt sse4} & {\tt avx2} & {\tt 512y} & {\tt 512z} \\
\hline
\emph{Time to compute 32 MEs}
\hspace*{1.5cm} &
13.8s & 8.91s & 3.26s & 3.21s & 2.18s \\
\emph{$\rightarrow$ Time in {\tt calculate\_jamps}} &
12.5s (90.1\%) & 8.24s (92.6\%) & 2.95s (90.3\%) & 2.89s (90.2\%) & 2.01s (92.2\%)\\
\emph{$\rightarrow$ Time in {\tt color\_sum\_cpu}} &
1.36s (9.8\%) & 0.66s (7.4\%) & 0.32s (9.7\%) & 0.32s (9.8\%) & 0.17s (7.8\%) \\
\emph{Throughput} & 
2.31/s & 3.59/s & 9.81/s  & 9.97/s & 14.7/s \\
\emph{Mean ME value / E-10} & 
3.057124 & 3.057124 & 3.057124 & 3.057124 & 3.057124 \\
\hline
\multicolumn{6}{c}{}\\
\hline
\multicolumn{6}{|c|}{\ggttgggg\ with 
16 diagram groups 
(1000 {\tt dpg})}\\
\hline
\emph{C++ SIMD build mode} &
{\tt none} & {\tt sse4} & {\tt avx2} & {\tt 512y} & {\tt 512z} \\
\hline
\emph{Time to compute 32 MEs} & 
13.3s & 6.08s & 2.55s & 2.46s & 1.43s \\
\emph{$\rightarrow$ Time in {\tt calculate\_jamps}} &
12.0s (89.8\%) & 5.46s (89.9\%) & 2.23s (87.7\%) & 2.14s (87.2\%) & 1.27s (88.7\%) \\
\emph{$\rightarrow$ Time in {\tt color\_sum\_cpu}} &
1.36s (10.2\%) & 0.62s (10.1\%) & 0.31s (12.2\%) & 0.31s (12.7\%) & 0.16s (11.3\%) \\
\emph{Throughput} & 
2.40/s & 5.27/s & 12.6/s  & 13.0/s & 22.4/s \\
\emph{Mean ME value / E-10} & 
3.057124 & 3.057124 & 3.057124 & 3.057124 & 3.057124 \\
\hline
\end{tabular}
\vspace*{1mm}
\caption{
Profiling of Feynman diagrams
and color sums 
for \ggttgggg\ on CPU in
different SIMD modes
and with different
numbers of diagrams per group
({\tt FPTYPE=m}).
C++ code generated
using a modified version of CUDACPP,
instrumented with {\tt rdtsc} timers
(hack\_ihel6p2\_ggtt4g branch).
Results for a single core
of an Intel Xeon Gold 6326 CPU
(using {\tt gcc} 11.5).
The tests are executed
with {\tt CUDACPP\_RUNTIME\_GOODHELICITIES=ALL} to bypass helicity filtering.
The average ME value
shown in the table,
computed from the same
random numbers in all cases,
is used as a basic validation 
of physics results.
}
\label{tab:ggtt4gcpuCS}
\vspace*{-4mm}
\end{table*}
}

\newcommand{\tabggttfourggpuCS}[1]{
\begin{table*}[#1]
\centering
\begin{tabular}{|l|c|c|c|}
\hline
\multicolumn{4}{|c|}{\ggttgggg\ with 1000 {\tt dpg} --- 
(1) diagrams: DCDIAG=0 ---
(2) color sum: kernel}\\
\hline
\emph{MEs per cycle (blocks * threads)} &
32 (1*32) & 128 (4*32) & 512 (16*32) \\
\hline
\emph{Time to compute one ME cycle}
\hspace*{1.5cm} &
5.16s & 5.20s & 5.66s \\
\emph{$\rightarrow$ Time in {\tt calculate\_jamps}} &
5.02s (97.3\%) & 5.09s (95.8\%) & 5.25s (92.7\%) \\
\emph{$\rightarrow$ Time in {\tt color\_sum\_gpu}} &
0.14s (2.7\%) & 0.22s (4.2\%) & 0.41s (7.3\%) \\
\emph{Throughput} & 
6.20/s & 24.1/s & 90.5/s \\
\emph{Mean ME value / E-10} & 
3.057124 & 5.683925 & 264.1109 \\
\hline
\multicolumn{4}{c}{}\\
\hline
\multicolumn{4}{|c|}{\ggttgggg\ with 100 {\tt dpg} --- 
(1) diagrams: DCDIAG=0 ---
(2) color sum: kernel}\\
\hline
\emph{MEs per cycle (blocks * threads)} &
32 (1*32) & 128 (4*32) & 512 (16*32) \\
\hline
\emph{Time to compute one ME cycle} &
9.54s & 11.8s & 12.4s \\
\emph{$\rightarrow$ Time in {\tt calculate\_jamps}} &
9.40s (98.6\%) & 11.6s (98.1\%) & 12.0s (96.7\%) \\
\emph{$\rightarrow$ Time in {\tt color\_sum\_gpu}} &
0.14s (1.4\%) & 0.22s (1.9\%) & 0.42s (3.3\%) \\
\emph{Throughput} & 
3.35/s & 10.9/s & --/s \\
\emph{Mean ME value / E-10} & 
3.057124 & 5.683925 & 264.1109 \\
\hline
\multicolumn{4}{c}{}\\
\hline
\multicolumn{4}{|c|}{\ggttgggg\ with 100 {\tt dpg} --- 
(1) diagrams: DCDIAG=1 ---
(2) color sum: kernel}\\
\hline
\emph{MEs per cycle (blocks * threads)} &
32 (1*32) & 128 (4*32) & 512 (16*32) \\
\hline
\emph{Time to compute one ME cycle} &
0.48s & 0.76s & 2.14s \\
\emph{$\rightarrow$ Time in {\tt calculate\_jamps}} &
0.32s (68.1\%) & 0.50s (66.1\%) & 1.71s (80.1\%) \\
\emph{$\rightarrow$ Time in {\tt color\_sum\_gpu}} &
0.15s (31.8\%) & 0.26s (33.9\%) & 0.43s (19.9\%) \\
\emph{Throughput} & 
66.1/s & 167.6/s & 239.7/s \\
\emph{Mean ME value / E-10} & 
3.057124 & 5.683925 & 264.1109 \\
\hline
\multicolumn{4}{c}{}\\
\hline
\multicolumn{4}{|c|}{\ggttgggg\ with 100 {\tt dpg} --- 
(1) diagrams: DCDIAG=1 ---
(2) color sum: cuBLAS}\\
\hline
\emph{MEs per cycle (blocks * threads)} &
32 (1*32) & 128 (4*32) & 512 (16*32) \\
\hline
\emph{Time to compute one ME cycle} &
0.34s & 0.54s & 1.76s \\
\emph{$\rightarrow$ Time in {\tt calculate\_jamps}} &
0.31s (90.9\%) & 0.50s (93.9\%) & 1.71s (97.0\%) \\
\emph{$\rightarrow$ Time in {\tt color\_sum\_gpu}} &
0.031s (9.1\%) & 0.032s (6.1\%) & 0.054s (3.0\%) \\
\emph{Throughput} & 
92.8/s & 238.7/s & 290.6/s \\
\emph{Mean ME value / E-10} & 
3.057124 & 5.683925 & 264.1109 \\
\hline
\multicolumn{4}{c}{}\\
\hline
\multicolumn{4}{|c|}{\ggttgggg\ with 100 {\tt dpg} --- 
(1) diagrams: DCDIAG=1 ---
(2) color sum: cuBLAS/TF32}\\
\hline
\emph{MEs per cycle (blocks * threads)} &
32 (1*32) & 128 (4*32) & 512 (16*32) \\
\hline
\emph{Time to compute one ME cycle} &
0.35s & 0.54s & 1.75s \\
\emph{$\rightarrow$ Time in {\tt calculate\_jamps}} &
0.32s (91.4\%) & 0.51s (94.3\%) & 1.71s (97.7\%) \\
\emph{$\rightarrow$ Time in {\tt color\_sum\_gpu}} &
0.030s (8.5\%) & 0.031s (5.7\%) & 0.039s (2.3\%) \\
\emph{Throughput} & 
91.5/s & 237.1/s & 293.4/s \\
\emph{Mean ME value / E-10} & 
{\em 3.057422} & {\em 5.684374} & {\em 264.1716} \\
\hline
\end{tabular}
\vspace*{1mm}
\caption{
Profiling of Feynman diagrams
and color sums 
for \ggttgggg\ on GPU in
different {\tt DCDIAG} modes
and with different
numbers of diagrams per group
({\tt FPTYPE=m}).
CUDA/C++ code generated
using a modified version of CUDACPP,
instrumented with {\tt rdtsc} timers
(hack\_ihel6p2\_ggtt4g branch).
Results for an NVidia A100 GPU
(using CUDA 12.4 and {\tt gcc} 11.5).
The tests are executed
with {\tt CUDACPP\_RUNTIME\_GOODHELICITIES=ALL} to bypass helicity filtering.
The average ME value
shown in the table,
computed from the same
random numbers in all cases,
is used as a basic validation 
of physics results.
}
\label{tab:ggtt4ggpuCS}
\vspace*{-3mm}
\end{table*}
}

\newcommand{\tabggttfourgcpu}[1]{
\begin{table*}[#1]
\hspace*{-0.5cm}
\centering
\begin{tabular}{|l|c|c|c|c|c|}
\hline
\multicolumn{6}{|c|}{\ggttgggg\ with 
15495 diagram groups 
(1 {\tt dpg}, 100 {\tt dpf})
--- build takes 2m20s}\\
\hline
\emph{C++ SIMD build mode} &
{\tt none} & {\tt sse4} & {\tt avx2} & {\tt 512y} & {\tt 512z} \\
\hline
\emph{Time to compute 16 MEs}
\hspace*{1.5cm} &
33.9s & 37.8s & 32.8s & 31.6s & 30.3s \\
\emph{Throughput} & 
0.47/s & 0.42/s & 0.49/s  & 0.51/s & 0.53/s \\
\emph{Speedup w.r.t. none} & 
x1 & x0.90 & x1.03 & x1.07 & x1.12 \\
\hline
\multicolumn{6}{c}{}\\
\hline
\multicolumn{6}{|c|}{\ggttgggg\ with 
1550 diagram groups 
(10 {\tt dpg}, 100 {\tt dpf})
--- build takes 3m10s}\\
\hline
\emph{C++ SIMD build mode} &
{\tt none} & {\tt sse4} & {\tt avx2} & {\tt 512y} & {\tt 512z} \\
\hline
\emph{Time to compute 16 MEs} & 
9.57s & 10.27s & 4.59s & 4.89s & 3.83s \\
\emph{Throughput} & 
1.67/s & 1.56/s & 3.48/s  & 3.27/s & 4.18/s \\
\emph{Speedup w.r.t. none} & 
x1 & x0.93 & x2.08 & x1.96 & x2.50 \\
\hline
\multicolumn{6}{c}{}\\
\hline
\multicolumn{6}{|c|}{\ggttgggg\ with 
155 diagram groups 
(100 {\tt dpg}, 100 {\tt dpf})
--- build takes 3m}\\
\hline
\emph{C++ SIMD mode} &
{\tt none} & {\tt sse4} & {\tt avx2} & {\tt 512y} & {\tt 512z} \\
\hline
\emph{Time to compute 16 MEs} & 
7.18s & 4.52s & 1.73s & 1.53s & 1.05s \\
\emph{Throughput} & 
2.23/s & 3.53/s & 9.22/s  & 10.0/s & 15.2/s \\
\emph{Speedup w.r.t. none} & 
x1 & x1.59 & x4.14 & x4.69 & x6.83 \\
\hline
\multicolumn{6}{c}{}\\
\hline
\multicolumn{6}{|c|}{\ggttgggg\ with 
78 diagram groups 
(200 {\tt dpg}, 200 {\tt dpf})
--- build takes 3m10s}\\
\hline
\emph{C++ SIMD mode} &
{\tt none} & {\tt sse4} & {\tt avx2} & {\tt 512y} & {\tt 512z} \\
\hline
\emph{Time to compute 16 MEs} & 
6.30s & 3.46s & 1.45s & 1.40s & 0.89s \\
\emph{Throughput} & 
2.54/s & 4.63/s & 11.0/s  & 11.4/s & 18.0/s \\
\emph{Speedup w.r.t. none} & 
x1 & x1.82 & x4.33 & x4.50 & x7.08 \\
\hline
\multicolumn{6}{c}{}\\
\hline
\multicolumn{6}{|c|}{\ggttgggg\ with 
16 diagram groups 
(1000 {\tt dpg}, 1000 {\tt dpf}) 
--- build takes 3m40s}\\
\hline
\emph{C++ SIMD build mode} &
{\tt none} & {\tt sse4} & {\tt avx2} & {\tt 512y} & {\tt 512z} \\
\hline
\emph{Time to compute 16 MEs} & 
6.23s & 3.08s & 1.29s & 1.22s & 0.72s \\
\emph{Throughput} & 
2.57/s & 5.19/s & 12.4/s  & 13.1/s & 22.2/s \\
\emph{Speedup w.r.t. none} & 
x1 & x2.02 & x4.81 & x5.11 & x8.65 \\
\hline
\multicolumn{6}{c}{}\\
\hline
\multicolumn{6}{|c|}{\ggttgggg\ with 
2 diagram groups 
(10000 {\tt dpg}, 10000 {\tt dpf})
--- build takes 28m}\\
\hline
\emph{C++ SIMD build mode} &
{\tt none} & {\tt sse4} & {\tt avx2} & {\tt 512y} & {\tt 512z} \\
\hline
\emph{Time to compute 16 MEs} & 
6.05s & 2.93s & 1.22s & 1.17s & 0.70s \\
\emph{Throughput} & 
2.64/s & 5.45/s & 13.1/s  & 13.7/s & 22.9/s \\
\emph{Speedup w.r.t. none} & 
x1 & x2.06 & x4.97 & x5.18 & x8.66 \\
\hline
\multicolumn{6}{c}{}\\
\hline
\multicolumn{6}{|c|}{\ggttgggg\ with 
1 diagram group 
(100000 {\tt dpg}, 100000 {\tt dpf}) --- build FAILS}\\
\hline
\multicolumn{6}{|c|}
{{\tt g++: internal compiler error: Segmentation fault}}\\
\multicolumn{6}{|c|}
{{\tt Please submit a full bug report,
with preprocessed source if appropriate}}\\
\hline
\end{tabular}
\vspace*{2.5mm}
\caption{
C++ run time performance
of \ggttgggg\ in
different SIMD modes
and with different
numbers of diagrams per group
({\tt FPTYPE=m}).
C++ code generated
using CUDACPP hack\_ihel6p2.
Results for a single core
of an Intel Xeon Gold 6326 CPU
(using {\tt gcc} 11.5).
The tests are executed
with {\tt CUDACPP\_RUNTIME\_GOODHELICITIES=ALL} to bypass 
helicity filtering, 
which is still 
performed
sequentially 
one helicity at a time.
As a basic validation 
of physics results,
it was only checked 
that all calculations
return exactly 
the same average ME value,
when starting from the
same random number seeds,
for all SIMD and
diagram-per-group modes.
}
\label{tab:ggtt4gcpu}
\vspace*{-4mm}
\end{table*}
}

\newcommand{\tabggttfourggpu}[1]{
\begin{table*}[#1]
\vspace*{-0.05cm}
\hspace*{-0.5cm}
\centering
\begin{tabular}{|l|c|c|c|c|c|c|c|c|c|}
\hline
\multicolumn{10}{|c|}{\ggttgggg\ with 
15495 diagram groups 
(1 {\tt dpg}, 100 {\tt dpf})}\\
\hline
\multirow{2}{*}{Build mode} & 
\multirow{2}{*}{Build time} & 
\multicolumn{8}{c|}
{Throughputs in MEs/s 
(blocks * threads)}\\
\cline{3-10}
&& 1*32 & 2*32 & 4*32 & 8*32 &
16*32 & 32*32 & 64*32 & 128*32 \\
\hline
{\tt DCDIAG=0} & 48m & 
3.2E-2 & 6.4E-2 & 1.3E-1 & 2.4E-1 & 
4.0E-1 & fail & fail & fail \\
{\tt DCDIAG=1} & 10m & 
6.5E-1 & 1.2E0 & 2.1E0 & 3.1E0 & 
4.2E0 & 4.9E0 & 5.4E0 & 5.7E0 \\
\hline
\multicolumn{10}{c}{}\\
\hline
\multicolumn{10}{|c|}{\ggttgggg\ with 
1550 diagram groups 
(10 {\tt dpg}, 100 {\tt dpf})}\\
\hline
\multirow{2}{*}{Build mode} & 
\multirow{2}{*}{Build time} & 
\multicolumn{8}{c|}
{Throughputs in MEs/s 
(blocks * threads)}\\
\cline{3-10}
&& 1*32 & 2*32 & 4*32 & 8*32 &
16*32 & 32*32 & 64*32 & 128*32 \\
\hline
{\tt DCDIAG=0} & 4m30s & 
6.6E-1 & 1.4E0 & 2.6E0 & 3.2E0 & 
4.8E0 & fail & fail & fail \\
{\tt DCDIAG=1} & 3m50s & 
1.7E1 & 2.8E1 & 4.2E1 & 4.7E1 & 
5.1E1 & 5.1E1 & 5.2E1 & 5.2E1 \\
\hline
\multicolumn{10}{c}{}\\
\hline
\multicolumn{10}{|c|}{\ggttgggg\ with 
155 diagram groups 
(100 {\tt dpg}, 100 {\tt dpf})}\\
\hline
\multirow{2}{*}{Build mode} & 
\multirow{2}{*}{Build time} & 
\multicolumn{8}{c|}
{Throughputs in MEs/s 
(blocks * threads)}\\
\cline{3-10}
&& 1*32 & 2*32 & 4*32 & 8*32 &
16*32 & 32*32 & 64*32 & 128*32 \\
\hline
{\tt DCDIAG=0} & 8m50s & 
2.9E0 & 5.9E0 & 1.1E1 & 2.1E1 & 
4.0E1 & fail & fail & fail \\
{\tt DCDIAG=1} & 6m20s & 
7.0E1 & 1.1E2 & 1.7E2 & 2.1E2 & 
2.4E2 & 2.4E2 & 2.4E2 & 2.4E2 \\
\hline
\multicolumn{10}{c}{}\\
\hline
\multicolumn{10}{|c|}{\ggttgggg\ with 
78 diagram groups 
(200 {\tt dpg}, 200 {\tt dpf})}\\
\hline
\multirow{2}{*}{Build mode} & 
\multirow{2}{*}{Build time} & 
\multicolumn{8}{c|}
{Throughputs in MEs/s 
(blocks * threads)}\\
\cline{3-10}
&& 1*32 & 2*32 & 4*32 & 8*32 &
16*32 & 32*32 & 64*32 & 128*32 \\
\hline
{\tt DCDIAG=0} & 15m & 
4.1E0 & 9.3E1 & 1.8E1 & 3.5E1 & 
6.6E1 & fail & fail & fail \\
{\tt DCDIAG=1} & 17m & 
fail & fail & fail & fail & 
fail & fail & fail & fail \\
\hline
\multicolumn{10}{c}{}\\
\hline
\multicolumn{10}{|c|}{\ggttgggg\ with 
16 diagram groups 
(1000 {\tt dpg}, 1000 {\tt dpf})}\\
\hline
\multirow{2}{*}{Build mode} & 
\multirow{2}{*}{Build time} & 
\multicolumn{8}{c|}
{Throughputs in MEs/s 
(blocks * threads)}\\
\cline{3-10}
&& 1*32 & 2*32 & 4*32 & 8*32 &
16*32 & 32*32 & 64*32 & 128*32 \\
\hline
{\tt DCDIAG=0} & 2h57m & 
6.5E0 & 1.2E1 & 2.4E1 & 4.7E1 & 
8.9E1 & fail & fail & fail \\
{\tt DCDIAG=1} & 
failed (killed by 
{\tt oom-kill}) &
N/A & N/A & N/A & N/A & 
N/A & N/A & N/A & N/A \\ 
\hline
\end{tabular}
\vspace*{1mm}
\caption{
CUDA run time performance
of \ggttgggg\ in
different {\tt DCDIAG} modes
and with different
numbers of diagrams per group
({\tt FPTYPE=m}).
CUDA/C++ code generated
using CUDACPP ihel6p2.
Results for an NVidia A100 GPU
(using CUDA 12.4 and {\tt gcc} 11.5).
The tests are executed
with {\tt CUDACPP\_RUNTIME\_GOODHELICITIES=ALL} to bypass 
helicity filtering, 
which is still 
performed
sequentially 
one helicity at a time.
As a basic validation 
of physics results,
it was only checked 
that all calculations
return exactly 
the same average ME value,
when starting from the
same random number seeds,
for all {\tt DCDIAG} and
diagram-per-group modes.
}
\label{tab:ggtt4ggpu}
\vspace*{-2.5mm}
\end{table*}
}

\newcommand{\tabggttfiveg}[1]{
\begin{table*}[#1]
\hspace*{-0.5cm}
\centering
\begin{tabular}{|l|c|c|c|c|c|}
\hline
\multicolumn{6}{|c|}{\ggttggggg\ (1000 diagrams per group and per file)
--- build takes 55m}\\
\hline
\emph{C++ SIMD mode} &
{\tt none} & {\tt sse4} & {\tt avx2} & {\tt 512y} & {\tt 512z} \\
\hline
\emph{Time to compute 16 MEs} & 
472.7s & 
337.3s & 
214.9s & 
176.2s &
179.8s \\
\emph{$\rightarrow$ Time in {\tt calculate\_jamps}} &
397.9s (84.2\%) &
305.6s (90.6\%) &
198.6s (92.4\%) &
160.4s (91.1\%) &
171.2s (95.2\%) \\
\emph{$\rightarrow$ Time in {\tt color\_sum\_cpu}} &
74.8s (15.8\%) &
31.7s (9.4\%) &
16.3s (7.6\%) &
15.7s (8.9\%) &
8.7s (4.8\%) \\
\emph{Throughput} & 
3.38E-2/s & 
4.74E-2/s & 
7.45E-2/s & 
9.08E-2/s & 
8.90E-2/s \\
\emph{Speedup w.r.t. none} & 
x1 & x1.40 & x2.20 & x2.68 & x2.63 \\
\emph{Mean ME value / E-13} & 
1.748038 &
1.748038 &
1.748038 &
1.748038 &
1.748038 \\
\hline
\end{tabular}
\caption{
Run time performance,
including profiling 
of Feynman diagrams
and color sums, 
of \ggttggggg\ in
different SIMD modes
({\tt FPTYPE=m}).
C++ code generated
using CUDACPP hack\_ihel6p1,
with 1000 diagrams 
per group and per file
(this is the \ggttggggg\ code described 
in Table~\ref{tab:complexity}).
The code was later
instrumented with {\tt rdtsc} timers
for color sum profiling.
Results for a single core
of an Intel Xeon Gold 6326 CPU
(using {\tt gcc} 11.5).
The tests are executed
with {\tt CUDACPP\_RUNTIME\_GOODHELICITIES=ALL} to bypass 
helicity filtering, 
which is still 
performed
sequentially 
one helicity at a time.
The average ME value
shown in the table,
computed from the same
random numbers in all cases,
is used as a basic validation 
of physics results.
}
\label{tab:ggtt5g}
\end{table*}
}

\newcommand{\figcspctall}[1]{
\begin{figure*}[#1]
\centering
\hspace*{-0.02\textwidth}
\includegraphics[width=0.96\textwidth,clip]{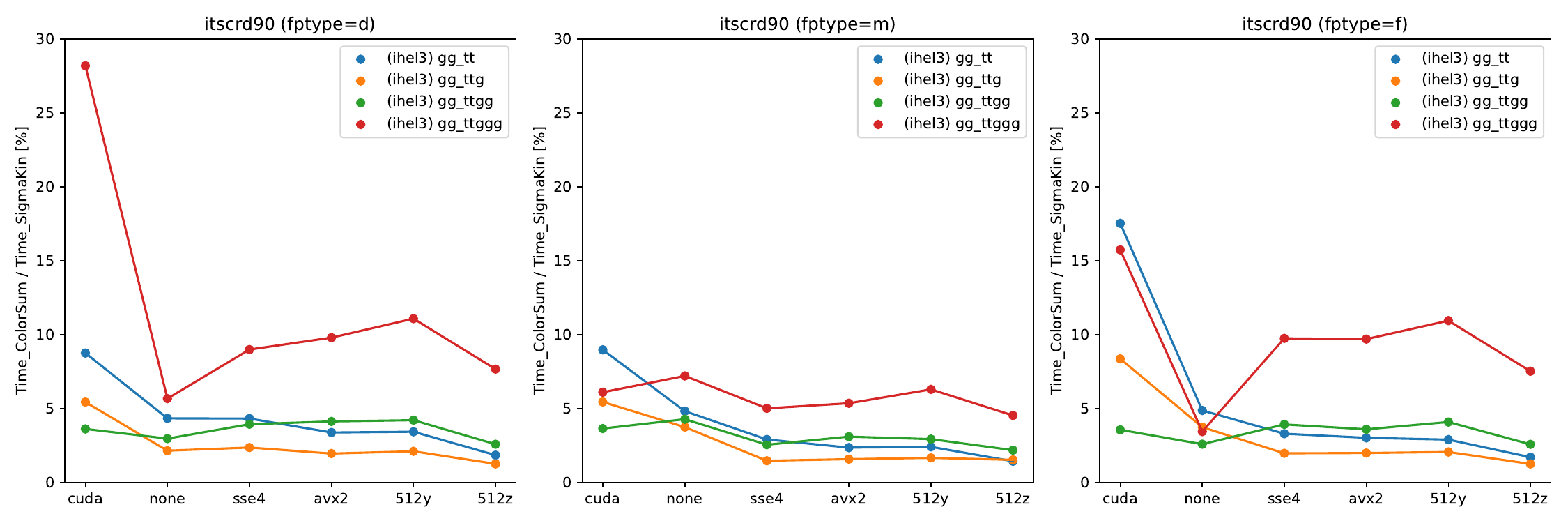}
\caption{
Time spent 
in the color sum
as a fraction 
of the total 
spent 
in the \sk\ ME engine,
for different 
physics processes
and software
configurations,
in the ihel3 
software.
Lower is better.
The 3 plots
correspond to 3
floating point precisions.
Measurements performed on 
an NVidia V100 GPU
and a single core of an Intel
Xeon Silver 4216 CPU,
both on the same node at CERN
(using CUDA 12.0 and {\tt gcc} 11.5).
The results 
for CUDA are based 
on the default ihel3
implementation
of color sums
using GPU kernels.
}
\label{fig:cspctall}
\end{figure*}
}

\newcommand{\figcspctblas}[1]{
\begin{figure*}[#1]
\centering
\hspace*{-0.02\textwidth}
\includegraphics[width=0.96\textwidth,clip]{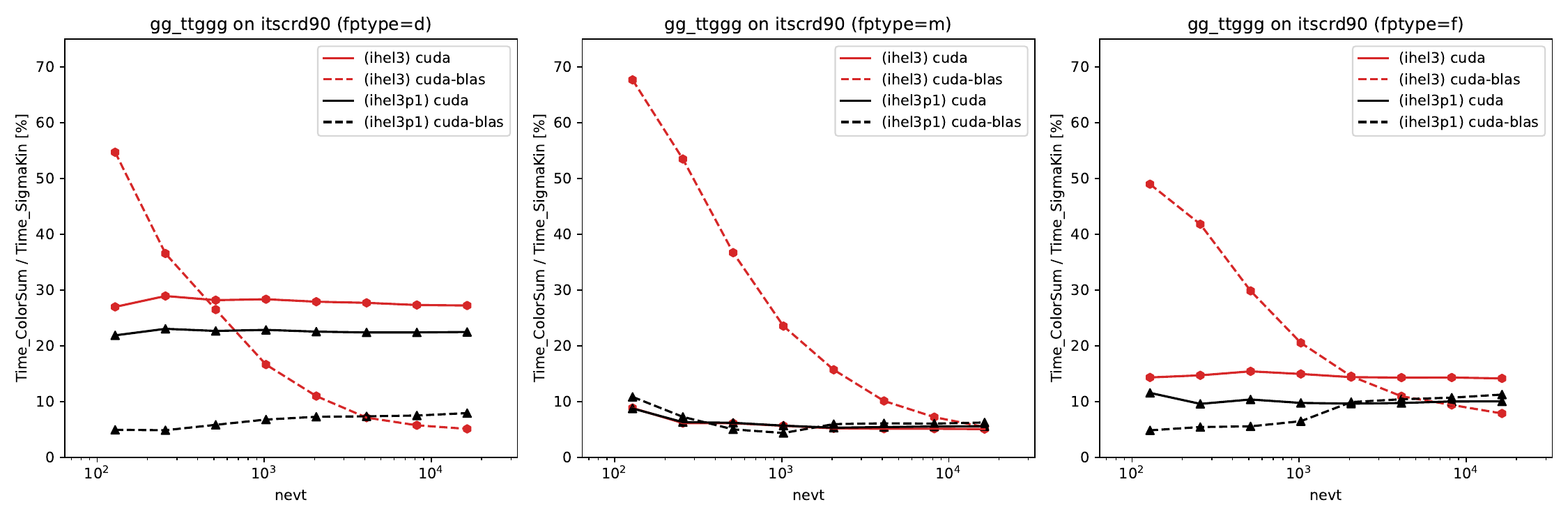}
\caption{
Time spent 
in the color sum
as a fraction 
of the total time
spent in the \sk\ ME engine
for the \ggttggg\ process,
as a function 
of the number 
of events {\tt nevt}
processed in parallel
in one GPU iteration.
Lower is better.
In each plot,
four curves are shown,
using the ihel3 (red) 
and ihel3p1 (black)
codebases,
in the CUDA kernel (solid)
and CUDA/BLAS (dashed)
implementations.
The 3 plots
correspond to 3
floating point precisions.
The handling of the
{\tt nGoodHel} good helicities
is different 
in the various scenarios.
In the two CUDA kernel
implementations,
{\tt nGoodHel} kernels 
are launched in separate
helicity streams,
each 
on a GPU grid
of size {\tt nevt}:
in particular,
{\tt nevt} is the product 
of the number of threads per block, 
which is fixed to 32,
by the varying number 
of blocks in the GPU grid.
For the CUDA/BLAS implementations:
in ihel3b, {\tt nGoodHel} 
separate BLAS color matrix 
multiplications are computed,
in different helicity streams,
using dual amplitude vectors
with {\tt nevt} events;
in ihel3p1b, 
a single BLAS color matrix 
multiplications is computed,
in the default stream,
using dual amplitude vectors
with {\tt nGoodHel*nevt} events.
Measurements performed
with an NVidia V100 GPU
(using CUDA 12.0 and {\tt gcc} 11.5),
on the same node
as in
Fig.~\ref{fig:cspctall}.
}
\label{fig:cspctblas}
\end{figure*}
}

\newcommand{\tabncu}[1]{
\begin{table*}[#1]
\centering
\begin{tabular}{|l|c|l|c|c|}
\hline
\emph{\makecell{Color sum\\implementation}} & 
\emph{\makecell{Blocks*Threads\\({\tt nevt})}} & 
\emph{Function} & 
\emph{FMA} & 
\emph{\makecell{Tensor\\cores}} \\ 
\hline
Kernel & 8*1 &
{\tt color\_sum\_kernel} &
1.9\% & 0.0\% \\
\hline
Kernel & 512*32 &
{\tt color\_sum\_kernel} &
9.0\% & 0.0\% \\
\hline
\multirow{2}{*}{cuBLAS} & 
\multirow{2}{*}{8*1} &
{\tt ampere\_sgemm\_32x32\_sliced1x4\_nt} &
22.7\% & 0.0\% \\
& &
{\tt gemvNSP\_kernel$<$\ldots
cublasGemvTensorStridedBatched\ldots$>$} &
7.3\% & 0.0\% \\
\hline
\multirow{2}{*}{cuBLAS} & 
\multirow{2}{*}{512*32} &
{\tt cutlass::Kernel2$<$cutlass\_80\_simt\ldots$>$} &
81.5\% & 0.0\% \\
& &
{\tt gemvx::kernel$<$\ldots
cublasGemvTensorStridedBatched\ldots$>$} &
20.6\% & 0.3\% \\
\hline
\multirow{2}{*}{cuBLAS/TF32} & 
\multirow{2}{*}{8*1} &
{\tt cutlass::Kernel2$<$cutlass\_80\_tensorop\ldots$>$} &
13.7\% & 16.4\% \\
& &
{\tt gemvNSP\_kernel$<$\ldots
cublasGemvTensorStridedBatched\ldots$>$} &
7.3\% & 0.0\% \\
\hline
\multirow{2}{*}{cuBLAS/TF32} & 
\multirow{2}{*}{512*32} &
{\tt cutlass::Kernel2$<$cutlass\_80\_tensorop\ldots$>$} &
14.1\% & 24.4\% \\
& &
{\tt gemvx::kernel$<$\ldots
cublasGemvTensorStridedBatched\ldots$>$} &
20.3\% & 0.3\% \\
\hline
\end{tabular}
\caption{Summary 
of the results
for the profiling
of the {\tt check.exe} 
application,
using the
NVidia NSight Compute 
({\tt ncu}) profiler
version 2024.1.1.0.
The results are obtained
on an NVidia A100 GPU
at CERN,
on a node equipped
with AMD EPYC 7313 CPUs,
using mixed floating precision 
({\tt FPTYPE=m})
in the ihel3 
version of CUDACPP.
Code built 
using CUDA 12.4 and {\tt gcc} 11.5.
Only the kernels
in one specific
helicity stream
were profiled.
The columns represent 
the following:
(1) color sum implementation
(three configurations are tested:
default using CUDA kernels;
cuBLAS; cuBLAS with TF32 math mode);
(2) GPU grid size configuration
for the kernel implementation
(or, equivalently, number of events
processed in one cuBLAS multiplication);
(3) simplified function name
printed out by the profiler;
(4) FMA (CUDA core) activity,
as per {\tt ncu} metric
{\tt sm\_\_pipe\_fma\_cycles\_active.avg.pct\_of\_peak\_sustained\_active};
(5) tensor core activity,
as per {\tt ncu} metric 
{\tt sm\_\_pipe\_tensor\_cycles\_active.avg.pct\_of\_peak\_sustained\_active}.
}
\label{tab:ncu}
\vspace*{-5mm}
\end{table*}
}

\renewcommand{\topfraction}{.9}
\renewcommand{\floatpagefraction}{.9}

\section{Introduction}

The MadGraph5\_aMC@NLO~\cite{bib:mg5,bib:mg5amc}
physics event generator
(in the following, \mgamc)
is an essential component 
of the data processing 
and analysis workflows
of many high-energy physics 
(HEP) experiments, 
notably those at CERN's 
Large Hadron Collider (LHC).
Using Monte Carlo (MC) techniques, 
\mgamc\ allows the calculation
of cross sections and the generation
of events to provide theoretical
predictions against which experimental
measurements can be compared.
The \mgamc\ software 
has been developed over more than 
two decades
and does not yet fully exploit
the potential of
modern computing architectures.
As is the case for other generators,
this is a source of concern because 
event generation 
in the LHC experiments has
a large computational cost,
which is predicted 
to further increase
during the High-Luminosity
LHC (HL-LHC) 
programme~\cite{bib:csbs}.

In this context,
a recent progress 
has been the 
port and optimization
of the \mgamc\ software 
for data parallel processing
on graphical processing units (GPUs)
and
on CPUs with 
SIMD (Single Instruction Multiple Data) 
vector instructions. 
After a five-year development programme,
which was initially facilitated by
the activities 
of the HEP Software Foundation (HSF) 
event generator working group~\cite{bib:csbs},
and which 
regularly reported
its progress through conference 
proceedings~\cite{bib:vchep2021,bib:ichep2022,bib:acat2022,bib:chep2023},
this project delivered
its first production release
of the \mgamc\ code
with GPU and SIMD support
for 
leading-order (LO) processes
in October 2024. 
This has been 
reported at the CHEP2024 
conference~\cite{bib:chep2024}
and in a more recent
journal submission~\cite{bib:zw2025}.
\mgamc\ is a 
code-generating framework,
largely written in Python,
which allows users 
to generate code,
by default using Fortran,
for any physics 
process of their choice.
For complex processes,
the computational bottleneck
of the 
\mgamc\ workflow is the 
calculation of the
matrix element (ME)
for each phase space
point 
from the momenta of
the initial and final
state particles 
in that event.
The main outcome 
of this project has been 
a new code-generating plugin,
which was
named ``CUDACPP'' because 
the generated physics code
uses C++ instead of Fortran
to execute 
the calculation of MEs
on vector CPUs,
and also includes CUDA extensions 
to run the code on NVidia GPUs
(as well as HIP extensions 
for AMD GPUs).
The new code speeds
up the ME computation
by performing it
for many events at the same time
using a data parallel 
approach~\cite{bib:hsf2020,bib:lhcc2020}.
To achieve this,
work was
also needed in the 
process-agnostic
core \mgamc\ event
processing algorithm
for LO processes, 
MadEvent,
which is written in Fortran
and with which the 
process-specific
C++ code generated by CUDACPP 
is interfaced and linked:
the main 
change 
is 
the move from
a sequential single-event API
to a parallel multi-event API,
and the refactoring 
of some functions
to make them stateless
and re-entrant~\cite{bib:ecfa2023}.
This 
effort
has been the result 
of a collaboration involving
many existing and new 
contributors of the \mgamc\ project,
as detailed in the references cited above~\cite{bib:vchep2021,bib:ichep2022,bib:acat2022,bib:chep2023,bib:chep2024,bib:zw2025}.

In this paper, 
I describe my work on
many additional enhancements
and optimizations of CUDACPP,
mainly but not exclusively 
targeting GPUs.
These are developments
for which the initial brainstorming
with some of my colleagues 
from the development team,
notably Olivier Mattelaer
who prototyped with me the
first concrete changes to the code,
took place during the CSCS GPU 
Hackathon that we attended
in Lugano in September 
2022~\cite{bib:hack2022}.
I then invested further effort
in this area
in February 2024 and 
especially 
October 2024~\cite{bib:ks2024}
after the first CUDACPP release, but 
had to pause this work 
again due to other constraints.
In September 2025,
I~was finally able to resume
this R\&D activity, 
which I now consider 
essentially complete
and which I am documenting
in detail in this paper
in view of its possible
integration in new
production versions
of the CUDACPP software.
A first batch of changes,
which I had described
in the preprint of a previous 
version of this 
paper~\cite{bib:preprintV1},
has already been merged
into a new production release of 
CUDACPP~\cite{bib:cudacpp364}, 
in collaboration with 
my colleagues.
Since then,
I have completed
a second batch of changes,
which were still work in progress
at the time 
of Ref.~\cite{bib:preprintV1},
but which I now also consider ready
to be merged in production.

\subsection{Kernel splitting}

The main idea behind 
these new
developments
can be referred to as
``kernel splitting''.
In short, this consists
in replacing 
the large monolithic kernel
previously used by CUDACPP
for the calculation
of MEs 
from particle momenta,
named \sk,
by many smaller kernels,
executed either in parallel
or sequentially,
to achieve a more efficient
and scalable use of the GPU.
The aim is not only that of
possibly achieving 
a higher peak throughput
(in terms of MEs computed per second)
with a large GPU grid,
i.e. with a large 
number of events 
processed in parallel
on the GPU 
during one offloading cycle,
but also that of 
achieving
a faster increase
of throughput already
with smaller GPU grids.
This was the main focus
of my first batch of changes,
which have since been released.
An important practical difference
between the traditional 
\mgamc\ workflow
on CPUs
and that using GPUs, in fact,
is that the former 
was designed to handle
iterations involving hundreds
of events at a time,
while the latter typically
needs many 
thousands 
of events to be efficient:
in this context,
achieving higher throughputs
with smaller grids 
would provide 
one way
to use 
more manageable,
but still reasonably efficient,
event generation jobs on GPUs.
An additional benefit
of kernel splitting,
which eventually became
the main focus of my
second batch of changes,
is that smaller kernels
are much more manageable
by humans and compilers alike
(both for CPUs and GPUs),
and make it possible 
to achieve complex calculations 
that are instead very difficult
if not impossible to perform
with large monolithic kernels,
such as the calculations
of QCD matrix elements
for physics processes
with a large number
of final state gluons.
In this paper, in particular,
I present 
the 
first 
performance results
from CUDACPP
for the \twotosix\ process 
\ggttgggg\ on CPUs and GPUs
and for the \twotoseven\ process 
\ggttggggg\ on CPUs,
which involve over 15k
and over 230k Feynman
diagrams, respectively.

\figtkz{t}

Schematically, in this paper 
I describe my work on
six sets of kernel splitting 
developments:
(1)~helicity streams;
(2)~color sum splitting 
into separate GPU kernels;
(3)~color sum host refactoring
with optional BLAS offloading;
(4)~Feynman diagram splitting,
with one separate GPU kernel
for each 
diagram;
(5)~Feynman diagram splitting,
with one GPU kernel
for each group of diagrams,
and separate source code files;
(6)~Feynman diagram splitting,
with a single GPU kernel
processing all diagram groups
as GPU device functions.
A brief description 
of these steps and of the
connection between them
is shown in Fig.~\ref{fig:tikz}.
The performance of 
each set of developments
was tested for 
different physics processes,
software configurations
and hardware architectures.
While these developments 
were initially aimed
at improving throughputs on GPUs,
the large code refactoring 
that they imply was also
propagated 
to the vectorized C++ code 
for SIMD CPUs,
as code generated 
by the CUDACPP plugin
is strictly the same
for CPUs and GPUs,
and the distinction between
the two cases is only done
at build time through
{\tt \#ifdef} directives.
I therefore analysed 
not only CUDA and HIP performance 
on NVidia and AMD GPUs,
but also C++ performance 
on vector CPUs
using different 
SIMD configurations.
The possibility to
split the ME calculation
into groups of Feynman diagrams
in different files,
in particular,
was the main enabling factor
for being able to
compute complex 
\twotoseven\ physics 
processes on CPUs.
In turn, this also 
pushed me to 
further analyse and optimize
a few specific details of
the SIMD implementation 
for vector CPUs,
notably the QCD color sum.

Considering the results
of all my tests,
my recommendation is to include 
the full set of these enhancements
in new production releases of CUDACPP.
This differs from the situation
that I had described 
in my first
preprint~\cite{bib:preprintV1},
where the diagram splitting
work had not yet been completed.
A first batch of changes
(green boxes in Fig.~\ref{fig:tikz})
has already been merged
into a new production release,
in collaboration with my colleagues:
this includes 
the first three steps
of this work, 
notably helicity streams,
the separation of color sums 
and Feynman diagrams,
and the optional 
BLAS offloading
of color sums
(which was 
significantly
improved with respect
to that described 
in Ref.~\cite{bib:preprintV1}).
The second batch 
of developments
(red and orange boxes 
in Fig.~\ref{fig:tikz}),
notably including the splitting 
of Feynman diagrams
into different groups
and source code files,
is in my opinion
now also ready 
to be merged upstream.
I note, in particular,
that the performance
of the code in its default
configuration would not
be negatively impacted
by these changes:
many of the features 
that I have added, in fact, 
are optional
and are only enabled
with specific flags 
at generation time,
build time or run time
(this includes 
the BLAS offloading of color sums,
the ME splitting 
into groups of Feynman diagrams,
their orchestration via CUDA graphs
and the choice to treat
diagram groups as separate kernels 
or as device functions
within a single kernel).

\subsection{Outline of this paper}

As these new developments
imply a significant
refactoring of the existing
CUDACPP codebase,
and 
rely on internal 
features that 
have not yet been documented,
I also found it useful
to start this article 
with a brief review
of some architectural
aspects of the software, 
focusing on those 
most relevant to the GPU
kernel splitting 
developments,
but also covering some
details
of the SIMD implementation.
In this context,
I should 
mention that 
the internal design of the 
CUDACPP plugin, 
which was largely my own work,
is far from perfect
and the code could certainly benefit
from an extensive cleanup.
In particular,
over time I have added
various switches
to support alternative
implementations
of some software components,
as I have done again
in many of the new
developments
described in this paper:
I believe 
that this 
has been
very useful,
and in some cases essential,
to try out
different approaches
and eventually converge
on well-defined solutions
to address specific 
problems, but 
I am also aware 
that this 
results
in code 
that might be 
more difficult to read,
and where some of this 
flexibility 
may be regarded as
an unnecessary complication.
I hope that this article
may help clarify the rationale
behind some of these 
design choices
and provide some 
guidance for possible future
developments and/or cleanups.
I stress, in any case,
that the opinions 
that I express in this paper
are only my own,
and may differ from those
of some of my colleagues.

The outline of this paper is 
the following.
In Sec.~\ref{sec:arch},
I~briefly review 
some aspects
of the CUDACPP software 
architecture design,
focusing on those 
relevant 
to GPU kernel splitting,
but also covering
some details of the
SIMD implementation
for vector CPUs.
In Sec.~\ref{sec:ksp},
I describe in detail 
the software design 
and implementation
of my first batch of
kernel splitting enhancements,
and I give 
the results
and an analysis of 
various performance tests:
this includes 
and partially extends
the material that I had
covered in my 
first 
preprint~\cite{bib:preprintV1},
with the exception 
of Feynman diagram splitting.
I also cover some 
more recent
optimizations
of SIMD color sums.
In Sec.~\ref{sec:diags},
I describe the various options
I introduced to split
the calculation of
Feynman diagrams
into different 
groups of diagrams 
and/or different kernels,
and different source code files.
I also give the first
CUDACPP performance results
for physics processes
with large numbers 
of final state gluons,
which could be computed
thanks to these new features.
In Sec.~\ref{sec:end},
I finally give my conclusions 
and an an outlook for this work.

\figanatomy{t}

\section{Software architecture review}
\label{sec:arch}

The overall 
software architecture design
of the CUDACPP plugin
for GPUs and vector CPUs,
its evolution over time, 
its interplay
with the \mgamc\ framework
and 
many specific 
implementation choices
have already been 
described
in the 
previous 
papers~\cite{bib:vchep2021,bib:ichep2022,bib:acat2022,bib:chep2023,bib:chep2024,bib:zw2025} from our team
and I will not repeat this here.
The main point of this section,
instead, is to
give further
information about some software
aspects which have not yet 
been documented,
mainly those relevant
to GPU kernel splitting,
such as memory layouts 
and memory access,
but also
some related to 
the SIMD implementation
for vector CPUs.
First, in any case, I will give 
a very high level overview
of the various areas and phases
of development on CUDACPP,
to put 
into context
the new kernel splitting
developments.

\subsection{Three software areas
and development phases}

The evolution of the work
on the CUDACPP plugin 
between 2020 and 2024
is represented schematically
in Fig.~\ref{fig:anatomy}
and was 
summarised 
in detail in the CHEP2024 
proceedings~\cite{bib:chep2024},
from which that plot is taken.
Many more details about
the progress 
of this work over time,
including snapshots and links
to the initial presentations
of some of these ideas at
working group meetings
of the Madgraph on GPU project,
can be found in the slides
of my CHEP2024 
presentation~\cite{bib:chep2024slides}.

The main aim of
the CUDACPP software,
from the very beginning,
has been to 
use event-level data parallelism
to try and speed up
the calculation of MEs,
which is the computational 
bottleneck
of the whole \mgamc\ framework.
This is
because, to a large extent,
the same mathematical functions
need to be numerically computed
for different events
during the ME calculation,
and this makes it possible
to efficiently compute the MEs
for many events 
{\em in lockstep}
at the same time
using GPUs and SIMD CPUs.
This is something that 
I had pointed out 
in two presentations
to the HSF generator working
group~\cite{bib:hsf2020} 
and the LHCC~\cite{bib:lhcc2020}
in 2020, using plots from which 
I later derived those 
in Fig.~\ref{fig:anatomy}.

Largely speaking,
the work on the CUDACPP plugin
has concerned three areas
of the software,
roughly corresponding
to three
phases of development:

\begin{enumerate}
\item 
{\em ME engines}:
the port 
to GPUs and SIMD CPUs,
using CUDA/HIP
and vectorized C++,
of the computational engines
to calculate MEs
from particle momenta,
and their backport
to the code-generating
Python framework,
for different physics processes.
\item 
{\em MadEvent integration}:
the injection 
of the new code 
into one {\tt madevent} 
application,
by modifying the existing
Fortran 
and linking the 
new ME engine
as a C++
library into it.
\item 
{\em Full workflow orchestration}:
the integration 
and testing of the full
\mgamc\ workflow
involving many {\tt madevent}
applications,
including the packaging 
and installation
of the CUDACPP plugin.
\end{enumerate}
Over time,
the focus of 
developments
has gradually shifted
from the first 
to the last of these 
software areas,
even if some
work on ME engines
and on 
the integration of C++ and Fortran
continued
up until the release
in October 2024.
This is reflected in the 
documentation
provided in 
our previous papers.
What I want to stress here,
to put it into context, is
that the new kernel splitting
work that I describe
in this paper
was all done at the
level of the ME engines alone,
even if its motivation
(as is the case for all
of the work in the
Madgraph on GPU project)
lies in the streamlining
of the full \mgamc\ workflow.
I mention this also
because I believe
that, amongst the many
previous documents
from this project,
those describing the
software features
most relevant to 
kernel splitting
in the ME engines
are actually the
very first ones,
namely the 
proceedings~\cite{bib:vchep2021}
and 
presentation~\cite{bib:vchep2021video}
from the vCHEP2021 conference.

More specifically,
in all of the work 
that I present in this paper,
I only developed code 
using the standalone application
{\tt check.exe},
where all three main components
of an event generation application
(random numbers, 
phase space sampling
and ME calculation)
are implemented in C++ and CUDA
using a data parallel approach.
This is because {\tt check.exe}
makes it possible 
to focus on the optimization
of the ME engine
without being constrained
by the MadEvent 
Fortran infrastructure,
where the legacy phase space sampling
and other sequential non-ME components
slow down the whole workflow.
I routinely built 
{\tt check.exe}
using the CUDACPP makefile
{\tt cudacpp.mk},
without even building 
MadEvent or 
other Fortran code
(which proceeds 
via 
a separate
makefile that internally delegates
the build of the CUDACPP library
to {\tt cudacpp.mk}).
I did however 
validate my changes
by 
running the full
MadEvent test suite
at the end of each
development cycle.

\begin{table*}[p] 
\begin{lstlisting}[frame=single,language=C++]
int main(...) // application check.exe (similar code in class Bridge)
{
  int nevt = m_gpuBlocks * m_gpuThreads; // number of events == GPU grid size
  DeviceBufferMomenta m_devMomenta( nevt ); // memory buffer for nevt events
  DeviceBufferMatrixElements m_devMEs( nevt ); // memory buffer for nevt events
  // Compute MEs for nevt events (m_gpuBlocks * m_gpuThreads) in one go
  MatrixElementKernelDevice mek( m_devMomenta, m_devMEs, m_gpublocks, m_gputhreads, ... );
  mek.computeMatrixElements( ... );
}

void MatrixElementKernelDevice::computeMatrixElements( ... )
{
  gpuLaunchKernel( sigmaKin, m_gpuBlocks, m_gpuThreads, m_momenta.data(), m_MEs.data(), ... );
}

__global__ void 
sigmaKin( const fptype* momenta, // input: momenta[nevt*npar*4]
          fptype* MEs, ... )     // output: MEs[nevt], final sum over ihel           
{
  const int ievt = blockDim.x * blockIdx.x + threadIdx.x; // one event <-> one thread
  //  Zero the running sum MEs[ievt] and add the contribution of each helicity ihel
  MEs[ievt] = 0;
  for( int ihel = ... )
     calculate_wavefunctions( ihel, momenta, MEs, ... );
  // Randomly select a color and a helicity for event ievt
  selcol[ievt] = ...; selhel[ievt] = ... // extra output (from extra input random numbers)
}

__device__ void
calculate_wavefunctions( int ihel,
                         const fptype* momenta, // input: momenta[nevt*npar*4]
                         fptype* MEs,  ... )    // output: MEs[nevt], running sum over ihel
{
  using M_ACCESS = DeviceAccessMomenta;        // non-trivial access (nevt events)
  using E_ACCESS = DeviceAccessMatrixElements; // non-trivial access (nevt events)
  using W_ACCESS = DeviceAccessWavefunctions;  // trivial access (one event)
  using A_ACCESS = DeviceAccessAmplitudes;     // trivial access (one event)
  fptype* wf[nwf] = ...; // local variable for one event (wavefunctions)  
  cxtype amp[1]; // local variable for one event (amplitude for one Feynman diagram)
  cxtype jamp[ncolor] = {}; // local variable for one event (dual amplitudes for color flows)
  // Feynman diagram 1 of 3
  vxxxxx<M_ACCESS, W_ACCESS>( momenta, ihel, ..., wf[0] ); // compute wf[0]
  vxxxxx<M_ACCESS, W_ACCESS>( momenta, ihel, ..., wf[1] ); // compute wf[1]
  oxxxxx<M_ACCESS, W_ACCESS>( momenta, ihel, ..., wf[2] ); // compute wf[2]
  ixxxxx<M_ACCESS, W_ACCESS>( momenta, ihel, ..., wf[3] ); // compute wf[3]
  VVV1P0_1<W_ACCESS, ...>( wf[0], wf[1], ..., wf[4] ); // compute wf[4]
  FFV1_0<W_ACCESS, A_ACCESS, ...>( wf[3], wf[2], wf[4], ... &amp[0] ); // compute amp[0]
  jamp[...] += ... * amp[0];
  ...
  // Feynman diagram 3 of 3
  ... // compute wavefunctions and amplitudes...
  // Add ME contribution for helicity ihel (from quadratic form on color amplitudes)
  fptype& ME = E_ACCESS::kernelAccess( MEs ); // ME for event ievt (from threadIdx.x etc)
  ME += ... sum_ij ( cxconj((cxtype2)(jamp[i])) * colormatrix[i][j] * (cxtype2)(jamp[j]) );
}
\end{lstlisting}
\caption{Pseudo-code 
of the \ggtt\ ME calculation 
in CUDACPP 
before any kernel splitting 
changes (``ihel0'').
Function {\tt main} represents
the {\tt check.exe} standalone application.
Note that \sk\ is the 
only GPU kernel
({\tt \_\_global\_\_} function)
in this workflow.
Only a simplified GPU version 
of the code is shown,
using scalar {\tt fptype} 
floating point 
and {\tt cxtype} 
complex number types:
the actual code that
supports both GPUs and SIMD CPUs
uses {\tt fptype\_sv}
and {\tt cxtype\_sv} data types
inside {\tt calculate\_wavefunction},
making it possible 
to use the same formal 
code for scalar data
and for SIMD vectors
(see Ref.~\cite{bib:vchep2021}
for details).
The CPU branch of the code
for {\tt main}, 
encapsulated by an {\tt \#ifdef}
and not shown here,
uses the {\tt Host} versions
of various classes instead
of the {\tt Device} versions.
The pseudo-code 
for the color sum
schematically indicates
that this proceeds
via a quadratic form
using a color matrix,
and that color amplitudes
are converted from
{\tt fptype/cxtype}
floating point precision
to a potentially different
{\tt fptype2/cxtype2} precision
(this enables the ``mixed''
precision mode, where
the former is based on {\tt double}
and the latter on {\tt float}).
}
\label{code:ihel0}
\end{table*}

\newcommand{\npar}{{\tt npar}}
\newcommand{\nevt}{{\tt nevt}}
\newcommand{\fptype}{{\tt fptype}}

\subsection{A closer look 
at the internals of the 
ME engine}
\label{sec:meinternals}

Consider a physics process
where \npar\ is the total number
of initial and final state particles.
Assume that the calculation
of Feynman diagram amplitudes
is performed
using floating point \fptype,
which is a {\tt typedef}
for either {\tt double} 
or {\tt float}.
Very schematically,
as shown in Fig.~\ref{fig:anatomy},
the ME computational engine
of CUDACPP is simply
a software component 
that takes as input
an array of particle momenta
for many events
(i.e. 4*\npar\ \fptype\ values
for each of \nevt\ events),
and returns as its output
an array of matrix elements
(i.e. one \fptype\ value
for each of \nevt\ events).
The calculation in CUDACPP 
is actually
more complex than this,
as 
the ME engine
also receives other inputs, 
such as: 
an event-by-event running 
coupling \as\ (at 
a scale which is computed 
from particle momenta
using an external Fortran module);
two event-by-event random numbers
for the selection 
of one specific helicity
and leading color 
in the generated event
(which are also returned
as outputs);
and an event-by-event channel identifier 
(for MadEvent single-diagram enhancement,
also resulting in additional outputs).
However, I will neglect this 
in the following,
except otherwise stated,
and focus only
on the calculation of output MEs
from input momenta.

In the 
production version 
of the CUDACPP plugin
before any kernel splitting changes
(hereafter ``ihel0''),
v1.00.02 
in \mgamc\ v3.6.3,
the ME computational engine 
on GPUs is essentially
a single and very large 
monolithic GPU kernel,
named \sk. 
This
performs 
all of the relevant operations
that are needed 
to compute the output ME
from the input momenta
for each event
(one notable exception
being the calculation
of other couplings that
depend on \as,
which is delegated
to a separate kernel).
Very schematically,
this 
is represented
in the pseudo-code 
in Table~\ref{code:ihel0}.
In particular,
\sk\ performs the following
operations for each event:
it keeps a running sum 
of the ME over helicities;
it loops on all possible 
helicities
of the external particles
(or, more precisely,
only on a pre-determined
set of ``good'' helicities
whose contribution is non-zero);
it adds to the running sum of the ME
the contribution 
from each helicity,
by calling a device function
{\tt calculate\_wavefunctions}
which internally involves 
for each QCD color flow
the calculation 
of particle and propagator
wavefunctions and that
of dual amplitudes,
followed by the sum 
over all color flows
using a quadratic form 
based on a color matrix;
after the end
of the helicity loop,
\sk\ randomly draws
a helicity and a color
for the generated event
based on the ME contributions
from individual
helicities and colors.
The calculation of 
helicity amplitudes 
for Feynman diagrams,
in particular,
is described in detail
in Ref.~\cite{bib:vchep2021}
(see Fig.~1 therein);
complementary details are
also given in Ref.~\cite{bib:zw2025}.
As shown schematically
in Table~\ref{code:ihel0},
this involves 
three types of elements
to compute: 
the wavefunctions of external
(initial/final state) particles,
via 
functions like 
{\tt VXXXXX};
the wavefunctions 
of internal propagators,
e.g. via
{\tt VVV1P0\_1};
the dual amplitudes
for a given color flow
(``jamps''),
e.g. via
{\tt FFV1\_0}.
The names and roles
of these functions 
in CUDACPP are the same
as in the original
HELAS~\cite{bib:helas}
and ALOHA~\cite{bib:aloha}
implementations.

One key aspect 
in the design 
of this software chain,
which has not been documented
in detail so far,
is the allocation of memory buffers
and their access and use 
in computational kernels,
in both the GPU and SIMD CPU code.
This was a core part 
of my design and implementation
work in the second half of 2020,
which was motivated 
mainly by the aim of achieving
SIMD speedups on CPUs,
but also by the aim of achieving
coalesced memory access on GPUs,
and turned out to be 
a key enabling factor for both.
The design principle 
which I adopted, 
in particular,
is a complete separation
of three key elements
of the software
(for details, see the backup 
slides 70--75 from December 2020 
in Ref.~\cite{bib:chep2024slides}):
\begin{enumerate}
\item {\em Memory allocation}.
In the {\tt check.exe}
standalone application
(or in the {\tt Bridge}
component that gets linked
with {\tt madevent}),
CUDACPP allocates 
host and device memory buffers
that are properly dimensioned
for the number \nevt\ of events
that will be processed 
in parallel in one offloading cycle.
On the GPU, \nevt\ is simply
the GPU grid size where kernels
are launched (i.e. the product
of the number of threads per block
and of the number of blocks),
as CUDACPP kernels use event-level
data parallelism where 
each GPU thread processes one event.
The classes responsible
to allocate and hold 
the pointers to
the memory buffers
have no knowledge 
of the internal memory layout
of the buffers;
for instance, 
{\tt DeviceBufferMomenta} 
in Table~\ref{code:ihel0}
allocates 4*\npar*\nevt\ \fptype\ values,
but it does not know the layout 
of the particle momenta arrays.
In particular,
as opposed to 
earlier versions of the code
that were using structured 
memory allocations
like {\tt std::vector} in C++
and {\tt cudaMalloc3D} in CUDA,
from the end of 2020
the code uses
unstructured
memory allocations
of raw buffers,
both on the host
(via {\tt malloc}) 
for the GPU 
and SIMD CPU code,~and 
on the device
(via {\tt cudaMalloc})
for the GPU code.
\item
{\em Memory layout and data access}.
The specific layout 
chosen for storing data
inside the raw memory buffers
allocated in the previous step
is encapsulated in a separate 
set of memory access classes.
All these 
classes
have methods named
{\tt kernelAccess},
or some variation of this.
On the GPU,
these methods take
as input the 
{\tt fptype*} pointer
associated to a 
raw memory buffer
for \nevt\ events,
and they 
return as output
a data item
for the single event
{\tt ievt}
processed by the GPU thread
where the code is executed,
typically indexed 
by the identifier
of the GPU thread,
{\tt blockDim.x*blockIdx.x+threadIdx.x}.
In some cases, these methods
have additional parameters:
method {\tt kernelAccessIp4IparConst}
in the {\tt DeviceAccessMomenta} class,
for instance,
takes two additional parameters
{\tt ip4} and {\tt ipar}
in order to retrieve 
a given 4-vector component
for a given particle.
It is only 
the memory access classes
that, internally,
are able to decode the 
memory layouts 
of a raw buffer:
this is done
by interpreting the buffers
as one-dimensional C-style array
or casting them as
multi-dimensional C-style arrays.
For most data buffers,
a Structure-Of-Array (SOA)
or an Array-Of-Structure-Of-Array (AOSOA)
layout is chosen,
where the data items of a given type
for different events are contiguous:
as explained in detail 
in the vCHEP2021 
proceedings~\cite{bib:vchep2021},
this is absolutely essential
for SIMD processing on vector CPUs,
and it is also beneficial 
-- but not at all a strict requirement --
on GPUs
to improve performance
through coalesced memory 
access.
In some cases, such as couplings,
the floating point raw arrays
include the real and imaginary 
components of complex data,
and it is the responsibility 
of the data access classes 
to return outputs with
the API of a complex number data type.
The case of CPU code
includes the additional
complication of returning
outputs whose data types
{\tt fptype\_sv} and {\tt cxtype\_sv}
can be scalar (for no-SIMD C++)
or SIMD vectors
{\tt fptype\_v} and {\tt cxtype\_v}
(through compiler vector
extensions~\cite{bib:vchep2021}, 
hereafter CVEs) of
{\tt fptype} or {\tt cxtype} data.
Concerning 
the use 
of memory access classes
in other CUDACPP software components,
notably 
the methods for computing
helicity amplitudes
like {\tt VXXXXX}, {\tt VVV1P0\_1}
or {\tt FFV1\_0},
I initially chose to implement
this via templates:
{\tt DeviceAccessMomenta}
in Table~\ref{code:ihel0},
for instance, 
is a template parameter 
{\tt M\_ACCESS} of {\tt VXXXXX}.
The idea behind the choice
of using templates
was to allow the flexibility
of easily switching between 
different memory layouts
at build time, to compare 
their performances;
in retrospective, the same
flexibility may have been achieved
with the same memory access classes
but without 
templates\footnote{
  As noted in 
  Sec.~\ref{sec:2to6},
  eventually I did decide to 
  remove those templates,
  even if it was not clear
  to me a priori
  if this 
  would make the code easier 
  to read and maintain,
  or faster to build:
  my main motivation
  was an attempt to allow
  some CUDA builds 
  in some configurations
  of particularly complex
  \twotosix\ and 
  \twotoseven\ processes,
  which were otherwise 
  crashing or taking too long.
  Unfortunately,
  my tests about the
  effectiveness of 
  template removal
  for simplifying CUDA builds
  were at best inconclusive,
  as those builds 
  continued to fail;
  however, removing templates
  did not have any negative
  impact on performance elsewhere,
  and I suggest that this
  may represent a useful
  cleanup to merge upstream
  and integrate 
  in a future production release.
}.
\item
{\em Arithmetic operations
and other computational functions}.
Finally, the decoupling
of data access from 
actual calculations
has probably been one of
the most important aspects
of the whole CUDACPP 
software design.
Just like 
the memory allocation classes,
also the helicity amplitude
functions like {\tt VXXXXX}, 
{\tt VVV1P0\_1} or {\tt FFV1\_0}
have almost\footnote{
  One notable exception 
  is that some assumptions
  are still made about
  the memory layout
  of wavefunctions,
  which for spin 1/2 and spin 1
  particles are complex 
  6-dimensional arrays
  as in the original
  HELAS~\cite{bib:helas}
  and ALOHA~\cite{bib:aloha}
  implementations, 
  from which CUDACPP is derived.
  This detail is
  relevant to the kernel splitting
  for different Feynman diagrams,
  as noted 
  in Sec.~\ref{sec:ihel5}.
} no knowledge 
of the memory layouts used
in the multi-event data buffers.
These functions also ignore
whether the calculations
are performed 
in single or double precision,
as I encapsulated this choice
in the {\tt fptype} type definition.
Even more, these functions
ignore whether the
arithmetic operations
within them are applied
to scalar values
on GPUs or CPUs
or to SIMD vectors on CPUs:
in the latter case,
this is possible because
simple operators like ``{\tt +}''
are automatically understood
by the compiler as vector
operations when applied
to CVE vectors of {\tt fptype},
and are also implemented
using SIMD CVE operations
in their definition 
for vectors of custom types,
notably the vectors
of {\tt cxtype} complex types.
This design is extremely powerful
because it has made it possible
to use formally 
the same exact lines 
of software 
in the code-generated
helicity amplitude
functions like
{\tt VVV1P0\_1} or {\tt FFV1\_0}.
I note in passing, however,
that the functions computing
wavefunctions for initial 
and final state particles,
like {\tt IXXXXX}, {\tt OXXXXX} 
and {\tt VXXXXX},
or their variations for massless
or beam-collinear particles,
are not code-generated
from a model Lagrangian,
but are instead hardcoded
and often required 
particular care 
and successive iterations.
Two complications 
that I had to address
for SIMD code,
in particular, are the 
following\footnote{
  In this context,
  I 
  find it important 
  to stress that, 
  in my experience,
  achieving a robust 
  and performant CUDACPP
  implementation 
  for SIMD CPUs
  was far from a trivial task,
  and was, in fact, much more complex 
  than achieving it for GPUs.
  The difficulties
  for implementing
  {\tt IXXXXX}-like functions
  are just two examples.
  I also already mentioned
  the fact that there are
  very strict constraints 
  in the memory 
  layouts for CPU SIMD,
  unlike those for GPUs;
  as described in 
  Ref.~\cite{bib:vchep2021},
  in particular,
  this also implied the
  development of wrappers 
  and adapters for vectors
  {\tt cxtype\_v}
  of complex numbers,
  as their real and 
  imaginary parts must
  be stored as an SOA
  using two contiguous 
  {\tt fptype\_v} arrays
  (RRRRIIII), rather than 
  as an AOS (RIRIRIRI).
  Another example is the 
  implementation 
  of mixed floating point mode,
  where I added
  a complex mechanism
  to merge/split two SIMD vectors
  of {\tt double}
  into/from one SIMD vector 
  of {\tt float}
  at the boundary between
  dual amplitude and color sum
  calculations, in order
  to use the widest
  possible SIMD vectors 
  in both cases
  and maximize efficiency
  (see Sec.~\ref{sec:csm}).
  More generally,
  in my opinion,
  it is fair to say that 
  achieving SIMD speedups 
  through vectorization
  is {\em always} much more
  complicated than achieving
  speedups on GPUs.
  Code acceleration on GPUs,
  in fact, can be achieved
  even without lockstep
  processing (at the cost
  of some inefficiency
  from warp divergence),
  or without optimized
  memory layouts 
  (at the cost of 
  some inefficiency
  from the lack of 
  coalesced data access),
  simply because a GPU
  has thousands of threads.
  One good feature of CUDACPP
  is that, by focusing on
  achieving C++ vectorization 
  on SIMD CPUs, this has also
  improved the efficiency
  of the code on GPUs.
}:
first, unlike {\tt VVV1P0\_1} 
or {\tt FFV1\_0},
the {\tt IXXXXX}-like functions
include some {\tt if/else} branching,
which I reimplemented
using vector masks 
in the SIMD case
in order to fully exploit
data parallel speedups
(even if, from a performance point 
of view, this is only
important for simple processes 
with few Feynman diagrams);
second, as I performed
all my functional testing
with Floating Point Exception 
traps (FPEs) enabled
in order to develop 
more robust code,
I came across some crashes
caused by the interplay
of compiler optimizations,
SIMD CVEs and FPE traps,
which I addressed
using {\tt volatile} keywords 
and which pushed me to develop
a large set of functional
tests specifically 
for these functions.
\end{enumerate}

On top of the separation
of the three software
concerns above
(data allocation,
data access and 
arithmetic calculations),
another important part 
of the design of 
the CUDACPP software
was the implementation of
kernel launching on GPUs
and event loops on CPUs.
My aim here was to keep 
as much as possible 
of the software logic
and of the actual code
identical for GPUs and CPUs,
in order to simplify 
the iterative addition
of new features 
and of bug fixes
for both 
cases.
In short, 
I addressed
this by keeping 
also for SIMD CPUs
the idea of a ``grid'' of events
that are processed 
in one given iteration,
and by explicitly 
subdividing this grid
into SIMD event vectors
and adding an explicit
loop over them.
Much more practically,
and more importantly
for the kernel splitting work
described in the 
following, 
the link between 
memory allocations
and the execution 
of the ME computational engines
is provided by two 
different incarnations of a 
{\tt MatrixElementKernel} 
(MEK) class, 
one for GPUs ({\tt Device})
and one for CPUs ({\tt Host}).
A singleton instance 
of the appropriate MEK class
is constructed and used
inside {\tt check.exe}
for standalone tests,
or inside the {\tt Bridge}
component that is linked
with Fortran in {\tt madevent} 
for the full \mgamc\ workflow.
As shown in Table~\ref{code:ihel0},
it is the 
{\tt MatrixElementKernelDevice} 
class that internally 
launches the monolithic \sk\ kernel
in the current 
CUDACPP.
This is the starting point
for the enhancements
presented in the 
following sections.

\newcommand{\none}{{\tt none}}
\newcommand{\ssef}{{\tt sse4}}
\newcommand{\avxt}{{\tt avx2}}
\newcommand{\foty}{{\tt 512y}}
\newcommand{\fotz}{{\tt 512z}}

\subsection{A few details 
on SIMD build modes}

As mentioned previously,
the new developments
that I present in this paper
are mainly optimizations
that target GPUs 
specifically,
but they
are also 
relevant and useful for 
the CUDACPP implementation
on vector CPUs.
In the following,
in particular, I will 
also show 
results
from the five different
SIMD build modes 
of CUDACPP, namely 
\none, \ssef, \avxt,
\foty\ and \fotz.
While these have already
been described 
to some extent 
in earlier publications
from our team~\cite{bib:vchep2021,bib:zw2025},
I find it useful 
to take 
this paper 
as an opportunity
to give a few additional
and previously undocumented
details.
This is also relevant
to the further 
optimizations of color sums
on vector CPUs that are
described in Sec.~\ref{sec:csm}.

The 
SIMD modes
of CUDACPP were 
initially introduced 
targeting Intel CPUs.
My design strategy
was to define 
separate 
build modes,
which on every
specific system
would allow not only
the fastest calculation
compatible with 
the hardware specs,
but also an evaluation
of the speedup
achieved through 
the SIMD implementation
of CUDACPP 
in vectorized C++.
Largely speaking:
\none\ does not 
explicitly
use any SIMD extensions;
\ssef\ uses {\tt SSE4.2}
instructions and 128-bit
{\tt xmm} registers; 
\avxt\ uses {\tt AVX2}
instructions and 256-bit
{\tt ymm} registers; 
\foty\ uses {\tt AVX512}
instructions and 256-bit
{\tt ymm} registers; 
\fotz\ uses {\tt AVX512}
instructions and 512-bit
{\tt zmm} registers.

The \foty\ mode
may appear surprising.
The rationale behind it
is, quite simply,
that it is around 10\%
faster than \avxt\ on
CPUs with the {\tt AVX512}
instruction set:
as can be seen by disassembling
CUDACPP objects
with the {\tt objdump} tool
(see slide 17 in 
Ref.~\cite{bib:vchep2021video}),
this is probably because
the compiler chooses to
process a few operations
using {\tt AVX512} 
instructions
rather than {\tt AVX2}
instructions.
In addition, 
\foty\ is also faster
than \fotz\ on 
{\tt AVX512} CPUs with
a single FMA unit
(such as Intel 
Xeon Silver 4216 CPUs),
making it the preferred
CUDACPP build mode
on this large family of CPUs.
Conversely, 
\foty\ is slower
than \fotz,
and the latter should be 
preferred, on 
{\tt AVX512} CPUs with
two FMA units
(such as Intel 
Xeon Gold 6326 CPUs):
this was first reported
for the CUDACPP software
in Ref.~\cite{bib:ichep2022}
and has been confirmed
in many tests since then.

More in detail,
each build mode
corresponds to 
a well-defined choice
of two parameters,
the CPU instruction set,
and the length of
CUDACPP SIMD vectors:
\begin{enumerate}
\item 
{\em CPU instruction set}. This
is configured via the
compiler flag
{\tt -march}
in the {\tt cudacpp.mk}
makefile.
First and foremost,
this must be well-defined
because an application
built 
for a given 
instruction set
would crash if it is
executed on a hardware
that does not support it
(for instance, 
if it is compiled
for {\tt AVX512} but executed
on a host that does
not support it).
While many different 
levels and sublevels
of vectorization
have been defined
over time, such as
{\tt AVX}, {\tt AVX2},
{\tt SSE4.1} 
or {\tt SSE4.2},
the choice I made
was to only support
a very limited subset
of these in CUDACPP.
In particular:
\none\ uses
{\tt -march=x86-64},
which 
is meant to disable
all SIMD extensions
(but does not completely
achieve this goal,
as discussed 
in Sec.~\ref{sec:csm});
\ssef\ uses
{\tt -march=nehalem},
which supports {\tt SSE4.2}
extensions with 128-bit
{\tt xmm} registers;
\avxt\ uses
{\tt -march=haswell},
which supports {\tt AVX2}
extensions with 256-bit
{\tt ymm} registers;
\foty\ and \fotz\ both 
use {\tt -march=skylake-avx512},
which supports 
a specific set of
{\tt AVX512}
extensions with 512-bit
{\tt zmm} registers.
In this context,
I chose not to
foresee a build mode
based on 
{\tt -march=native},
which would offer the
greatest set of
hardware features
available on a CPU,
but is somewhat
poorly defined.
This is similar
to what is done 
in some LHC experiments:
LHCb, for instance,
uses well-defined
software architecture
builds based on {\tt x86\_64} 
Micro-Architecture Feature
Levels~\cite{bib:phoronix}
such as {\tt x86\_64-v2}.
In retrospective,
indeed a better choice 
would have been 
to configure {\tt -march}
based on {\tt x86\_64} 
Feature Levels,
but these were not yet
widely used when 
I developed the SIMD
implementation of CUDACPP 
in 2020.
\item 
{\em SIMD vector length}. This
is the core of 
the CUDACPP SIMD
implementation,
where it is encapsulated 
by a preprocessor
variable 
{\tt MGONGPU\_CPPSIMD}
and an equivalent 
C++ variable {\tt neppV}.
As described more 
in detail in 
Ref.~\cite{bib:vchep2021},
in fact,
the basic strategy
of the CUDACPP SIMD
implementation
is the definition,
using compiler vector
extensions,
of vectors {\tt fptype\_v}
of {\tt neppV} floating
point variables of
type {\tt fptype}.
The five build modes
are defined as follows:
\none\ uses
{\tt neppV=1},
\ssef\ uses
{\tt neppV=2} and
{\tt neppV=4}
in double and
single precision,
respectively;
\avxt\ and \foty\ both use
{\tt neppV=4} and
{\tt neppV=8}
in double and
single precision
(i.e. this is 
the same source code
compiled with different flags);
\fotz\ uses
{\tt neppV=8} and
{\tt neppV=16}
in double and
single precision.
\end{enumerate}
The SIMD instruction set
and vector length
in the five CUDACPP
build modes are,
of course, tightly related.
The choice of the
appropriate 
{\tt MGONGPU\_CPPSIMD}
and {\tt neppV} 
depends on the 
compiler options
defined in the
{\tt cudacpp.mk} makefile.
In most cases,
notably \none,
\ssef\ and \avxt,
this depends on
the preprocessor
variables defined
by the specific 
{\tt -march} choice.
In the case 
of \foty\ and \fotz,
however, which 
use the same
{\tt -march},
an additional
compiler flag
{\tt -DMGONGPU\_PVW512}
informs CUDACPP
that it should
``prefer vector width 512''
in the \fotz\ mode,
i.e. {\tt zmm} registers.
In addition,
the {\tt -mprefer-vector-width=256} flag
is passed in the \foty\ mode,
to enforce the use
of {\tt ymm} registers
alone, although this
is probably unnecessary.

Finally,
I note that the
names of the CUDACPP
SIMD build modes 
are presently used
also on non-Intel CPUs.
For instance,
on ARM CPUs
the \ssef\ mode uses 
NEON vector instructions
and 128-bit registers;
this is defined via
preprocessor
directives that are
currently being reviewed.
This simplifies
the user interfaces
of the software,
but is clearly
a language abuse:
eventually, it may 
be appropriate to
switch to a better
terminology.
Apart from a
rationalization
of build modes,
in any case,
the port to new
architectures
should not pose
any big challenge.
Some opportunities
for larger SIMD
speedups may arise 
if new hardware architectures
with larger vector 
registers became available,
although this seems
unlikely at the moment.
From the CUDACPP point
of view, 
this would only involve
the definition of
new build modes\footnote{
  As a proof of concept,
  I already 
  prepared
  an implementation
  with the values
  of {\tt neppV}
  that would be
  appropriate to 1024-bit 
  and 2048-bit registers,
  and tested its
  functionality.
  This is currently 
  of no practical utility,
  also because,
  not surprisingly,
  it is slower than
  the existing 
  CUDACPP SIMD modes.  
} 
with larger values
of {\tt neppV},
but the overall
implementation
based on compiler
vector extensions
would remain 
essentially 
the same.

\section{Kernel splitting I:
helicity streams, color sum}
\label{sec:ksp}

As discussed 
extensively
above,
a key parameter of the
CUDACPP ME computational engine
is the number
{\tt nevt} of events 
that are processed in parallel
in one offloading cycle
(in practice: in one call
of the MEK {\tt computeMatrixElements}
function in Table~\ref{code:ihel0}).
On a GPU,
this is simply equal
to the GPU grid size, 
i.e. to 
the product of the 
number of blocks {\tt gpuBlocks}
and of the number of threads 
per block {\tt gpuThreads}.
All relevant multi-event arrays,
both in the Fortran
{\tt madevent} application
and in the CUDACPP 
{\tt Bridge} component
(or in the {\tt check.exe}
standalone application),
must be large enough 
to contain {\tt nevt} events:
the larger the GPU grid size,
in particular, the larger
the RAM 
footprint of 
the application
(which can become 
very large~\cite{bib:ichep2022}
for {\tt madevent}).
This parameter
can be configured at build time 
(and partly at runtime)
in the {\tt madevent} application,
where it is referred 
to as {\tt VECSIZE}~\cite{bib:zw2025},
and is highly configurable in
{\tt check.exe}, 
where {\tt gpuBlocks}
and {\tt gpuThreads}
are 
defined independently
at runtime.

\tabcomplexity{t}

As I briefly hinted above,
one of the main aims 
of the new developments 
presented in this section
is the fact that the 
ME calculation throughput (in MEs/s) 
on a GPU 
is generally quite low 
for small grids,
i.e. small values of {\tt nevt},
and only reaches a peak plateau
for relatively 
large grids,
i.e. large values of {\tt nevt}.
This is an issue that exists 
in CUDACPP since the very beginning 
of our developments:
in Fig.~5 of our vCHEP2021 proceedings~\cite{bib:vchep2021}),
for instance, we had shown
that the ME throughput
for the \ggttgg\ process
only ramps up significantly
with at least 16k events in the grid,
and reaches the peak plateau even later, 
with 128k events in the grid.
In the same paper,
we had also already suggested 
that higher throughputs,
possibly because of lower ``register pressure'',
might be achieved by splitting
monolithic \sk\ kernel into smaller kernels:
two specific ideas that we had mentioned,
in particular, were 
the possible use of 
different GPU threads 
to process different helicities
in the same event,
or the possible use of CUDA graphs
to orchestrate a much larger
number of smaller GPU kernels.
I will come back to both 
of these ideas in the following.

Since the ramp-up 
of ME throughputs
with increasing GPU grid sizes
is one of the main issues
that this new work aims at addressing,
or in any case a very good test
of the effectiveness 
of these code changes,
the results that I 
present
in this paper
will mainly 
consist of 
plots of that sort.
One limitation of this work
is that, due to lack of time, 
I will only show
plots for a single CPU process,
and only for the standalone
application {\tt check.exe}.
This is unfortunate, because
the practical benefits
of any progress in this area
would mainly be
for the full \mgamc\ workflow,
using several {\tt madevent}
applications accessing 
the GPU in parallel.
In the ACAT2022 
presentation~\cite{bib:acat2022}
(see Fig.~1 therein),
I~had also studied 
the ramp-up of throughput
as a function of the GPU grid size
when several {\tt check.exe}
applications are launched in parallel,
which seemed to indicate
that the use of the GPU 
is more efficient in that case.
Those results had been obtained
using software containers
that I had
prepared for the HEP-SCORE
benchmarking project~\cite{bib:bmk};
it would be useful to repeat 
similar tests
in the future,
when updated HEP-SCORE
containers are prepared
using more recent 
versions of CUDACPP,
notably
those presented in this paper.

Concerning my
software development process,
I note that for this new research
I used the same methodology
that I have been following
for all of my work on MG5aMC
during the last four years,
since the 
completion~\cite{bib:ichep2022}
of a fully functional CUDAPP 
code-generating plugin 
in the ``epochX'' 
development workflow
at the end of 2021.
In particular,
I always prototyped,
tested, fixed and optimized
my changes using as baseline
the code-generated software
for one specific physics process,
typically \ggtt\ or \ggttgg.
In the case of this work 
on kernel splitting,
I focused
on the CUDA implementation
but also tested
the SIMD C++ version.
I then committed all changes 
for the given process to git,
until these reached a state 
that I considered reasonably complete.
At that point, I backported
my changes to the code-generating
Python framework,
and regenerated all physics processes.
I then performed larger-scale
functional and performance tests,
also using different hardware
implementations like AMD GPUs,
and iterated until completion.
This development process
has been possible only because,
while the git repository 
of \mgamc~\cite{bib:mg5amc-github}
mainly contains 
the hardcoded components
of the framework and
its code-generating engine,
the git repository of the
{\tt madgraph4gpu}
project~\cite{bib:mg4gpu-github}
contains not only 
the CUDACPP plugin but also 
the code-generated software 
for several physics processes.
In my experience,
this has been a key ingredient
in the development 
of these latest enhancements
to CUDACPP, but also much
more generally of the whole plugin.
In particular, I stress
that I would never have been able
to design and implement
such large code changes directly 
in the code-generating Python code.

\figihelzott{p}

In the following subsections, I 
will 
describe
the first batch of three sets 
of kernel splitting changes 
that I developed,
and which have already
been integrated
in a new production release 
of CUDACPP:
(ihel1) helicity streams;
(ihel2) color sum splitting 
into separate GPU kernels;
(ihel3) color sum host refactoring
with optional BLAS offloading,
including some further
refinements (ihel3p1).
In the final subsection,
I also describe a more recent
set of color sum optimizations
in the SIMD implementation (csm),
which instead are 
not yet merged upstream.
As shown in Fig.~\ref{fig:tikz},
these developments
are sequential,
e.g. 
ihel3 includes ihel2,
which includes ihel1.
The software architecture 
of the \sk\ ME engine
for the ihel1, ihel2 
and ihel3 scenarios
is represented 
in Fig.~\ref{fig:ihel0123},
where they are 
also compared
to the situation before
any kernel splitting, ihel0.
I will not provide
new pseudo-code listings
for these 
kernel splitting
developments, but 
in some cases
I will refer 
to the pseudo-code
for 
version ihel0
in Table~\ref{code:ihel0}
to point out what I changed
or which technical
issues I had to address.
The results of my tests
for these different
versions of the software,
as well as for those
in my second batch of changes
described in the next section,
are given in 
Fig.~\ref{fig:rd90dmf}
for an NVidia V100 GPU at CERN,
in Fig.~\ref{fig:lumidmf}
for an AMD MI200 GPU at LUMI,
and in
Fig.~\ref{fig:rd90dmfsimd}
for an Intel
Xeon Silver 4216 CPU,
for three 
floating point precisions
(double, mixed, float) and
for four 
physics processes 
of increasing complexity
(\ggtt, \ggttg, 
\ggttgg, \ggttggg),
whose relevant parameters
are described in detail
in Table~\ref{tab:complexity}.

\figrddmf{p}

\figlumidmf{t}

\figrddmfsimd{p}

\subsection{Helicity streams (``ihel1'')}

The internal substructure
of the \sk\ ME engine
in the previous version 
of CUDACPP before my kernel
splitting changes (ihel0)
is illustrated schematically
in the top-left diagram
in Fig.~\ref{fig:ihel0123}.
This gives a visual
representation of the
pseudo-code
in Table~\ref{code:ihel0}.
As mentioned previously
and as visible in the diagram,
the two main components
of the calculation
for each event,
namely the computation 
of wavefunctions
and dual amplitudes
for a given color flow
from Feynman diagrams,
and their squaring 
and sum over all color flows,
are performed sequentially 
for all helicities.
This workflow makes it
impossible to split
these two components
into separate kernels,
because the loop over helicities
is effectively inside
the loop over events.

The very first step
in my kernel splitting 
developments 
was therefore, quite naturally,
to reverse this situation
and make the event loop
the innermost loop,
inside an outermost loop 
on helicities.
This was the focus of my work
with Olivier Mattelaer
during the 2022 GPU hackathon
(and is quite likely an idea
that he suggested, in fact).
In practice,
the main change to achieve here
was to turn \sk\ from a
{\tt \_\_global\_\_}
device kernel into a host function,
and to turn instead 
{\tt calculate\_wavefunction}
from a {\tt \_\_device\_\_}
function callable by a kernel
into a kernel itself.
Other computations
also had to be modified:
for instance, the
selections of event-by-event
colors and helicities,
which the \sk\ kernel
was performing outside
the helicity loop
(see Table~\ref{code:ihel0})
have now also been turned 
into separate GPU kernels.
As discussed in Ref.~\cite{bib:ks2024},
by themselves these changes 
already provide a moderate
increase of throughputs
for small grids.

The real breakthrough, however,
came when the parallel
calculations for different
helicities were moved 
to separate CUDA Streams.
In this ``ihel1'' version
of the software,
which is illustrated
in the top-right diagram
in Fig.~\ref{fig:ihel0123},
ME throughputs 
on NVidia GPUs reach
their peak performance
with much smaller grids
than in the previous 
ihel0 version,
for all physics processes
and floating precisions I tested.
The peak throughput themselves
are also increased
by around 10-20\%.
This can be seen
in Fig.~\ref{fig:rd90dmf},
by comparing the blue (ihel0)
and orange (ihel1) curves.
The improvement 
is especially impressive
for complex processes
like \ggttggg,
where peak throughputs
are reached with 
${\cal O}$(100) events 
per grid in ihel1,
as opposed to 
${\cal O}$(10k) 
in the current ihel0.
These improvements from ihel1
are essentially the
single most important
progress described 
in this paper.
Most likely,
the improvement 
comes from the 
increase in 
parallelism of the workflow:
instead of 
a single kernel launch
in each offloading cycle, 
which runs for a long time
because it internally 
loops over helicities,
there are now several,
much shorter kernels 
launched {\em in parallel},
one for each helicity,
in a separate GPU stream
for each helicity.

For AMD GPUs,
where the same solution
was implemented 
using HIP Streams,
the benefits are much
less clear,
as shown 
in Fig.~\ref{fig:lumidmf}:
throughputs increase
with both small and large grids
for complex processes
like \ggttgg,
but for simpler processes
the opposite effect is 
observed\footnote{
  I show no 
  results for \ggttggg\ on AMD GPUs, 
  because
  the code failed to build
  on the system I used
  (``{\tt error: unhandled 
  SGPR spill to memory}'')
  with the ihel0 software.
  It is possible that this
  issue could be addressed
  using the Feynman diagram
  splitting techniques
  in ihel5-ihel6, but 
  I did not attempt this.
  More generally, I stress
  that, while in this paper
  I present results also
  for AMD GPUs, I never attempted 
  any serious optimization
  targeting them 
  specifically, and it is perfectly
  possible that CUDACPP performance
  on this platform
  could be significantly improved
  with relatively little effort.
}.
The code refactoring in ihel1
was also propagated
to the 
C++ code,
and tested on a reference
Intel CPU; 
these showed that the 
throughputs are essentially 
unchanged for all SIMD builds.

\newcommand{\ncolor}{{\tt ncolor}}
\subsection{Color sum as a separate 
GPU kernel (``ihel2'')}

The next logical step
in my kernel splitting developments
consisted
in separating the two main
components of the ME calculation,
namely, (1) the calculation
from Feynman diagrams
of the dual amplitudes
for {\ncolor}
color flows
({\tt jamp}: a vector $J$ 
with {\ncolor} complex elements)
and (2) the color sum.
Initially,
I simply split the
{\tt calculate\_wavefunction}
kernel, which was doing
both computations in ihel1,
into two separate kernels:
{\tt calculate\_jamps},
which calculates the {\tt jamp} $J$,
and {\tt color\_sum},
which 
computes
the quadratic form $J^H(C)J$
for a symmetric real color matrix~$C$
using the vector~$J$ and
its conjugate transpose~$J^H$.
This ``ihel2'' version
of the software
is illustrated
in the bottom-left diagram
in Fig.~\ref{fig:ihel0123}.
One important difference
in this case is that 
the {\tt jamp} variable
is no longer a local variable
inside {\tt calculate\_wavefunction},
as it was in ihel0
(see Table~\ref{code:ihel0})
and ihel1:
instead, it is
now a GPU global
memory buffer
that is 
allocated
outside the MEK component
and is 
accessed,
with the appropriate layout decoding
provided by a new class
{\tt DeviceAccessJamps},
in both the {\tt calculate\_jamps}
and {\tt color\_sum} kernels.

As the plots I prepared
are already packed, 
I decided not to present
the ihel2 results explicitly:
for reasons that will become
clear in the following,
in fact, these
are essentially 
identical to the ``ihel3''
results shown
by the solid-red curve
in Fig.~\ref{fig:rd90dmf}.
As can be seen
by comparing this to
the orange curve for ihel1,
on Nvidia GPUs
the further change
in the software 
from the split
of Feynman diagrams
and color sums
yields
an additional
increase 
of the peak throughput
by 10-20\% for complex processes
like \ggttgg\ and \ggttggg.
For simpler processes,
conversely, it results
in a minor decrease 
of peak throughputs;
in my opinion, 
this is a moderate cost
that can be tolerated,
as speeding up 
complex processes
is more important.
On AMD GPUs,
as shown
in Fig.~\ref{fig:lumidmf},
the additional change
results 
in a minor increase
in peak throughputs
for \ggttgg,
but 
for simpler processes
ihel2 is almost 
indistinguishable from ihel1.
In the CPU
implementation,
throughputs are again 
essentially unchanged 
for all SIMD builds.

\subsection{Color sum dispatcher
to kernel or BLAS (``ihel3'')}
\label{sec:ihel3}

The next step 
of my developments
consisted in
investigating the
possible use 
of the cuBLAS
linear algebra library
for computing 
color sums on GPUs,
instead of using
a CUDA kernel.
One of the main
motivations for this work
was that the current
CUDACPP code only
uses the traditional
CUDA cores,
but a large part 
of the computing power
on recent NVidia GPUs
comes from specialized
tensor cores
designed  
for the matrix algebra
operations used in AI,
and cuBLAS may provide 
a way for CUDACPP
to exploit them.
Developing code for
tensor cores, 
in fact, 
is challenging because
it requires the use of
programming APIs
other than CUDA:
an easier alternative
consists in using
specialized libraries
for AI or linear algebra
that internally use
the tensor core APIs,
cuBLAS being one of them.

The BLAS implementation
I developed is fully integrated
in CUDACPP,
and its functionality
has been extensively
tested both 
in the standalone use case
and in the 
full \mgamc\ workflows.
It was originally
developed for cuBLAS
on Nvidia GPUs,
but it has also been 
ported to AMD GPUs
using the hipBLAS
wrapper for rocBLAS.
Studies
of standalone color sums
with cuBLAS and tensor cores
had already been 
done~\cite{bib:srblas}
in previous years
by 
my colleagues
in the Madgraph 
on GPU project:
the work that I present
in this section, however,
is not based on the code
developed for those studies
and represents a restart
from first principles.

\figcspctall{t}

In practice,
my work on this
``ihel3'' version
of the software
was the following.
To start with,
I encapsulated
the color sum
calculation on GPUs 
in a host function
{\tt color\_sum\_gpu}
(in parallel,
I also created
a {\tt color\_sum\_cpu}
function for the
vectorized C++
version on SIMD CPUs).
To make the software
more modular 
and more manageable,
I also took this
opportunity 
to clean up
the color sum code
and move it to
a separate 
source code file.
The {\tt color\_sum\_gpu}
host function 
is just a wrapper
that may dispatch
the color sum
calculation to
two different 
GPU implementations:
\begin{enumerate}
\item {\tt color\_sum\_kernel}.
By default,
the calculation 
is performed using
a kernel
{\tt color\_sum\_kernel}.
This is essentially
the same code
as in the ihel2 software,
with one minor difference:
the GPU global memory 
layout of the dual amplitudes,
which was an SOA
{\tt jamp[ncolor][2][nevt]}
in ihel2,
is now an SOA
{\tt jamp[2][ncolor][nevt]},
because this makes 
it easier to separate
the real and imaginary
parts of the dual amplitudes
for the BLAS calculation,
and the same layout
is used for simplicity
in both the kernel
and BLAS implementations.
The achieved performance
is represented 
by the ``ihel3''
solid-red curve
in Fig.~\ref{fig:rd90dmf}.
As mentioned above,
these results are
indistinguishable
from those 
of ihel2,
which I therefore decided not
to show this explicitly
to simplify the plot;
the comparison, however,
is available in a 
previous 
preprint of this 
article~\cite{bib:preprintV1}.
\item 
{\tt color\_sum\_blas}.
The second 
implementation
of the color sum
consists in a host
function {\tt color\_sum\_blas}
that internally 
calls the BLAS library.
Specifically,
since the color matrix~$C$
is real and symmetric,
the color sum $J^H(C)J$
over the vector 
of dual amplitudes
$J=A+iB$
may be decomposed as 
\begin{equation}
(A^t-iB^t)(C)(A+iB)
= A^t(C)A + B^t(C)B,
\label{eq:jcj}
\end{equation}
i.e. as the sum
of two quadratic forms
$V^t(C)V$,
where the real 
vector $V$
may represent either
the real part $A$
or the imaginary part $B$
of the complex 
dual amplitude 
vector $J$.
This decomposition 
is also used in
GPU kernel color sums.
Each calculation
involves two steps.
In 
mixed or float
precision,
for instance,
the vector
$(C)V$ is computed
using {\tt cublasSgemm},
while its dot product
with $V^t$ is then
computed using
{\tt cublasSgemmStridedBatched}.
These two functions,
as well as their
double-precision
and their HIP
counterparts,
are called through
their abstractions
via {\tt \#define} 
directives,
in the header-only
approach described
in Ref.~\cite{bib:chep2024}.
The intermediate
results of the first
calculation, $(C)V$,
are stored in GPU
global memory
using an additional
buffer allocated
in the MEK component
(the allocation is done
at runtime
after determining
the number 
of ``good'' helicities
in each physics process,
since the amount
of memory allocated
is proportional
to the number
of good helicities).
As shown 
in the bottom-right diagram
in Fig.~\ref{fig:ihel0123},
the BLAS calculations
for different helicities
are performed 
in separate GPU streams:
technically,
many BLAS handles
are used, each 
associated to a
different stream.
Since the performance
of the BLAS implementation
of the color sum
is generally worse
than that using kernels,
as discussed below,
this is only
available as an option,
which must be explicitly
enabled at runtime
by setting an 
environment
variable\footnote{Set 
environment variable {\tt 
CUDACPP\_RUNTIME\_BLASCOLORSUM}
at runtime to use 
the cuBLAS (on Nvidia GPUs) 
or hipBLAS (on AMD GPUs)
implementation
of color sums.}.
Finally, 
as the BLAS library
contains several 
switches
targeting tensor cores,
I~also added
another environment 
variable\footnote{Set
environment variable {\tt
CUDACPP\_RUNTIME\_CUBLASTF32TENSOR}
at runtime
to encourage cuBLAS 
to use tensor cores
in color sums.
This sets math mode
{\tt CUBLAS\_TF32\_TENSOR\_OP\_MATH}.
}
to encourage BLAS 
to use tensor cores
in the color sum
(TF32 math mode).
\end{enumerate}
The performances
of the cuBLAS
and kernel implementation
of color sums 
on NVidia GPUs
using the ihel3 codebase
are compared
in Fig.~\ref{fig:rd90dmf},
where they are referred
to as ``ihel3b'' (dashed-red)
and ``ihel3'' (solid-red),
the only difference 
between them being
an environment variable
set at runtime.
The picture clearly shows that 
the BLAS implementation 
in the ihel3 codebase
performs much worse 
than CUDA kernels
for the simpler \ggtt, 
\ggttg\ and 
\ggttgg\ physics processes.
For the more complex
\ggttggg\ process,
the situation is
less clear-cut:
for small grids,
the kernel implementation
is faster for all 
floating point precisions;
for large grids, however,
the ihel3 BLAS implementation
eventually becomes
as fast as the kernel
implementation,
and in double and float
precision (but not 
in mixed precision)
it is eventually faster
for very large grids.
This is interesting,
but of not much 
practical relevance,
as most production workflows
would use small grids
to keep event generation jobs
more manageable.
This observation, however,
was the starting point
of my further refinement
of BLAS color sums
in the ``ihel3p1''
version of the code,
described in a later subsection.
On AMD GPUs,
Fig.~\ref{fig:lumidmf}
shows 
that
the kernel version 
of color sums
is always much better
than the corresponding
hipBLAS implementation,
in the ihel3 codebase.

\figcspctblas{t}

\subsubsection{Color sum profiling}
To put 
this work
on color sums into context,
and to better understand
the relative merit
of the cuBLAS and kernel
implementations
in \ggttggg,
I found it useful
to perform some
more detailed profiling
of this calculation.
One of my aims was
to measure the time taken 
by the color sum
as a fraction 
of the total time
taken by 
the ME calculation in \sk, 
in different situations.
In fact,
the motivation for 
many recent 
efforts 
to speed up
the color sum calculation,
such as our development
of the mixed precision 
mode~\cite{bib:acat2022,bib:chep2024,bib:zw2025}
or the work 
I present here on BLAS,
is that the 
profiling~\cite{bib:kiran}
of earlier versions
of \mgamc\ had
shown that this could 
represent up to 60\%
of the total ME
computation for \ggttggg;
however,
more recent
versions of the software,
notably CUDACPP,
have not yet been
profiled in detail.

Initially,
I profiled the code 
by a sampling approach,
using {\tt perf},
but this did not allow 
detailed
color sum profiling
on the GPU.
I therefore made
some additional
modifications
to the ihel3 version
of the code
to instrument it 
with dedicated timers.
Specifically,
I used some timers
based on the {\tt x86\_64}
(Read Time-Stamp Counter)
{\tt rdtsc} 
instruction,
which I had developed
for some previous
profiling work 
on \mgamc~\cite{bib:prof2024}.
This approach 
can provide relatively
accurate results
with a limited ($<$10\%)
overhead.

The results 
of my analysis,
which are shown in 
Fig.~\ref{fig:cspctall}
for the CUDA 
and SIMD C++ backends,
are somewhat surprising:
for \ggttggg,
in the ihel3 version
of the software,
the color sum
implementation in CUDACPP
represents
only 5 to 10\%
of the total time 
to compute the ME
in all SIMD CPU modes,
while in the CUDA 
kernel implementation
this fraction
is around 28\%, 
6\% and 15\%
in double, mixed and 
single precision modes,
respectively.
These results
are particularly
good for the 
mixed precision mode,
which may
indicate that
our 
previous 
optimizations
have managed 
to reduce 
the time taken 
by the color sum 
to a level where
this is no longer
a bottleneck
of the ME calculation.

In Fig.~\ref{fig:cspctall},
the results for 
BLAS color sums
are not shown:
this is 
because they 
heavily depend
on the
number of events 
{\tt nevt}
processed
in each GPU cycle,
which corresponds 
to the size of 
the event vector in BLAS
and to the GPU grid size
for the {\tt calculate\_jamp}
kernels.
These results are 
instead shown
by the two red curves
in Fig.~\ref{fig:cspctblas},
which show that 
in ihel3
the efficiency 
of BLAS color sums
improves
as {\tt nevt} increases,
while for the 
kernel implementation
it is essentially constant.
This plot is consistent
with the trend shown
in Fig.~\ref{fig:rd90dmf}
and is useful 
to better understand it.
The two other black curves
in Fig.~\ref{fig:cspctblas}
refer to the further
improvements in 
the ``ihel3p1'' version
of the software,
described 
below.

In the C++ implementation,
one reason 
for the relatively
low time footprint
of the color sum
may be a set of 
optimizations
that I introduced
in October 2022
after some discussions in the team
(and notably some suggestions
by Olivier Mattelaer),
and never documented
in detail.
To start with,
as discussed above
for BLAS
in Eq.~\ref{eq:jcj}, 
I simplified 
the color sum
on the complex vector
of dual amplitudes
as the sum of two 
quadratic forms
on real vectors.
In addition,
I also rewrote
the multiplication
as one involving
a triangular matrix,
to ensure that
the non-diagonal terms
of the symmetric matrix
are only used once
in the calculation,
with a factor 2.
While writing this paper,
I realized that 
the latter
optimization had 
been applied 
to color sums in C++,
but not yet in GPU kernels:
this is another point
that I later fixed in
the ``ihel3p1'' version,
described further below.

\tabncu{t}

\subsubsection{Tensor core profiling}

As the use of tensor cores
was one of the main motivations
for prototyping
cuBLAS for color sums,
I present 
in Table~\ref{tab:ncu}
the results
of some quick studies
using the NVidia NSight Compute
({\tt ncu}) profiler.
Unlike the results
shown so far in this paper,
these were obtained
using an 
NVidia A100 GPU,
which has more advanced
tensor core features
than the older V100 model.
This test,
performed 
using CUDACPP ihel3
with mixed floating precision,
clearly shows that tensor cores
are used by cuBLAS color sums.
In particular,
the {\tt cublasSgemm} multiplication
$(C)V$ makes significant use
of tensor cores, but only
if TF32 math mode
is explicitly enabled,
as otherwise cuBLAS
seems to prefer 
traditional SIMT kernels.
As for the  
multiplication 
of $V^t$ by $(C)V$
in {\tt cublasSgemmStridedBatched},
two different implementations
are used by cuBLAS
for small and large 
numbers of events,
but neither of them
makes a significant use
of tensor cores.
Other tests
in different configurations,
whose results are not 
presented here in detail,
indicate that cuBLAS color sums
make a significant
use of tensor cores
(irrespective of whether 
TF32 math mode is enabled 
or not) when double
precision is used.
This is somewhat surprising,
as tensor cores
are designed and optimized
for lower precision
calculations in AI.
A more detailed investigation
of tensor cores in the 
CUDACPP cuBLAS implementation
is beyond the scope 
of this paper,
but could be useful
for instance 
in the context 
of the HEP-SCORE
benchmarking project.
Further investigations
of cuBLAS performance
for CUDACPP color sums 
on more recent NVidia GPUs,
with much more advanced 
tensor core features,
could also be useful.

\tabggttthreegcpuCS{t}

\subsection{GPU color sum
further improvements (``ihel3p1'')}

As mentioned above,
while writing the first preprint
for this paper~\cite{bib:preprintV1}
I realised that two 
optimizations
of color sums on GPUs
could still be attempted.
I have implemented both
of them in a later version
``ihel3p1'' of the code,
the results for which 
are represented by
the black-solid
(ihel3p1, BLAS disabled)
and black-dashed lines
(ihel3p1b, BLAS enabled)
in Figures~\ref{fig:rd90dmf}
and~\ref{fig:lumidmf}
for NVidia and AMD GPUs.
These patches have 
already been merged
in production
with the rest 
of my first batch
of code changes: 
the black lines
in the figures,
in other words,
represent the performance
of the current 
latest release
CUDACPP v1.01.01.
As shown in 
Fig.~\ref{fig:rd90dmfsimd},
I also note
that the C++ implementation
in this ihel3p1 release
has essentially the 
same performance
as the previous 
ihel0 release
for all SIMD builds.

The first and simplest optimization
was that, a previously done in C++,
I rewrote the color sum kernel
using a triangular matrix, 
to ensure that the non-diagonal 
terms of the symmetric matrix
are only used once 
in the calculation 
with a factor 2,
rather than twice.
As shown by the bottom-left
and bottom-right plots
in Fig.~\ref{fig:rd90dmf},
this small change
does provide a minor
but visible increase
in peak throughput
in \ggttggg,
for double and single precision,
but not for mixed precision
where the time footprint
of the color sum was already small.
This is also confirmed
by comparing 
the solid-red (ihel3) 
and solid-black (ihel3p1)
curves in
Fig.~\ref{fig:cspctblas},
which shows
that the fraction of time
spent in color sum kernels
is reduced by around 5\%
in both double and single
precision.

The second optimization,
conversely, 
is specific to BLAS
and was motivated
by the previous
observations,
based
on Figures~\ref{fig:rd90dmf}
and~\ref{fig:cspctblas},
that BLAS throughputs
significantly increase
when large vectors are used.
The basic idea, therefore,
was to group together
the color sums for all helicities
as a single BLAS multiplication,
rather than perform a separate
BLAS multiplication for each helicity.
The changes that I implemented
are therfore the following:
first, I removed the
color sum dispatcher
from the helicity loop,
implementing a single
dispatcher for all helicities
rather than one per helicity;
in the default case where
color sum kernels are used,
the dispatcher takes
care of launching separate
kernels for the different 
helicities, in the
corresponding helicity streams;
if BLAS is selected at runtime,
however, the dual amplitudes
from the different helicities
are concatenated 
to form a single much 
larger array of events,
as if the different helicities
for each event were treated
as different events,
and a single BLAS 
multiplication is performed
in the default stream.
This is represented
in the lower diagrams
of Fig.~\ref{fig:ihel6}
(this refers to a later
ihel6 version 
of the software,
which however only
differs from ihel3p1
in the handling 
of Feynman diagrams).
As shown in Fig.~\ref{fig:rd90dmf},
the BLAS throughputs of this
improved ihel3p1b implementation
(dashed-black) are 
much higher than
those of the previous
ihel3b implementation
(dashed-red).
For \ggttggg\ (bottom row),
in particular,
the performance
receives a massive boost:
BLAS color sums are now
better than 
kernels
in double precision,
and have equivalent
performance in mixed 
and single precision.
This can also be seen in
Fig.~\ref{fig:cspctblas},
which shows that 
the fraction of time
spent in the color sum
in \ggttggg\ is
now consistently
in the 5--10\% range
when BLAS is used,
independently of the 
number of events 
in each GPU iteration.
For simpler processes,
however, color sum kernels
still have 
a higher
performance: taking this
into account,
it seems appropriate
to keep color sum kernels
as the default
and BLAS as an optional
alternative.

\subsection{CPU color sum
SIMD optimizations
(``csm'')}
\label{sec:csm}

In this section,
I describe 
some
recent studies
and 
optimizations
of color sums
specifically 
targeting
their 
implementation
on vector CPUs.
This work, which
corresponds to the
orange box
in Fig.~\ref{fig:tikz},
represents my most recent
and final addition
to this paper,
and was only included
after my second 
preprint~\cite{bib:preprintV2}
of it.
Its
main motivation,
in fact,
was a desire 
to better understand
the SIMD speedups
of \ggttgggg\ that
I initially reported
in Ref.~\cite{bib:preprintV2}
and which 
are listed
later on
in Table~\ref{tab:ggtt4gcpuCS}
in Sec.~\ref{sec:2to6cs}.
In particular,
in mixed precision 
mode,
where helicity amplitudes
are computed in double precision
and color sums in single precision,
the SIMD speedups of both
calculations with respect
to the {\tt none} build mode
seem consistent 
with that expected
for double precision
(e.g. 
a factor~2 
from \none\ to \ssef),
while a larger speedup
(a factor~4 
for \ssef)
would be expected
for color sums.
Based on previous observations,
I initially suspected
that this may be due 
to a suboptimal SIMD implementation
of {\tt FPTYPE=m} color sums,
which led me to
investigate them
in more detail.
Eventually, 
I found 
that mixed-precision color 
sums on CPUs
already have a reasonable 
performance,
except in some corner cases 
which I improved.
I understood, instead,
that the reason for 
the lower-than-expected
SIMD speedups
with respect to the \none\ mode
is that the latter 
does use some SIMD 
through auto-vectorization, 
and cannot always 
be taken as a reference 
for no-SIMD performance.
I give more details 
in the following.

To start with,
I analysed 
the absolute time spent
in the Feynman diagram
and color sum calculations
in \ggttggg\ for
different
SIMD build modes 
and floating point precisions,
in the current 
version of CUDACPP (ihel3p1).
The results, which are 
listed 
in the top half of 
Table~\ref{tab:ggtt3gcpuCS},
are consistent with
and complementary to 
those
previously given 
in Fig.~\ref{fig:cspctall},
where only 
the fraction 
of time 
taken by 
color sums
was shown.
The main takeaway 
is that, for 
all SIMD modes
(\ssef\ to \fotz),
the time spent in 
color sums
in mixed precision
is only 10\% to 20\% larger
than in single precision,
and almost a factor 2 smaller 
than in double precision.
This is an indication
that the SIMD implementation
of {\tt FPTYPE=m} color sums
does have 
a reasonably good
performance,
and there is only 
limited room to
further optimize it
to bring it closer
to that of {\tt FPTYPE=f}.
Two other 
observations concern
the \none\ build mode:
to start with,
\none\ color sums 
are twice slower
in {\tt FPTYPE=f}
with respect to
{\tt FPTYPE=f},
while they should
normally be just as fast;
in addition,
{\tt FPTYPE=f}
color sums are
only 30\% slower
for \none\ with
respect to \ssef,
while one would
expect them
to be a factor 4 slower.
It should be noted,
in particular,
that the last two 
observations 
only concern color sums:
in fact,
the Feynman diagram
calculations 
in \none\ mode
all take the same time
for {\tt FPTYPE=d,m,f},
and are around a factor
2,2,4 slower than \ssef,
respectively,
as expected.
Based on 
these observations,
I 
performed
three 
further studies
and optimizations
in a new ``csm'' 
branch of the code,
as described below.

A first optimization
consisted 
in streamlining
the way the actual
multiplication
$V^t(C)V$ is performed
for a given precision
(i.e. using double
precision vectors $V$
and matrices $C$
for {\tt FPTYPE=d}
and single precision
$V$ and $C$ for
for {\tt FPTYPE=m,f}).
As shown in the
bottom half of
Table~\ref{tab:ggtt3gcpuCS},
this change 
speeds up 
all build modes,
and especially
the \none\ build mode,
where {\tt FPTYPE=m}
color sums are twice faster
than they were
and are now as fast
as for {\tt FPTYPE=f}.

A second test
consisted in explicitly
disabling 
auto-vectorization
in the CPU color sum,
using {\tt \#pragma GCC 
optimize("no-tree-vectorize")}.
The results of this test,
shown
in Table~\ref{tab:ggtt3gcpuCS},
clearly show that 
color sums 
are
significantly slower
in this modified
``{\tt novec-cs}'' \none\ mode
than in the default 
\none\ build mode,
and are now
around a factor
2,2,4 slower than \ssef,
respectively,
for {\tt FPTYPE=d,m,f}.
This is an indication 
that the {\tt -march=x86-64}
flag in the \none\ mode
does not completely
disable SIMD operations:
in fact, 
the {\tt gcc} manual~\cite{bib:gccx86}
states that 
{\tt -march=x86-64}
specifies
``a generic CPU 
with 64-bit extensions, 
{\tt MMX}, {\tt SSE}, 
{\tt SSE2}, and {\tt FXSR} 
instruction set support'',
which does allow 
{\tt SSE} and 
{\tt SSE2} 
SIMD instructions
on 128-bit {\tt xmm}
vector registers.
In retrospective,
I recognize that
it may have been
a mistake to base the definition
of the \none\ build mode
on {\tt -march=x86-64} alone,
and it might have
been better to also 
disable auto-vectorization,
if the only purpose was that
of having a no-SIMD
reference scenario. 
On the other hand,
the main point of
the \none\ build mode
is that of testing
software performance
without any explicit
event-level vectorization,
i.e. when all operations
are performed using
scalar {\tt fptype} variables
rather than vectors
{\tt fptype\_v} 
of {\tt neppV} variables.
It is also very interesting
that a significant level
of auto-vectorization
in the \none\ build mode
only seems to happen
in color sums,
not in Feynman diagram
calculations.
This is an indication
that the auto-vectorization
of color sums in
the \none\ mode
exploits color-level
SIMD parallelism,
as it is the multiplication
of the color matrix 
and dual color amplitudes
for a single event
that gets auto-vectorized.
This is different
from what happens
in the \ssef, \avxt,
\foty\ and \fotz\ modes,
where the speedup
comes --- by design ---
from event-level 
SIMD parallelism.

Finally, a third set of
tests and attempted optimizations
I performed,
which is very specific
to {\tt FPTYPE=m}
color sums on CPUs,
concerns not the 
actual calculation
$V^t(C)V$ of the
color sum in single
precision,
but the mechanisms
to interface the 
double-precision outputs
Feynman diagram calculation
as the single-precision
inputs to the color sum,
and the single precision
outputs of the color sum
as the double-precision
inputs to the further
processing of matrix elements.
As I briefly hinted
in a footnote 
of Sec.~\ref{sec:meinternals}, 
the algorithm 
that I implemented
for the calculation
of mixed-precision
SIMD color sums
is the following:
the Feynman diagram
calculation is performed
sequentially 
for two vectors
of {\tt neppV} events,
producing two
{\tt fptype\_v} SIMD vectors
of {\tt neppV} 
double precision 
color amplitudes ({\tt jamp});
an ad-hoc function
{\tt fpvmerge}
converts these two
{\tt fptype\_v} vectors
into one 
SIMD vector
{\tt fptype2\_v}
of 2*{\tt neppV} 
single precision 
color amplitudes;
the color sum is performed 
in single precision,
using event-level
SIMD parallelism;
its output is one
SIMD vector
{\tt fptype2\_v}
of 2*{\tt neppV} 
single precision 
squared matrix elements;
two 
functions
{\tt fpvsplit0}
and {\tt fpvsplit1}
extract the lower
and higher halves
of this vector
as two {\tt fptype\_v} 
SIMD vectors of {\tt neppV} 
double precision 
squared matrix elements.
The current ihel3p1 version
of CUDACPP 
still uses my original
implementation of
the {\tt fpvmerge}
and {\tt fpvsplit0/1}
functions, which is 
based 
on CVE constructors
from hardcoded
initializer lists.
Over time, I 
often suspected that this
implementation was
suboptimal and may
cause a large overhead
of {\tt FPTYPE=m}
color sums over {\tt FPTYPE=f}.
My third optimization
in the 
csm branch 
simply consists
in two new implementations
of the {\tt fpvmerge} function,
one using Intel 
intrinsics~\cite{bib:intrinsics}
and another using
the {\tt std::experimental::simd}
proposed C++ extension.
The results of some tests
of these new implementations
in \ggttggg,
which I do not list in detail,
are actually quite reassuring:
my original implementation
of {\tt fpvmerge} is
just as fast, within 1 or 2\%,
as that using intrinsics,
and both are up to 30\%
faster than that using
experimental~SIMD.

In summary, 
this work was
very useful to better understand
the SIMD performance
of CPU color sums,
clarifying some issues
that had previously 
appeared in \ggttgggg\ tests,
and it 
also resulted
in a few 
optimizations
of SIMD color sums.
The new code is available
in a github tag~\cite{bib:csmtag}
of the csm branch, 
from which I created
a pull request
that I recommend 
to merge upstream.
In this pull request,
the original implementation
of the {\tt fpvmerge} function
based on initializer lists
is kept as the default;
the new versions 
based on intrinsics 
and experimental SIMD
are disabled but 
are still available in the code,
where they may provide
a useful reference
for future developments.

\section{Kernel splitting II:
Feynman diagram groups}
\label{sec:diags}

After implementing
helicity-level parallelism,
separating Feynman diagrams
from color sums and 
porting the latter to BLAS,
the next step 
in my kernel splitting
developments consisted
in trying to further decompose 
the {\tt calculate\_jamps} kernel,
which in ihel3
computes dual amplitudes
from all Feynman diagrams,
into many shorter kernels.
My initial motivation
for attempting this
was to investigate
whether this may lead to 
further throughput increases,
possibly thanks to the 
lower number of registers
used by smaller GPU kernels
(lower ``register pressure'').
Unfortunately,
this strategy did not
allow me to 
achieve further 
throughput increases;
this is probably
because moderately small
Feynman diagram kernels
are still quite complex
and especially come 
at the cost 
of a 
large increase
in GPU global memory access,
which massively 
slows down the calculation.
However,
there is also a second
motivation for
splitting Feynman diagrams,
namely making the code
easier to build and manage 
for complex 
physics processes
with very large numbers
of diagrams.
As described in
this section,
my work in this direction
did achieve 
some interesting progress,
such as the possibility 
to calculate MEs for
\twotosix\ and 
\twotoseven\ processes 
that were previously
non-computable 
in the CUDACPP software.

In the following subsections,
I describe my 
second batch 
of three sets 
of kernel splitting changes,
which I now
also consider ready
to be integrated
in new 
releases of CUDACPP:
(ihel4) one kernel 
per Feynman diagram
in each helicity stream;
(ihel4p1) ditto, with optional 
CUDA graph orchestration;
(ihel5 --- and ihel6 
with {\tt DCDIAG=0}) 
one kernel 
per diagram group
in each helicity stream,
with separate source 
code files;
(ihel6 with {\tt DCDIAG=1}) 
a single kernel
in each helicity stream,
processing all diagram
groups as device functions.
The performance
of the intermediate 
ihel4 step
and 
of the final ihel6p2 step,
for \ggtt\ production
with 0, 1, 2 or 3 
extra final state gluons,
are given in 
Figures~\ref{fig:rd90dmf}
and~\ref{fig:lumidmf}
for NVidia and AMD GPUs.
These plots show
that the single-diagram kernel
strategy of ihel4
is very inefficient,
but that throughputs that
are completely equivalent
to those of the earlier
ihel3p1 software
are recovered
in the newer ihel6p2
if the code is generated
without diagram splitting,
i.e. with a single kernel
processing all Feynman
diagrams as in ihel3p1.
In other words,
the usefulness of ihel6p2
is not that it yields
higher throughputs
for our standard candle processes;
instead, 
its main interest
is that it allows to compute
the more
complex \ggttgggg\ and 
\ggttggggg\ with 4 and 5 extra 
final state gluons,
the results for which
are presented
at the end of this section.

\subsection{Feynman diagrams
as individual 
kernels (``ihel4'')}

\figihelfour{t}

The first step
in my attempt to split the 
calculation 
of all Feynman diagrams in
{\tt calculate\_jamps} kernel
into smaller GPU kernels
consisted, 
not surprisingly,
in encapsulating each
diagram
in an individual kernel.
In practice, rather
than launching
one {\tt calculate\_jamps}
kernel per helicity stream,
which internally
processes 
all Feynman diagrams,
the new code
launches many kernels
{\tt diagram1},
{\tt diagram2},
\ldots {\tt diagramN}
sequentially in
each helicity stream.
In both cases,
the processing 
of Feynman diagrams
must follow
a predefined order,
because for every 
diagram
the algorithm
knows which intermediate
wavefunctions have already
been computed,
and which wavefunctions
must still be computed.
This ``ihel4'' version
of the software
is illustrated
in Fig.~\ref{fig:ihel4}.

Unfortunately,
this version 
of the software 
has a consistently
worse performance
than the previous ones,
both on NVidia and
AMD GPUs,
as shown in 
Figures~\ref{fig:rd90dmf}
and~\ref{fig:lumidmf}.
With respect to ihel3,
for instance,
the ihel4 throughputs
for \ggttggg\ on 
an NVidia V100 GPU
are a factor 3-4
worse than ihel3p1
at their peak
for large grids,
and worse by
much larger factors
for small grids.
In addition, 
ihel4 crashes
for very large grids,
in configurations
where ihel3p1
performs very well.
As shown in 
Fig.~\ref{fig:rd90dmfsimd},
the ihel4 software
in vectorized C++ 
for SIMD CPUs
also performs worse
than ihel3p1,
with degradations
in throughput
up to 40\%
for \ggttggg.

There are at least two
possible explanations 
for the worse GPU 
performance of ihel4:
heavier access to
GPU global memory, and 
kernel launch overhead.
At the time of writing
my first 
preprint~\cite{bib:preprintV1}
for this article,
I wrongly assumed
that the latter was
the most likely culprit.
With hindsight 
and after many additional
tests and developments,
it is now clear 
that the bottleneck
is instead the former.
Kernel launch overhead
can be reduced 
by adopting the CUDA 
graphs~\cite{bib:graphsblog}
technology: this is
what I implemented
in the ihel4p1 software,
described below.
Reducing the impact
of GPU global memory access,
conversely,
required a much more 
radical rethink
of Feynman diagram
splitting,
described 
in the two following
subsections:
first of all, 
by launching kernels
which include groups
of many Feynman diagrams
(ihel5), rather than
individual diagrams;
later on, by moving
back to a single kernel,
which however 
is no longer monolithic
as it internally
calls diagram groups
as GPU device 
functions (ihel6).

\figgraphs{t}

\subsubsection{
Small kernels in
CUDA graphs (``ihel4p1'')}

As discussed
in Ref.~\cite{bib:preprintV1},
I investigated the impact 
of kernel launch overhead
with a quick profiling analysis
of the \ggttggg\ process
using the NVidia Nsight 
Compute tool ({\tt ncu}).
This showed that
the {\tt calculate\_jamps} 
kernel in ihel3,
which computes 1240 diagrams,
has an average 
duration\footnote{{\tt ncu} metric 
{\tt gpu\_\_time\_duration.avg}} 
of 2.6 ms,
while the average 
duration of a kernel
computing a single
diagram in ihel4
is 
between 5 $\mu$s and 40 $\mu$s
depending on the diagram.
This is a problem,
because kernel launch
overheads of the order
of 10 $\mu$s are not 
uncommon~\cite{bib:graphsblog}:
in other words,
the total overhead
for launching 1240
sequential kernels
may easily be
comparable to
or much larger
than the total
kernel execution time.
This observation
was what led me
to investigate
the use of CUDA graphs
to orchestrate
all these
small kernels,
with the idea
that this might
result in a large
performance boost.

The switch to CUDA graphs
was relatively easy
and was achieved
in the ihel4p1 software.
As I did for many other
components of CUDACPP,
I added this feature
as an option, which 
is disabled by default
and must be explicitly 
enabled at runtime 
if desired\footnote{Set 
environment variable {\tt 
CUDACPP\_RUNTIME\_GPUGRAPHS}
at runtime to enable
CUDA graphs (Nvidia GPUs) 
or HIP graphs (AMD GPUs)
orchestration 
of Feynman diagram kernels.}.
In practice, 
I did the following:
the {\tt calculate\_jamps}
host function 
was restructured
so that
it delegates kernel
launching of Feynman diagrams
to a {\tt gpuDiagrams} 
wrapper function;
{\tt calculate\_jamps}
also 
keeps a static set
of pointers that 
are needed to configure
graphs if required;
if graphs are disabled
(default), {\tt gpuDiagrams}
is 
a pass-through
function that simply 
launches each kernel
directly in the relevant
helicity stream;
if graphs are enabled,
however, {\tt gpuDiagrams}
configures
graphs on its first
execution (i.e. if the
static configuration
in {\tt calculate\_jamps}
is empty), 
while on
subsequent executions
it runs kernels
via 
graphs rather
than launching them
individually.

The performance results
for this implementation
are shown 
in Fig.~\ref{fig:graphs},
which compares
three grey (ihel4, no graphs),
cyan (ihel4p1, graphs enabled)
and red (ihel3, single kernel)
curves,
for a single execution (left)
or ten executions (right)
of the workflow.
As expected,
these plots indicate 
that graphs slow down
the workflow when kernels 
are executed only once
(ihel4p1 is worse than 
ihel4 in the left plots),
but they speed it up
when the workflow is 
executed several times
(ihel4p1 is worse than 
ihel4 in the right plots).
However,
Fig.~\ref{fig:graphs} 
also shows that,
even if graphs are used,
launching Feynman
diagrams as individual
kernels is still
much slower
than keeping them
together in a single kernel
(ihel4p1 is much worse
than ihel3 in both plots).
This is a clear indication
that the main problem
in this approach is not 
the kernel launch overhead,
but global memory access:
this is precisely the issue
addressed using diagram groups
in the ihel5 and ihel6
versions of the software,
which are
described in the two next
subsections.

\figihelsix{t}

\subsection{Feynman diagram groups
as kernels (``ihel5'')}
\label{sec:ihel5}

After showing that the impact 
of kernel launch overhead
is relatively limited,
it is thus clear that
the most likely
explanation for 
the worse GPU performance of 
ihel4 with respect to ihel3
is 
memory access.
One technical challenge
in this case,
in fact,
is that the wavefunctions
for all external and internal
particles, 
which are a local variable
{\tt wf} in the 
{\tt calculate\_wavefunctions} 
and {\tt calculate\_jamps} 
kernels in versions ihel0--ihel3
of the code
(see Table~\ref{code:ihel0}),
must now also be stored 
into and retrieved 
from GPU global memory.
In addition,
as previously mentioned
in a footnote,
the existing implementation
of helicity amplitude functions 
like {\tt FFV1\_0}
essentially constrains
the wavefunction layout
in GPU global memory to be
an Array-Of-Structures (AOS)
{\tt wf[nevt][6][2]},
which is suboptimal
as it does not allow
coalesced memory access
as an SOA would do.
This further increases the
pressure on memory
bandwidth, 
and may represent an additional
cause of slowdowns.

\figdpgscan{t}

The next step in my kernel
splitting developments,
as a consequence,
was an attempt to reduce 
the frequency 
of GPU global memory access,
while still 
partitioning the full
list of Feynman diagrams
into several smaller 
and more manageable kernels.
My 
``ihel5'' implementation,
in particular,
consisted in
grouping together
many Feynman diagrams
in kernels 
of intermediate sizes.
In doing this,
I kept unchanged
the order in which
Feynman diagrams
are sequentially computed.
As I did not reorganise
the calculation
to allow some level 
of parallelism,
I did not expect that
these developments could
yield any throughput increases.
Their interest, instead,
was mainly that they
could make the software
more modular and manageable,
possibly leading
to faster build times
and possibly allowing
ME calculations
with a large number
of final state particles.
This goal was indeed
achieved, as detailed below.

My ihel5 implementation 
of diagram groups is 
configurable
according to two parameters:
one ``{\tt dpg}'' parameter
defines the maximum
number of Feynman diagrams
in one diagram group,
while another 
``{\tt dpf}'' parameter
defines the minimum
number of Feynman diagrams
in one source code file.
Currently these parameters
are only managed
through special scripts,
but eventually they 
could become standard
\mgamc\ runcard parameters.
In practice,
diagram group kernels
are implemented as follows:
for each diagram group,
the code generator keeps 
track of which wavefunctions
the diagram group needs
in its calculations,
and of which wavefunctions
the diagram group updates;
at the beginning 
of all diagram groups
except the first one,
all necessary
wavefunctions
are read from global memory
and copied 
to local variables
in GPU registers;
incidentally,
this makes it possible
to define a better SOA
layout in global memory
for coalesced memory access,
while maintaining 
the legacy AOS layout
{\tt wf[nevt][6][2]}
in local variables
to avoid the need
to restructure all calculations;
at the end 
of all diagram groups
except the last one,
any updated wavefunctions
are written back 
to GPU global memory;
at the end of all
diagram groups,
the {\tt jamps} 
dual amplitudes 
in GPU global memory
are also updated
from the results
of those computed
using local variables.
This design implies that,
the lower the number
of diagram groups,
the lower the frequency
of GPU global memory access.
In particular,
in the special case where
the {\tt dpg} is larger
than the total number 
of diagrams 
in the physics process,
only one diagram group
is created:
this is executed
as a single kernel,
which does not need
to read wavefunctions
at the beginning
or to write them back
at the end,
so effectively
this is a single-kernel
implementation
equivalent to that in ihel3,
and indeed 
has the same performance,
as previously noted.

For reasons 
clarified
below, 
ihel5 is equivalent
to the default
{\tt DCDIAG=0}
implementation
in the ihel6 version 
of the software, 
which is 
illustrated
in the upper-left
part of Fig.~\ref{fig:ihel6}.
As a consequence,
in this paper
I do not show explicit
results for 
the ihel5 software,
but only those from ihel6.
The impact of the 
{\tt dpg} parameter
on ihel5 performance
for \ggttggg,
in particular,
is shown by the 
left plot in 
Fig.~\ref{fig:dpgscan},
derived
from the default
{\tt DCDIAG=0}
version  of ihel6.
The plot is 
interesting
as 
it shows 
that peak performance,
equivalent to that of ihel3,
can only be achieved
with a single kernel
that processes all diagrams.
Even a minimal splitting
of the 1240 diagrams
into two diagram groups
({\tt dpg}=1000),
in fact, comes
at the cost of 
a very large
performance hit
for small GPU grids
and of a 10\% reduction
in peak performance 
for large grids
(with the additional
problem that the
program crashes 
due to limited memory
when the GPU grids 
are too large).
This clearly shows
that GPU global memory 
access is very expensive.

\subsection{Diagram groups 
as device functions (``ihel6'')}

Having understood that
processing all
Feynman diagrams
in a single kernel
is better for performance,
but still wishing
to split up complex
physics processes
into smaller
and more manageable
software chunks,
my next step 
of development
was a relatively minor,
but rather consequential,
modification of ihel5
diagram groups.
As I had 
already prepared
the code generator
to define diagram groups,
I simply 
transformed
diagram groups 
into
GPU device functions
and 
turned
{\tt calculate\_jamps}
into
a single 
GPU kernel that calls 
them all sequentially.
In this ``ihel6''
software version,
illustrated in the 
upper-right part 
of Fig.~\ref{fig:ihel6},
the big difference
is that all wavefunctions
are 
defined 
as local variables
of GPU device functions,
stored in GPU registers,
and do not need to
be read from or stored
into GPU global memory.
Sticking to my general
strategy of allowing
the choice between 
different alternatives,
the code generated
in ihel6 supports
both the ihel5-like mode,
where diagram groups
are GPU kernels,
and the new ihel6-like mode,
where diagram groups
are device functions.
The choice between the
two implementations
can be made at build time,
by passing {\tt DCDIAG=1} 
to specify that 
device code diagrams
should be used,
or by keeping the default
value {\tt DCDIAG=0}.
The name of this variable
is also meant 
as a reminder that,
for technical reasons
related to the
handling of coupling
constants and other 
physics parameters,
GPU relocatable device code
(i.e. the CUDA 
{\tt rdc} option)
is needed in the
{\tt DCDIAG=1} 
implementation.

The impact on performance
of the {\tt DCDIAG=1}
approach is quite interesting,
as shown in the 
right plot in 
Fig.~\ref{fig:dpgscan}.
As in the case
of {\tt DCDIAG=0},
performance is quite poor
for low values of {\tt dpg},
i.e. when there
are many diagram groups.
Unlike {\tt DCDIAG=0},
however, peak performance,
only slightly lower
than that of ihel3,
is reached for 
intermediate values
of {\tt dpg} between
100 and 1000.
For {\tt dpg}=2000,
instead, i.e. with a
single diagram group,
performance 
degrades again.
In other words,
there seems to be 
a sweet spot
for diagram groups
of intermediate sizes,
where {\tt DCDIAG=1} 
performance 
is reasonably good.
This is especially
interesting because
it is the same 
sweet spot 
where source code files
and their builds
appear to be easier
to manage.
This is further
discussed 
in the following,
in the case
of \ggttgggg\ and~\ggttggggg,
which can now be computed
by splitting the
calculation into 
several diagram groups.

\tabggttfourgcpu{t}

\tabggttfourggpu{t}

Before moving on
to these more complex physics
processes,
I add one final comment
about our standard candles,
\ggtt, \ggttg,
\ggttgg\ and~\ggttggg.
As shown in
Figures~\ref{fig:rd90dmf}
and~\ref{fig:lumidmf}
for NVidia and AMD GPUs
and in 
Fig.~\ref{fig:rd90dmfsimd}
for an Intel CPU,
what I want to stress is that
the ihel6 line of
software versions,
and more specifically
the ``ihel6p2'' version 
described below,
performs just as well
as the current ihel3p1
release,
or even slightly better,
when these processes
are generated 
using a single
diagram group.
In other words,
the new diagram splitting
functionality in ihel6p2
offers the benefit
of allowing computations
of more complex processes,
without compromising
the well established
performance for
simpler processes.

\tabggttfourgcpuCS{t}

\subsection{Studies of 
a \twotosix\ process:
\ggttgggg}
\label{sec:2to6}

Using the new features
in CUDACPP ihel6,
notably its diagram
splitting capabilities
and the possibility
to treat diagram groups
either as independent kernels
or as device methods on GPUs,
I was able to test
and obtain some 
throughput results for
complex \twotosix\ and 
\twotoseven\ physics processes 
such as \ggttgggg\ and \ggttggggg.
I note that some studies
of \ggttgggg\ had already
been performed in the past
using the previous ihel0
software with a monolithic
\sk\ kernel, by myself and 
other members
of the development team,
but they had immediately 
appeared extremely challenging.
In my recollection,
in 2023 I had been able
to build the C++ code
for \ggttgggg,
but only using 
the {\tt clang} compiler,
and this 
had taken 32 hours.
The C++ build using {\tt gcc},
conversely, had crashed
with an internal error,
while the CUDA build
had gone on for more than
one week before 
I decided to stop it.

\newcommand{\dpg}{{\tt dpg}}
Diagram splitting
dramatically improves
this situation,
not only for the GPU
implementation that I was
initially targeting,
but also for the vectorized
C++ implementation.
As shown 
in Table~\ref{tab:ggtt4gcpu},
the C++ build 
of \ggttgggg\ with {\tt gcc}
successfully completes
in less than 4 minutes
if the number \dpg\ of 
diagrams per group
is kept below 1000,
and it takes longer
but still completes
in 30 minutes
if \dpg\ is increased 
to 10000, which implies
only two diagram groups
for this specific process.
Interestingly, instead,
the {\tt gcc} build 
fails
with an internal
compiler error if
all 15495 diagrams
are included in the
same function:
this is exactly the same 
situation that I had faced
earlier on with 
the ihel0 software.
Table~\ref{tab:ggtt4gcpu}
is also very interesting 
because it shows 
that C++ runtime
throughputs are significantly
degraded with fewer than
100 \dpg, 
but they are essentially
at the their peak plateau
with 1000 \dpg: 
the runtime
performance with 10k 
\dpg\ is 
only negligibly
better than that 
with 1k \dpg,
at the cost 
of a much longer
build time.
This observation,
which seems to fall in line
with that previously
made about \ggttggg\ for GPUs,
is especially interesting
because it seems to suggest
that, if the C++ build of the 
100k \dpg\ code (i.e. that
with a monolithic function)
had succeeded, its runtime
performance might have been 
only negligibly better
than that with 10k 
or 1k \dpg.
In other words,
also for the C++
implementation of
\ggttgggg\ (similarly to what 
was previously 
observed for the GPU
implementation of \ggttggg),
there seems to be 
a sweet spot around
1k \dpg\ where 
software builds 
are fast and manageable,
and runtime performance
is satisfactory.
It should 
be noted,
in particular,
that the SIMD speedup
observed with \dpg\ 1k 
or 10k is consistent 
with the maximum
achievable in this 
configuration,
using mixed 
floating point precision:
on this Intel Xeon Gold 6326 CPU
with two FMA units,
one would expect
a maximum SIMD speedup
of x8 for Feynman diagrams
and x16 for the color matrix,
and an overall speedup
somewhere in between,
which is consistent
with the observed
speedups around x9.
This very good 
scaling seems to suggest
that this calculation
is not limited 
by memory access
on the CPU.

\tabggttfourggpuCS{t}

For what concerns 
the GPU implementation,
diagram splitting
is also a game changer,
allowing the first 
(to my knowledge)
CUDACPP calculation 
of matrix elements on GPUs
for \ggttgggg.
As shown 
in Table~\ref{tab:ggtt4ggpu},
CUDA builds
succesfully complete
in less than 10 minutes,
in both modes
{\tt DCDIAG=0} 
and {\tt DCDIAG=1},
if the number of 
diagrams per group
is kept below 100.
If \dpg\ is increased to 1000, 
the {\tt DCDIAG=0} 
build succeeds in 3h,
while the {\tt DCDIAG=1}
build fails because
it requires too much memory.
It was specifically this
point that motivated me
to get rid of templates,
in an attempt to simplify
the task of the CUDA compiler,
but unfortunately 
the {\tt DCDIAG=1}
build continues to fail
after removing templates
(ihel6p2).
This is 
a pity, 
because the 
{\tt DCDIAG=1} build
clearly gives higher
peak throughputs,
and reaches them 
with smaller grid sizes,
compared to {\tt DCDIAG=0}
(this is not too surprising,
as the latter
involves much heavier
read/write access
of wavefunctions 
and dual amplitudes
in GPU global memory,
while the former
only uses GPU registers
in most cases).

\subsubsection{Profiling 
of color sums in \ggttgggg}
\label{sec:2to6cs}

The results I have presented
for \ggttgggg\ in
Tables~\ref{tab:ggtt4gcpu}
and~\ref{tab:ggtt4ggpu}
refer to the overall
throughput for 
computing MEs,
which includes the
calculations of 
both Feynman diagrams
and QCD color sums.
Since kernel splitting
into diagram groups
and the two different
{\tt DCDIAG} options
are only relevant 
to the former
and not the latter,
I 
found it useful
to separately profile
these two components.
In doing this, I also
tested the impact
of the various color sum
options in \ggttgggg.
I only studied a limited
number of scenarios,
including in particular
those which had achieved
the highest overall
throughputs in the 
tests presented above.

\tabggttfiveg{t}

The results for
the SIMD implementation,
shown in 
Table~\ref{tab:ggtt4gcpuCS},
show that the 
choice of \dpg\ only
affects the calculation
of Feynman diagrams,
as expected.
The fraction of time 
spent in color sums
is around 10--12\%
in the \dpg\ scenario
with the fastest
calculation of
Feynman diagrams:
this number is 
somewhat lower than
I had expected,
but it is still 
significantly larger
than the 5--10\%
observed for the simpler
\ggttggg\ process.
The SIMD speedups
of the color sum alone,
which is computed 
here in single precision
({\tt FPTYPE=m}),
are also lower than
I had expected:
for instance, a factor 
around 9 is observed for the 
512z implementation,
whereas a x16 speedup
could in principle be
expected for this 
single precision calculation.
This observation
was essentially 
the main motivation
for the additional
tests and optimization
of SIMD color sums
that I have described
in Sec.~\ref{sec:csm}.
As discussed therein,
it is now clear that 
this lower-than-expected
speedup does not indicate
a suboptimal implementation
of SIMD color sums,
but rather the fact 
that the \none\ build mode
does not represent 
a good no-SIMD 
reference scenario
for this type of
comparisons,
as the color sum
actually benefits from
auto-vectorization
in that case.

The results for
the GPU implementation,
listed in 
Table~\ref{tab:ggtt4ggpuCS},
also show that
the large throughput
variation previously noted
in Table~\ref{tab:ggtt4ggpu}
for different choices 
of \dpg\ and {\tt DCDIAG}
only come from the 
calculation of 
Feynman diagrams.
In the fastest scenario
(100 \dpg\ and {\tt DCDIAG=1}),
in particular,
color sums for 
relatively large 
grids with 512 events
represent 20\% 
of the time needed
for the ME calculation
in their default implementation 
with GPU kernels.
What is very interesting,
however,
is that using cuBLAS
in this case speeds
up color sums by almost 
one order of magnitude,
reducing them to only 
3\% of the total time
taken to compute MEs.
This provides additional
evidence that cuBLAS
may be the best choice
for the color sums
of particularly complex
processes, even if 
it should not be used
as the default as
its performance
for simple processes
is very poor.

In this context,
it should also be noted
that cuBLAS color sums
provide physics results
with the same accuracy
as those of GPU kernels,
except in TF32 math mode.
In Tables~\ref{tab:ggtt4gcpuCS}
and~\ref{tab:ggtt4ggpuCS},
in particular,
I explicitly included
the average ME value
of my simple tests:
this is 
the same to seven 
significant digits
in all scenarios
I considered
both on CPUs and GPUs,
when starting from the
same random numbers;
the only exception is
cuBLAS/TF32,
where discrepancies
of order 10$^{-4}$
are observed.
As discussed in
Sec.~\ref{sec:ihel3},
the TF32 option
is a switch that
I introduced
to encourage cuBLAS
to use tensor cores.
Table~\ref{tab:ggtt4ggpuCS}
shows that
this does provide 
a moderate speedup
of color sums
by approximately 30\%,
but it seems quite clear 
that the overall benefit
(reducing the color sum
fraction from 3\% to 2.3\%)
is not worth the 
cost of a loss
in physics precision.

\subsection{Studies of 
a \twotoseven\ process:
\ggttggggg}
\label{sec:2to7}

Taking this one step further,
I used the diagram splitting
functionality of ihel6p1
to also perform
a first quick
analysis of the 
\twotoseven\ process \ggttggggg.
Since code generation 
for this process takes 30h,
I only attempted one
configuration that I thought
could be promising,
using 1000 diagrams per group.
It turns out that,
both for the C++ 
and CUDA builds,
the main challenge 
in this case came not
from Feynman diagrams
but from the size 
of the \mbox{5k x 5k} color matrix,
which in CUDACPP
is defined as a {\tt constexpr}.
In the case of C++,
I was able to address
the issue by modifying
the {\tt constexpr} limit
in the compiler:
this was enough for 
the build to succeed
(in less than 1h),
and I was then able
to obtain the 
first
CUDACPP results from C++
for a \twotoseven\ process,
which are shown 
in Tab.~\ref{tab:ggtt5g}.
In the case of CUDA,
I used a similar trick
to successfully complete
compilation
(replacing {\tt constexpr}
compile-time evaluations
by runtime {\tt const}
evaluations),
but linking failed with the 
{\tt relocation truncated to fit}
error, which signals
that the executable
would be too large 
for the linker to handle.
I tried several options
to bypass this issue,
but none of them succeeded.

Concerning runtime performance
of the C++ code, the results
in Tab.~\ref{tab:ggtt5g}
are somewhat puzzling,
as the observed SIMD speedups
are much lower than one 
would expect.
Unlike the \ggttgggg\ calculation, 
this seems
to suggest that CPU memory
access may be a limiting factor
for this calculation.
In the table,
I also provide 
the detailed profiling 
of the Feynamn diagram
and color sum components,
derived with the same tools
used for simpler processes:
the fact that color sums
have a much better SIMD speedup
than Feynman diagrams,
and that their overall footprint
is relatively low 
(around 5 to 15\%)
is an interesting
and somewhat surprising
indication that the Feynman
diagram calculation
is the bottleneck
for such a complex
calculation.

Finally, I should
note that this might well
be regarded 
as an academic exercise.
To achieve the 
calculation of 
processes with 
a large number of
final state gluons,
I believe that 
Berends-Giele recursion 
relations~\cite{bib:bg}
represent a better approach
to explicit Feynman diagrams,
which has already been
successfully ported
to GPUs by different generator 
teams~\cite{bib:bg-bothmann,bib:pepper,bib:martinez}.
I still find it interesting, 
in any case, to explore
how far one can push
calculations based 
on Feynman diagrams,
to identify some
practical bottlenecks
in the software 
and hardware we use,
but also because
Feynman diagram calculations 
may provide a complementary
tool to cross-check 
other approaches.
Diagram splitting,
in particular, appears
to be a very promising
approach to achieve
more complex calculations
using vectorized C++ code.

\subsection{Discussion}

To conclude,
taking into account
that the ihel6p2 software
has the same 
performance 
without diagram splitting
as the latest 
CUDACPP release
based on ihel3p1
for our 
standard candles
such as \ggttggg,
but that it also allows
the calculation of 
more complex
physics processes
such as \ggttgggg\ and
\ggttggggg\ when
diagram splitting 
is enabled, 
my recommendation 
is that this work
should also be 
merged upstream and integrated 
in a new CUDACPP release.
I believe that the 
new features offer 
many opportunities
for further studies,
certainly from a software 
point of view but 
also from a physics
perspective.
A few technical details
would still need to be sorted
out, such as the visibility
of the \dpg, {\tt dpf}
and {\tt DCDIAG} parameters 
through standard runcards,
or the preparation
of new CI tests
including diagram splitting.
To ease future maintenance,
one may also choose to exclude
some of the features
I developed:
in particular, I think
that testing CUDA graphs 
has been a very useful
exercise, but the 
benefits of this 
technology for 
Feynman diagrams
appear quite limited,
and it might therefore
be appropriate 
to remove this specific
feature before merging
this work upstream.

In addition,
in case of stronger
interest in the 
calculation of
more complex processes.
one issue that 
may deserve some
minor additional 
development
is the following:
currently,
helicity streams
are used only
for computing MEs
after helicity filtering,
but helicity filtering
itself, i.e. the
selection of the good
helicities with
non-zero ME contributions,
is done sequentially
for the various helicities.
As a consequence,
helicity filtering
is extremely long
for the \twotosix\ and
\twotoseven\ processes
I studied, and for this
reason I added 
a hack\footnote{
  Set {\tt CUDACPP\_RUNTIME\_GOODHELICITIES=ALL}
  to bypass 
  helicity filtering.
  The ME calculation
  will then 
  use all helicities.
} to bypass
helicity filtering
(this made sense
for \ggttgggg\ and
\ggttggggg\ as
all helicities do
contribute to the
ME calculation 
in that case).
A simple solution,
which would be very
simple to develop
but I skipped for 
lack of time,
would be to use parallel 
helicity streams
also during the
helicity filtering step.
An even better 
approach might be
to study 
how many streams 
are actually useful,
and limit 
them
to a fixed number,
properly chosen
taking into account the 
GPU hardware specs.

\section{Summary and outlook}
\label{sec:end}

In summary,
in this paper
I have described
my work on two large 
batches of 
optimizations
of the CUDACPP plugin 
for the Madgraph5\_aMC@NLO 
(\mgamc) generator
on GPUs and vector CPUs.
The new approach mainly
consists in splitting
the calculation 
of matrix elements (ME),
which had been so far performed
using a single large GPU kernel,
into smaller kernels.
The software I developed 
and the studies I presented
have all been been performed
using the CUDACPP 
standalone application,
but are expected to
provide benefits 
also and especially
in the full MadEvent-based
\mgamc\ generator workflow.
This work has mainly 
targeted NVidia GPUs, 
but benefits
the CUDACPP implementation
for CPUs, too. 
It has also been
ported to AMD GPUs,
but without any attempt
to understand or improve
its performance
on this platform.
I have also taken 
the opportunity of this paper
to describe more in detail 
some features of the 
CUDACPP software that 
had not yet been documented,
especially those most
relevant to these new developments.

The first large batch of changes,
which includes three sets
of optimizations,
has already been merged upstream
and integrated 
in a new production 
release of the CUDACPP
software~\cite{bib:cudacpp364},
in collaboration with 
my colleagues.
The first optimization,
which parallelizes the
calculation for different
helicities of the initial
and final state particles
to different 
CUDA Streams,
achieves a reduction
by one to two orders
of magnitude in the number
of events that must be
computed in parallel
to make an efficient use 
of the GPU.
This is interesting
from a user perspective,
as it should
allow event generation 
jobs on GPUs using
smaller numbers 
of events than
the current CUDACPP.
The second and third 
optimizations reorganize
the calculation of
color sums as
a separate GPU kernel,
which is taken as
the default implementation,
and also offer the possibly 
to perform it using 
GPU-specific BLAS libraries.
On NVidia GPUs,
in particular,
the cuBLAS implementation
has been optimized 
to a level where
it is faster than kernels
for moderately 
complex processes
such as \ggttggg;
this is also interesting
as it may offer a mechanism
for CUDACPP to exploit
tensor cores on NVidia GPUs,
which are otherwise unused.

The second large 
batch of enhancements,
which includes 
three further sets
of software changes,
was still work in progress
at the time of my 
first 
preprint~\cite{bib:preprintV1}
of this paper,
but it has now been
completed and I also
recommend its inclusion
in a new production
release of the software.
It is available 
as a github
tag~\cite{bib:ihel6tag},
and a pull request 
for its merge upstream
has been created.
These three 
sets of changes
all consist in different
ways to split up
the computation of
Feynman diagrams
into separate GPU kernels.
The fourth development,
which uses a different
GPU kernel 
for each Feynman diagram,
has a very poor performance,
even when the workflow
is orchestrated
by CUDA graphs,
because it implies
very heavy access 
to GPU global memory.
The fifth development
allows
the definition
of kernels that
compute larger groups 
of Feynman diagrams,
possibly defined
in separate source
code files. 
The sixth development
goes back to a single
GPU kernel for 
all Feynman diagrams,
which however delegates
their calculation 
to different device functions
for separate diagram groups.
All these changes
are fully functional and
have been extensively tested
on all platforms.

While this second
line of R\&D
started off as an attempt
to further improve
throughputs 
for our standard candles
such as \ggttggg,
it now seems clear to me
that these 
processes are best 
computed
using the traditional
CUDACPP approach
of a single GPU kernel
and a single diagram group
for all Feynman diagrams,
since they are simple
enough to afford this:
the same approach 
and the same throughputs
are still achievable
using the new software,
which can be configured
to produce a single
diagram group.
The interest of the new 
feature, conversely,
is in much more complex
physics calculations
such as those required
by \twotosix\ and
\twotoseven\ processes.
I have shown,
in particular, that 
diagram splitting 
makes it
possible to compute
matrix elements
for \ggttgggg\ (which
involves over 15k 
Feynman diagrams) in
both the GPU and vector CPU
implementations of CUDACPP,
and for \ggttggggg\ (over 230k
diagrams) in
the latter.
The code I used 
for the \ggttgggg\ tests 
that I 
presented 
is also available as a github
tag~\cite{bib:ihel6tag4g},
which I created only 
for documentation purposes
and not as a patch
to integrate upstream.
Another github 
tag~\cite{bib:csmtag}, 
which instead
I recommend 
to merge upstream,
includes a few minor 
optimizations of SIMD color sums
that I developed
in the context of these tests
of \twotosix\ and
\twotoseven\ processes.

Further optimizations
of the internal CUDACPP engine
for computing matrix elements,
possibly extending the work 
I presented here,
are of course possible.
The clean separation
of Feynman diagrams
and color sums 
in this new implementation,
for instance,
makes it possible
to imagine scenarios
where the former
is performed on vector CPUs
and the latter on GPUs
using cuBLAS for particularly 
complex processes.
In my opinion, however,
the ME calculation
for LO processes
in CUDACPP
has now been optimized
to a level where
there are no obvious
opportunities for
further significant
throughput speedups.
Other areas of work 
have higher priority,
such as in particular 
the integration 
of LO CUDACPP into 
the upcoming Madgraph7
with an enhanced
phase space sampling based 
on MadNIS~\cite{bib:madnis},
or the extension
of the CUDACPP
infrastructure
to NLO physics 
processes~\cite{bib:zwnlo}.
The work I presented
in this paper will
be relevant in both
of these activities.

\section*{Acknowledgements}

I thank the 
\mgamc\ and the Madgraph on GPU
development teams for our fruitful
collaboration on these projects.
In particular,
I warmly thank 
Olivier Mattelaer for our initial
discussions and our coding together
for the new developments 
shown in this paper
during the 2022 GPU Hackathon,
and more generally for our
effective
collaboration 
over many years on 
the Madgraph on GPU 
developments.
This project would not have been
possible without him.
I thank
Daniele Massaro
for useful discussions 
about SIMD support in CUDACPP
and for his help in the
integration of this work
in new production releases
of the CUDACPP software.
I warmly thank Domenico Giordano
for many useful discussions
about hardware benchmarking.
I am grateful to
the organisers 
and our mentors 
at the CSCS GPU Hackathon 
in Lugano in September 2022,
and I thank 
the other participants from our team,
Stephan Hageb\"ock, Stefan Roiser 
and Zenny Wettersten,
for 
useful discussions. 
I kindly 
acknowledge the use 
of LUMI HPC resources 
under project 
465001592 
(CERN / Madgraph GPU porting,
EHPC-DEV-2024D11-007)
to derive the results I showed for
AMD GPUs.

\newcommand{\arxiv}[1]{\href{https://arxiv.org/abs/#1}{arxiv:#1}}
\newcommand{\doi}[1]{\href{https://doi.org/#1}{doi:#1}}
\newcommand{\doiref}[2]{\href{https://doi.org/#2}{#1}}
\newcommand{\Href}[2]{\href{#2}{#1}}
\renewcommand{\url}[1]{\href{#1}{#1}}

\end{document}